\documentclass[usegraphicx,usenatbib,useAMS]{mn2e} 
 
\usepackage{times}
\usepackage{graphicx,natbib}
\usepackage{amssymb}
\newcommand{\be}{\begin{equation}}
\newcommand{\ba}{\begin{eqnarray}}
\newcommand{\ee}{\end{equation}}
\newcommand{\ea}{\end{eqnarray}}

\def\lesssim{\mathrel{\hbox{\rlap{\hbox{\lower4pt\hbox{$\sim$}}}\hbox{$<$}}}}
\def\gtrsim{\mathrel{\hbox{\rlap{\hbox{\lower4pt\hbox{$\sim$}}}\hbox{$>$}}}}

\def\simless{\mathbin{\lower 3pt\hbox
   {$\rlap{\raise 5pt\hbox{$\char'074$}}\mathchar''7218$}}}   
\def\simgreat{\mathbin{\lower 3pt\hbox
   {$\rlap{\raise 5pt\hbox{$\char'076$}}\mathchar''7218$}}}   

\newcommand{\url}{\tt}%

\begin{document}

\title[Observability of Ly-$\alpha$ Emitters during Reionization]{The 
Effect of the Intergalactic Environment on the Observability of 
Ly-$\alpha$ Emitters During Reionization} 

\author[I. T. Iliev, et al.]{Ilian T. Iliev$^{1,2}$\thanks{e-mail:
    iliev@physik.uzh.ch}, Paul R. Shapiro$^3$,
Patrick McDonald$^2$, Garrelt Mellema$^4$, Ue-Li Pen$^2$ 
\\
$^1$ Universit\"at Z\"urich, Institut f\"ur Theoretische Physik,
Winterthurerstrasse 190, CH-8057 Z\"urich, Switzerland\\
$^2$ Canadian Institute for Theoretical Astrophysics, University
of Toronto, 60 St. George Street, Toronto, ON M5S 3H8, Canada\\
$^3$ Department of Astronomy, University of Texas, Austin, TX 78712-1083,
  U.S.A.\\
$^4$ Stockholm Observatory, AlbaNova
University Center, Stockholm University, SE-106 91 Stockholm, Sweden}

\date{\today} \pubyear{2008} \volume{000} \pagerange{1}
\twocolumn \maketitle\label{firstpage}

\begin{abstract} 
Observations of high-redshift Ly-$\alpha$ sources are a major tool for
studying the high-redshift Universe and are one of the most promising ways 
to constrain the later stages of reionization. The understanding and
interpretation of the data is far from straightforward, however. We discuss
the effect of the reionizing intergalactic medium on the observability of
Ly-$\alpha$ sources based on large simulations of early structure formation
with radiative transfer. This takes into account self-consistently the
reionization history, density, velocity and ionization structures and
nonlinear source clustering. We find that all fields are highly anisotropic
and as a consequence there are very large variations in opacity among the
different lines-of-sight. The velocity effects, from both infall
and source peculiar velocity are most important for the luminous sources,
affecting the line profile and depressing the bright end of the luminosity
function. The line profiles are generally asymmetric and the line centers of
the luminous sources are always absorbed due to the high density of the local
IGM. For both luminous and average sources the damping wing effects are of
similar magnitude and remain significant until fairly late, when the IGM is
ionized between 30\% and 70\% by mass. 

The ionizing flux in the ionized patch  surrounding a high density peak is 
generally strongly dominated, particularly at late times, by the cluster of
faint sources, rather than the central massive galaxy. Our results reproduce
well the observed mean opacity of the IGM at $z\sim6$. The IGM absorption does
not change appreciably the correlation function of sources at high redshift.   
Our derived luminosity function assuming constant mass-to-light ratio provides
an excellent match to the shape of the observed luminosity function at $z=6.6$
with faint-end slope of $\alpha=-1.5$. The resulting mass-to-light ratio 
implies that the majority of sources responsible for reionization are too
faint to be observed by the current surveys.  
\end{abstract}

\begin{keywords}
  cosmology: theory --- diffuse radiation --- intergalactic medium ---
  large-scale structure of universe --- galaxies: formation --- radio lines:
  galaxies
\end{keywords}

\section{Introduction}

The reionization of the universe was the last global transition of the
Intergalactic Medium (IGM), from fully-neutral after cosmic recombination at
$z\sim1100$ to fully-ionized as we see it today, caused by the radiation from
the first stars. Currently there are still only very few direct observational
constraints on this epoch. The lack of Gunn-Peterson trough in the spectra of
high-redshift sources indicates a low mean neutral fraction 
$x_{\rm HI}\lesssim10^{-4}$ out to redshift $z\sim6$, which implies overlap
was achieved sometime before that, at $z_{\rm ov}>6$ . 

On the other hand, the WMAP 3-year data \citep{2007ApJS..170..377S} yielded a
fairly high value for the integrated Thomson electron scattering optical depth
to the surface of last scattering, at $\tau_{\rm es}=0.09\pm0.03$. This
requires a significant ionized fraction out to high redshifts, $z>12$,
and thus implies an extended reionization. The optical depth by itself does
not put very stringent constraints on the possible reionization histories,
however. The reason for this is the self-regulated nature of the reionization
process \citep{2007MNRAS.376..534I}, whereby the Jeans-mass filtering of 
low-mass sources in the ionized regions results in the $\tau_{\rm es}$ and 
$z_{\rm ov}$, the overlap redshift, being only loosely related. The overlap
redshift $z_{\rm ov}$ is determined by the abundances and efficiencies of the
high-mass sources, whose formation is not suppressed by reionization, while
$\tau_{\rm es}$ depends on both high- and low-mass sources. Thus, varying the
ionizing efficiencies of the small sources yields a wide range of $\tau_{\rm
  es}$ values for the same value of $z_{\rm ov}$. 

This relative lack of observational data is set to change dramatically in the
near future due to a number of large observational projects which are
currently under way. The 21-cm data from high redshifts contains potentially
the richest set of information since the signal is inherently
three-dimensional, on the sky and in redshift/frequency \citep[e.g.][see
\citet{2006PhR...433..181F} for a detailed recent
review]{1997ApJ...475..429M,2000ApJ...528..597T,2002ApJ...572L.123I, 
2003MNRAS.341...81I,2004ApJ...608..622Z,2006MNRAS.372..679M,
2006ApJ...646..681S,wmap3}.   
The features that could be derived include the full reionization history, 
geometry, statistics and individual bright features \citep{2006MNRAS.372..679M,
2006PhR...433..181F,wmap3}. There are significant challenges to be overcome,
however, particularly related to precise subtraction of the very strong 
foregrounds present at low frequencies
\citep[e.g.][]{2006PhR...433..181F,2006ApJ...648..767M,wmap3}. 

The patchiness of reionization creates secondary temperature anisotropies 
in the CMB through the kinetic Sunyaev-Zel'dovich effect 
\citep[][]{1998ApJ...508..435G,2000ApJ...529...12H,2001ApJ...551....3G,
2003ApJ...598..756S,2005ApJ...630..643M,kSZ}, as well as polarization
anisotropies \citep{2000ApJ...529...12H,2003ApJ...595....1H,
2003ApJ...598..756S,mortonson06,cmbpol}. Unlike the 21-cm signal,
the reionization signatures in the CMB are integrated over the reionization
history and contain no frequency information. However, the typical scales 
of reionization are reflected in a characteristic peak of the kSZ 
anisotropy signal \citep{kSZ}, and the shape of the power spectrum is 
dependent on the reionization parameters (source efficiencies and small-scale
gas clumping). CMB anisotropy observations can therefore provide us with key
information about the process of reionization and since its systematics are
different it would be an important complement to the 21-cm studies. One could
also combine these observations more directly, by using 21-cm observations to
derive the Thomson optical depth fluctuations \citep{pol21}. Small-scale CMB
anisotropy and polarization measurements would be quite difficult due to the
weakness of these signals, but are within the abilities of modern detectors
\citep{kSZ,cmbpol}.    

Narrow-band searches for high-redshift Ly-$\alpha$ emitters have been very 
successful at finding sources at ever higher redshifts, currently up to $z\sim7$
\citep[e.g.][]{2002ApJ...568L..75H,2005pgqa.conf..363H,2003PASJ...55L..17K,
2005PASJ...57..165T,2003AJ....125.1006R,2004ApJ...604L..13S,2004ApJ...617L...5M,
2006NewAR..50...94B,2006PASJ...58..313S,2006ApJ...648....7K,2008ApJ...677...12O}. 
Together with studies of the Ly-$\alpha$ resonant absorption 
in the IGM \citep[e.g.][]{2002AJ....123.1247F,2003AJ.126..1W,2004AJ....128..515F,
2006AJ....132..117F} they provide important independent approaches for studying
reionization \citep[see][for a recent review]{2006ARA&A..44..415F}. The
optical depth for Ly-$\alpha$ resonant absorption in neutral IGM at high
redshifts is quite large \citep{1964AZh....41..801S,1965ApJ...142.1633G}, thus
absorption studies are mostly sensitive to very low hydrogen neutral
fractions, typically below $x_{\rm HI}\sim10^{-4}$, otherwise the absorption
saturates. This technique is thus best suited for studying highly-ionized
regions and the very end of reionization and has been very successful for
setting low limits for the redshift at which reionization was completed. On
the other hand, Ly-$\alpha$ emitter surveys do not require a very low average
hydrogen neutral fraction and thus in principle can probe further into the 
reionization epoch \citep{2004ApJ...617L...5M}. Other related probes include
damping wing measurements \citep[e.g.][]{1998ApJ...501...15M,2004ApJ...611L..69M,
2006PASJ...58..485T}, quasar HII region sizes and their evolution 
\citep[e.g.][]{2007ApJ...660..923M,2004ApJ...610..117W,2006AJ....132..117F,
2007MNRAS.374..493B,2007MNRAS.376L..34M}, and sizes of dark gaps in high-z 
spectra \citep[e.g.][]{2006AJ....132..117F,2008MNRAS.386..359G}.

The correct interpretation of these data for reionization
is far from straightforward, however. The strong dependence of Ly-$\alpha$
absorption on the neutral fraction and gas density means that both should be
modelled with certain precision. High-redshift sources are rare and strongly 
clustered \citep[see e.g.][]{2004ApJ...609..474B,2004ApJ...613....1F,2007MNRAS.376..534I}.
As a result, H~II regions generally contain multiple ionizing sources and grow
much larger than the ones created by individual sources. This minimizes the
effect of the damping wings and increases the transmission, allowing the
detection of more and fainter sources than would be naiively expected
\citep{2005ApJ...625....1W}. However, while this is the generic expectation,
the actual effect would be dependent on the exact geometry of reionization,
e.g. sources close behind a neutral region will be damped even if they are
inside a very large H~II region. Simplified models typically assume spherical
ionized regions and either ignore source clustering or assume linear bias 
\citep[e.g.][]{2000ApJ...542L..75C,2004MNRAS.349.1137S,2004MNRAS.354..695F,
2005ApJ...623..627H,2005ApJ...628..575W,2006ApJ...649..570K,2007MNRAS.374..960W}. 
In practice the large ionized regions form by local percolation of multiple
smaller ones and as a consequence are highly nonspherical
\citep{2006MNRAS.369.1625I,mellema06,2007MNRAS.376..534I}. The inhomogeneous 
cosmological density fields and non-equilibrium chemistry effects 
(particularly in recently-ionized gas)
further complicate the picture and point to the need of following the
cosmological structure formation and the reionization history of a given
region. A proper account of all these effects can only be done through
detailed cosmological radiative transfer simulations.


A number of radiative transfer methods have been developed in recent
years and now they are reaching a certain level of maturity and are
producing fairly reliable results \citep{comparison1}. However,
performing large-scale reionization simulations, as required for
Ly-$\alpha$ studies is still technically very challenging. Recently
\citet{2007MNRAS.381...75M} used large scale structure formation
numerical simulations postprocessed with radiation transfer to study
the observability of Lyman-$\alpha$ emitters at high redshifts and what 
these can tell us
about reionization. In order to achieve high dynamic range, these
authors employed a subgrid model for the collapse of the smallest
halos. Another, semi-numerical approach was used by
\cite{2008MNRAS.385.1348M}, who took the linear density and velocity
fields at early time (essentially the initial conditions for an N-body
simulation) and used an excursion-set approach combined with a 
first-order Lagrangian theory to ``paint'' the H~II regions on the 
density field. This procedure provides large dynamic range at a low cost, 
but at the expense of making significant approximations and thus cannot
fully replace full numerical simulations.

In this paper we use the results of large scale numerical simulations
to study the transfer of Ly-$\alpha$ through the IGM. In \S~2 we
briefly describe our simulation method. In \S~3 we describe the evolution 
and environment of a rare, massive source, similar to the ones that are 
currently observed. In \S~4 we address the observability
of Lyman-$\alpha$ emitters, considering the reduction of the transferred 
line flux due to the absorption in the IGM and luminosity functions. 
In \S~5 we describe the effects of the patchiness of reionization on the 
angular correlation function of Lyman-$\alpha$ emitters, and, finally, in 
\S~6 we sum up our conclusions.

\section{Simulations}

Our simulation results follow the full, self-consistent reionization
history in a large volume of $(100\,\rm h^{-1}Mpc)^3$ and were
described in detail in \citet{2006MNRAS.369.1625I,2006MNRAS.372..679M}
and \citet{2007MNRAS.376..534I}. While this volume is too small to
allow us to consider the rarest, most luminous sources like the SDSS
QSO's \citep{2002AJ....123.1247F,2006AJ....132..117F}, we have
sufficient resolution to locate the majority of sources responsible
for reionization and take explicit account of their radiation, and to
derive good quality absorption spectra.

Of the range of simulations presented in \citet{2007MNRAS.376..534I} we
here consider one specific run, labelled f250C.
Our simulations were performed using a combination of two very efficient
computational tools, a cosmological particle-mesh code called PMFAST
\citep{2005NewA...10..393M} for following the structure formation, whose
outputs are then post-processed using our radiative transfer and
non-equilibrium chemistry code called C$^2$-Ray \citep{methodpaper}.
The parameter $f_\gamma=250$ characterizes the
emissivity of the ionizing sources - how many ionizing photons per gas atom in
the (resolved) halos are produced and manage to escape from the host halo
within $\sim20$~Myr, which is the time between two consecutive density
slices, equal to two radiative transfer timesteps, while 'C' indicates that
this run models the gas clumping at small (sub-radiative transfer grid)
scales based on a fit given by
\be
C_{\rm sub-grid}(z)= 26.2917e^{-0.1822z+0.003505\,z^2}.
\label{clumpfact_fit3}
\ee
based on the used WMAP3 cosmology (a good fit for $6<z<30$). This fit to
the small-scale clumping factor is a more precise version of the one we
presented in \citet{2005ApJ...624..491I}. We derived it based on a PMFAST
simulation with a computational mesh of $3248^3$ and with particle number
of $1624^3$, with computational volume of $(3.5\,\rm h^{-1}~Mpc)^3$. These
parameters correspond to particle mass of $10^3M_\odot$, minimum resolved
halo mass of $10^5M_\odot$, and a spatial resolution of $\sim1$~kpc comoving.

Our $(100\,\rm h^{-1}~Mpc)^3$ volume simulation resolves all halos with mass
above $2.2\times10^9M_\odot$, higher
than the mass above which atomic-line cooling of hydrogen becomes effective,
which is $\sim10^8M_\odot$.  As a consequence, our treatment does not include
the contribution of low-mass sources to reionization. Higher-resolution,
smaller-box simulations which do include all ionizing sources above the
atomic-line cooling limit \citep{2007MNRAS.376..534I} indicate that the
effects of
low-mass sources are primarily confined to the earliest stages of reionization,
when such sources are dominant. Throughout most of the reionization process,
and especially during its late stages, the low-mass sources, which are strongly
clustered around the high density peaks, are heavily suppressed due to
Jeans-mass filtering in the ionized regions and thus have limited effect on
the reionization progress and large-scale geometry. Since the Ly-$\alpha$
observations largely probe the later stages of reionization, where the
neutral gas fraction is $\sim30\%$ or less \citep{2004ApJ...617L...5M}, we do
not expect that our conclusions will be strongly affected by the absence of
low-mass sources. Larger simulations, currently in progress
\citep{2008arXiv0806.2887I,2008arXiv0806.3091S}, which resolve all
atomically-cooling halos in $\sim100$~Mpc boxes will settle these uncertainties.

Throughout this work we assume a flat ($\Omega_k=0$) $\Lambda$CDM cosmology 
($\Omega_m,\Omega_\Lambda,\Omega_b,h,\sigma_8,n)=(0.24,0.76,0.042,0.73,0.74,
0.95)$ based on WMAP 3-year results \citep{2007ApJS..170..377S}, hereafter
WMAP3. Here $\Omega_m$, $\Omega_\Lambda$, and $\Omega_b$ are the total matter,
vacuum, and baryonic densities in units of the critical density, $\sigma_8$ is
the rms density fluctuations extrapolated to the present on the scale of $8
h^{-1}{\rm Mpc}$ according to the linear perturbation theory, and $n$ is the
index of the primordial power spectrum of density fluctuations.


\section{Luminous high-redshift sources and their environment: properties,
  evolution and reionization history} 

The luminous sources at high redshift, the ones typically seen in current 
surveys, are hosted by rare, massive halos which form at the location of 
the highest peaks of the density field. The statistics of Gaussian fields 
predicts that such high density peaks are rare and highly clustered in space,
more strongly so at high redshifts. As a consequence, each high-redshift, 
massive galaxy should be surrounded by numerous smaller ionizing sources. The
self-consistent reionization history simulations of such regions require
following a sufficiently large volume, in order to obtain the correct statistics 
and biasing of the rare peaks, while at the same time resolving all the low-mass 
halos which are the main drivers of the reionization process. Our current radiative
transfer simulations are able to achieve this. We also note that correctly 
modelling the nonlinear bias of the rare peaks in semi-analytical modes is a
difficult and still unsolved problem. As a consequence, the halo clustering in 
semi-analytical models is typically underestimated, and in some cases even ignored. 
This often yields incorrect results, e.g. in estimates of the suppression of 
low-mass sources by Jeans-mass filtering \citep[e.g.][]{2007MNRAS.376..534I}.

\begin{figure}
  \includegraphics[width=3.5in]{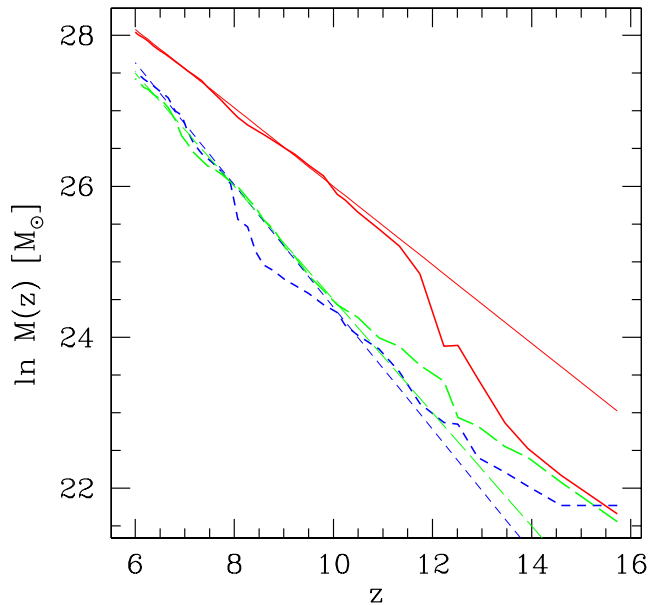}
\caption{Mass accretion history of the three most massive 
  halos found in our computational volume at redshift $z=6$.  The mass 
  growth is roughly exponential in redshift, and is well-approximated by
  $M(z)[M_\odot]=\exp(A-\alpha z)$, where $A=31.2, \alpha=0.52$ for most 
  massive halo (solid, red), $A=32.5, \alpha=0.81$ for the next most massive 
  halo (short-dashed, blue) and $A=32.0, \alpha=0.75$ for the third most
  massive halo (long-dashed, green). Fits are shown by the thin straight lines
  (with corresponding line types and colors).
\label{mass_accr}}
\end{figure}

\subsection{Mass accretion history of massive halos at high redshift}

In Figure~\ref{mass_accr} we show the mass-growth history of the three
most massive (at redshift $z=6$) halos found in our computational volume, 
with final masses of $1.5\times10^{12}M_\odot$, $9\times10^{11}M_\odot$
  and $8.2\times10^{11}M_\odot$, respectively. All correspond to very rare,
$\approx4.5-5-\sigma$ peaks of the density field. The first progenitors of 
these halos resolved in our simulation ($M_{\rm halo}>2\times10^9M_\odot$) 
form very early, at $z\sim16$. Thereafter, the halo masses grow roughly 
exponentially with redshift, $M\propto\exp(-\alpha z)$. This behaviour is 
in good agreement with previous results on the growth of dark matter halos 
in hierarchical $\Lambda$CDM models \citep{2002ApJ...568...52W}, as well 
as with similar results obtained in more generic, idealized situations 
\citep{2004astro.ph..9173S}. The slopes $\alpha$ for our halos exhibit much 
smaller scatter ($0.52<\alpha<0.81$; see Figure~\ref{mass_accr} caption for 
the complete fitting formulae) than the ones found by
\citet{2002ApJ...568...52W} ($0.4<\alpha<1.6$). The exponential fits are 
excellent during periods when no major mergers (i.e. mergers with halos of 
similar mass) occur. This is the case e.g. at late times ($z<11$) by which
time these halos have grown enough to dominate their surroundings and no 
similar-mass halos remain in their vicinity. The exception here is the second 
most massive halo (short-dashed line). Its last major merger occurs at
$z\sim8$. At early times ($z>12$) the mass evolution of all three halos does 
not follow the overall exponential trend. Instead, the most massive halo is 
growing faster than its long-term trend, through series of rapid major
mergers, while the other two halos initially grow slower than their long-term
trend. 

\begin{figure*}
\begin{center}
  \includegraphics[width=2.in]{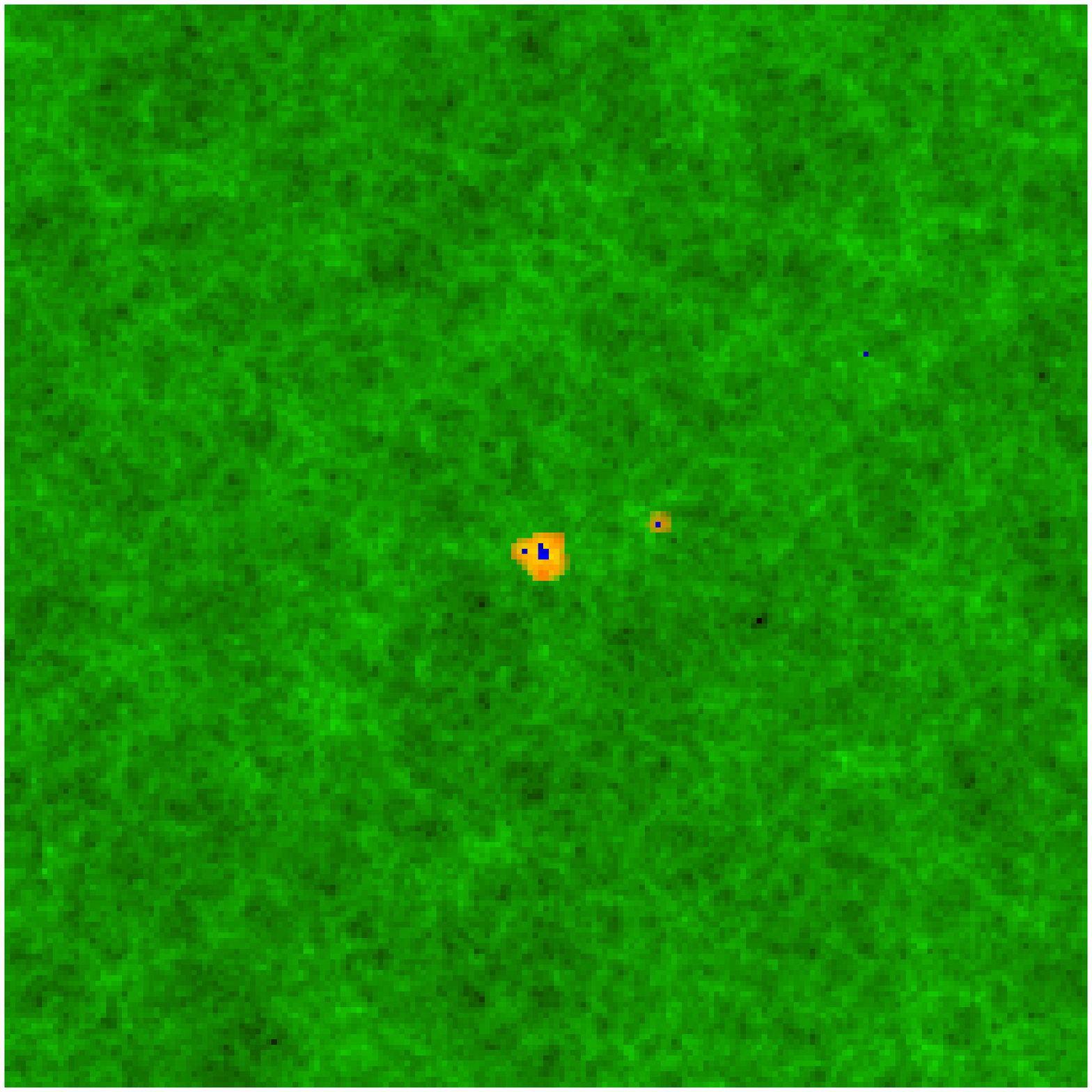}
  \includegraphics[width=2.in]{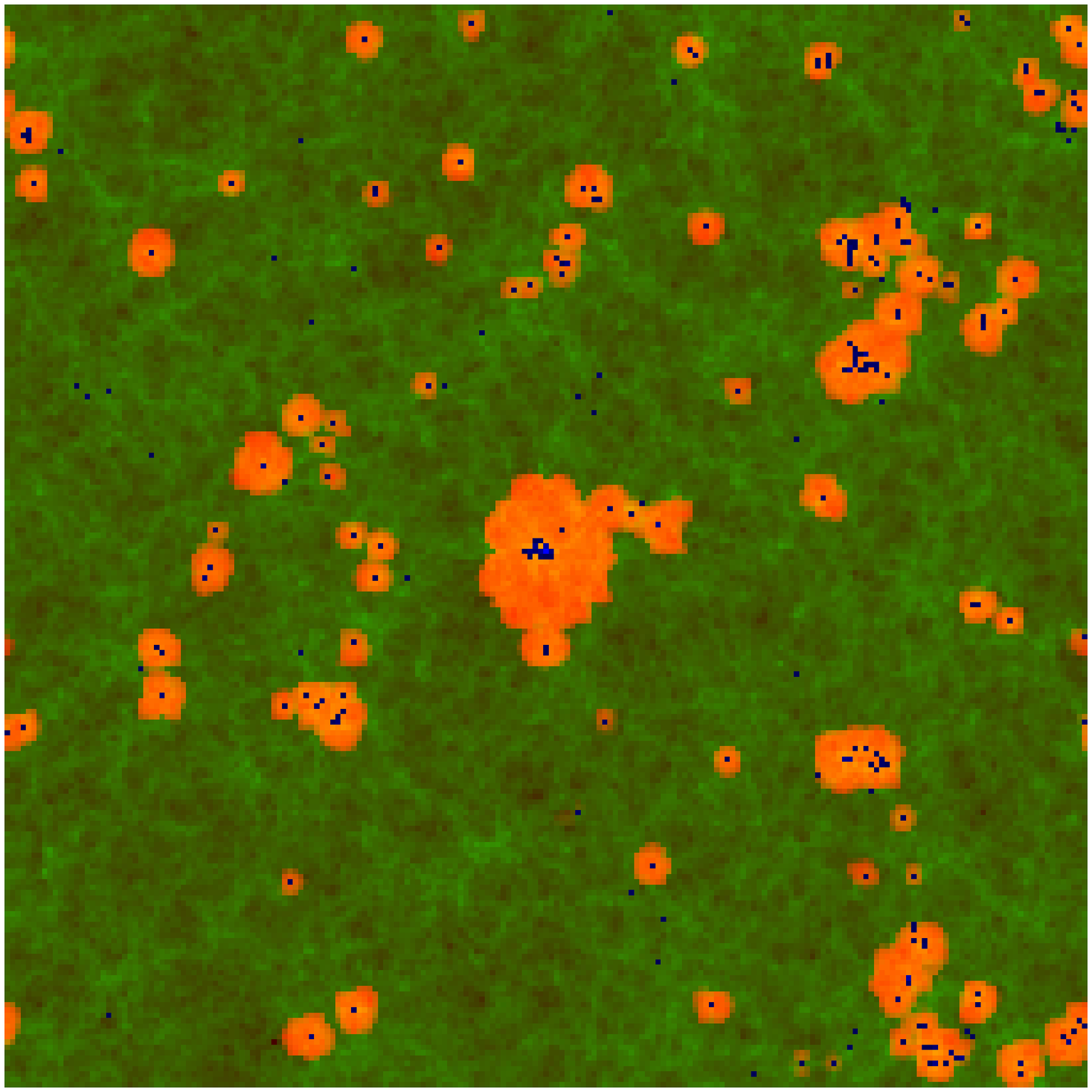}
  \includegraphics[width=2.in]{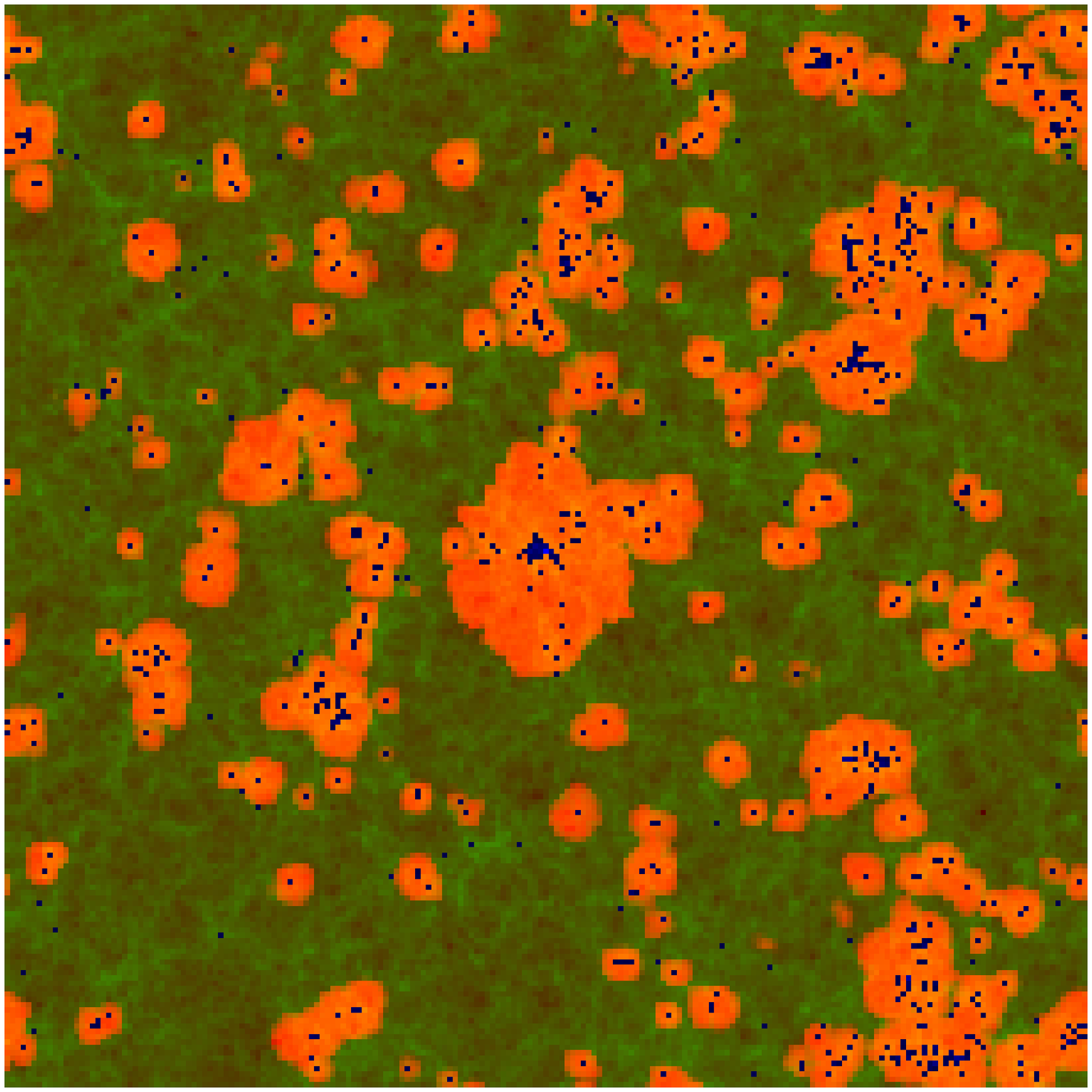}
  \includegraphics[width=2.in]{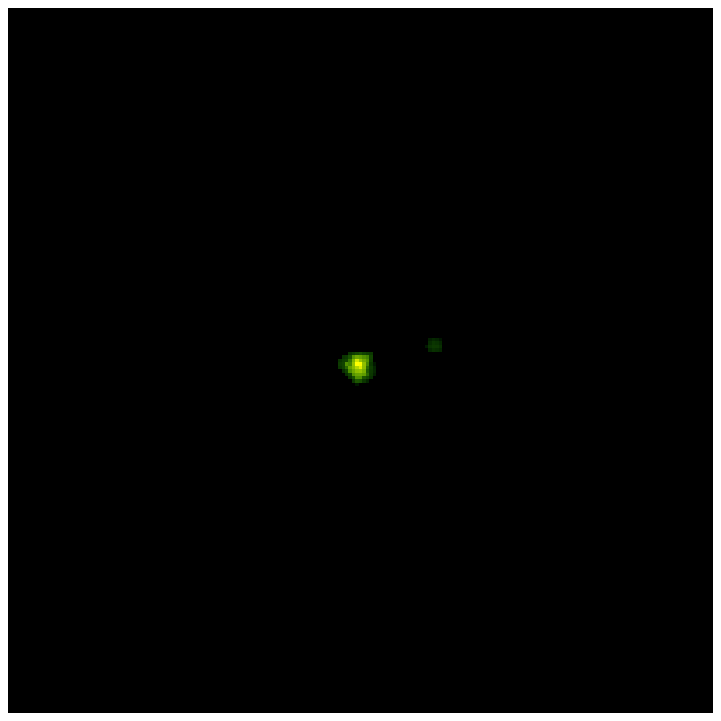}
  \includegraphics[width=2.in]{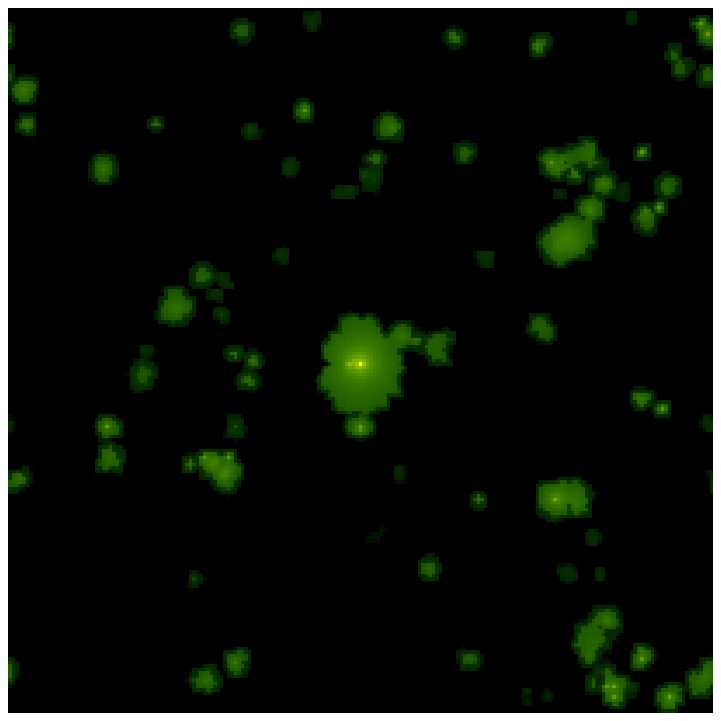}
  \includegraphics[width=2.in]{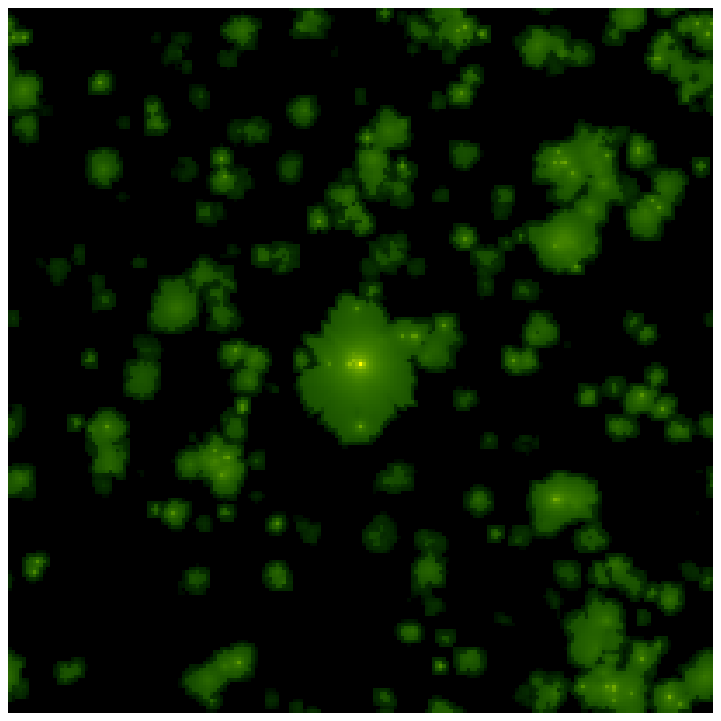}
  \includegraphics[width=2.in]{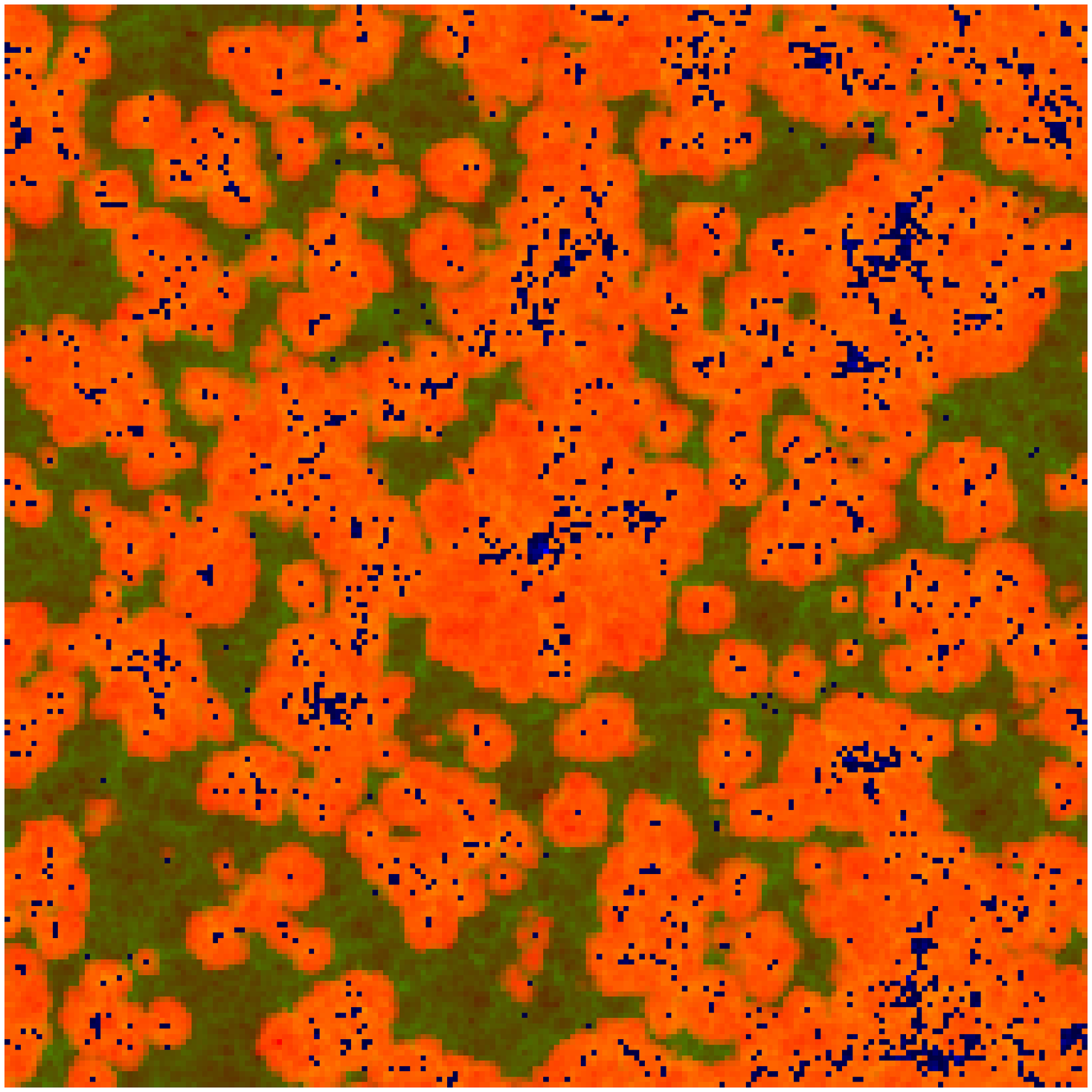}
  \includegraphics[width=2.in]{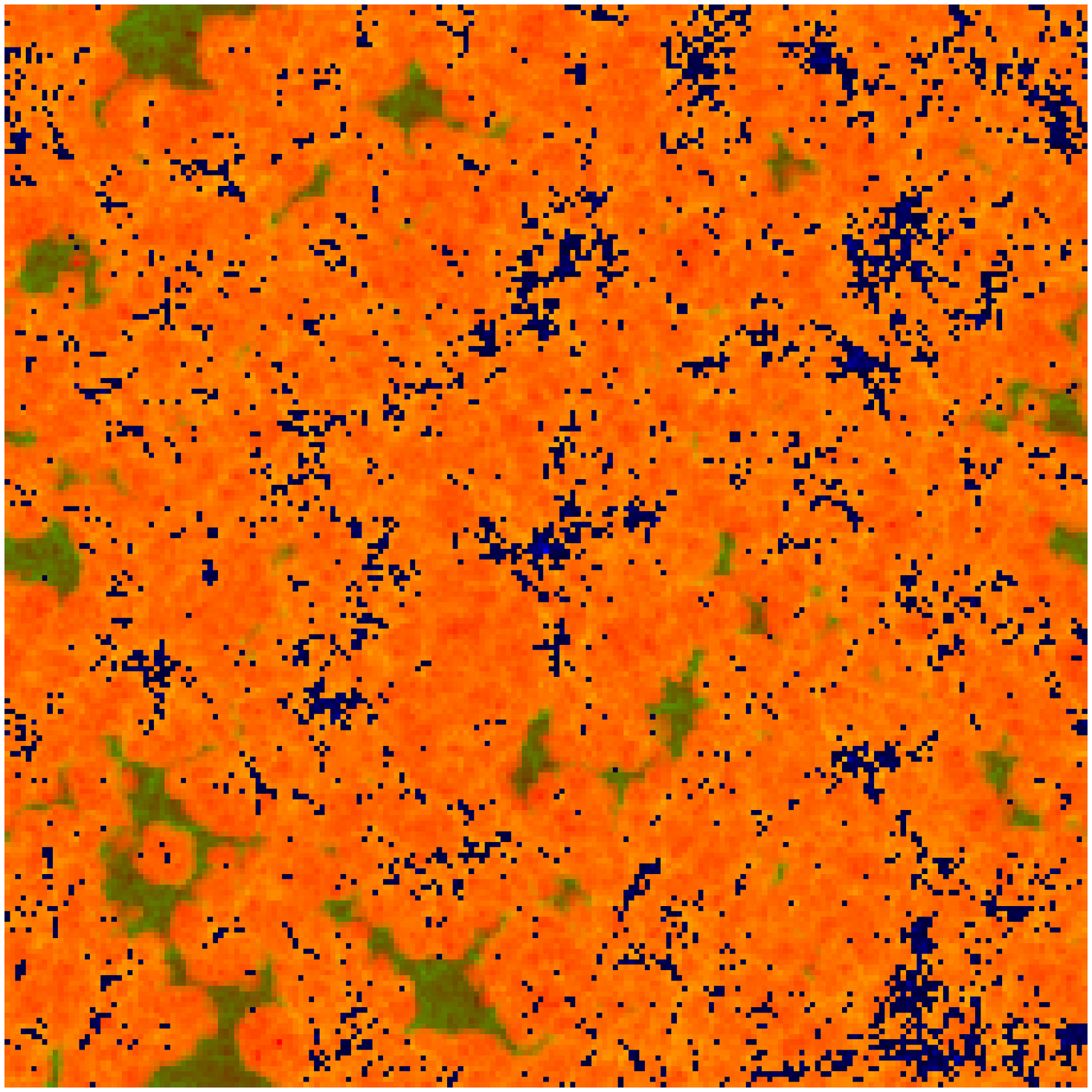}
  \includegraphics[width=2.in]{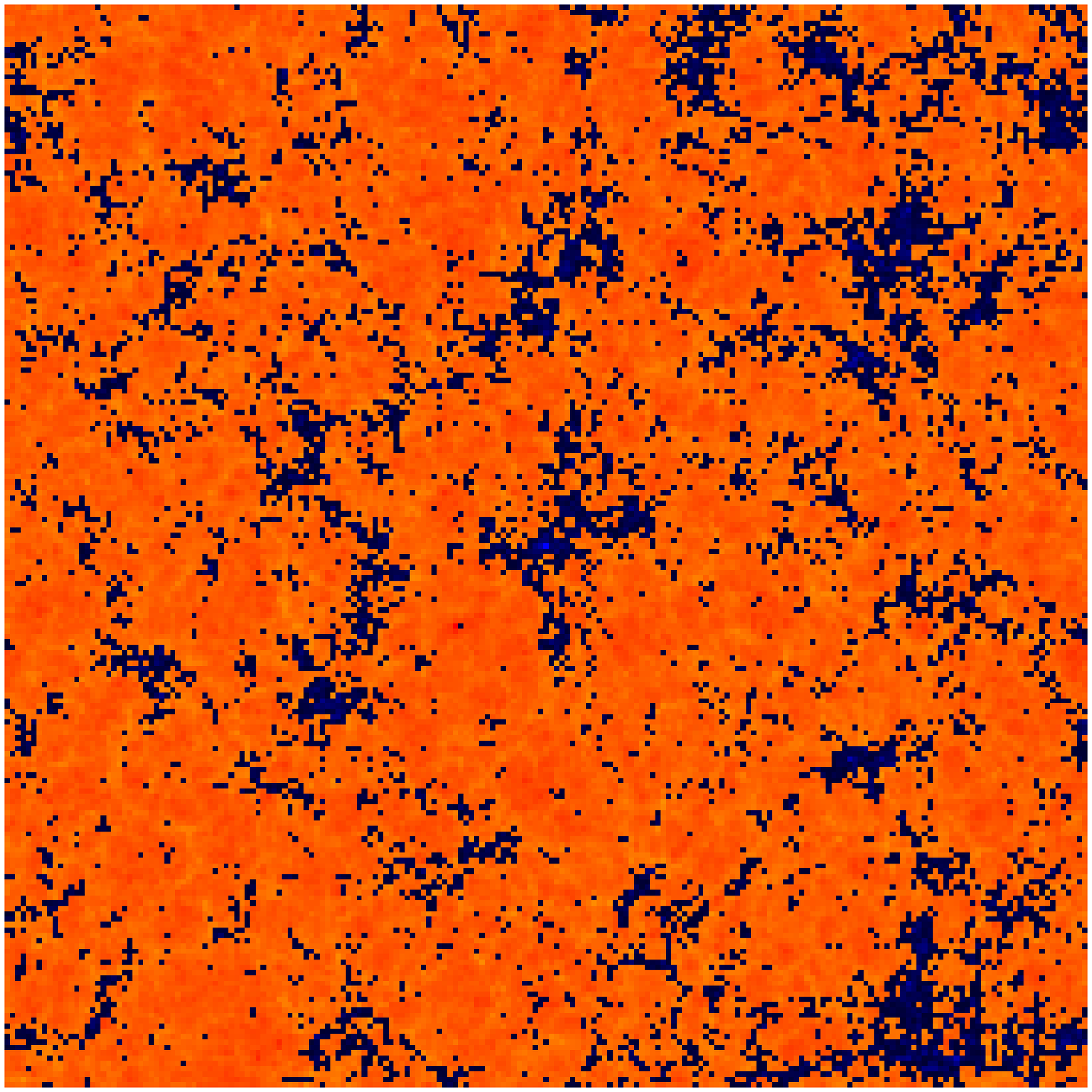}
  \includegraphics[width=2.in]{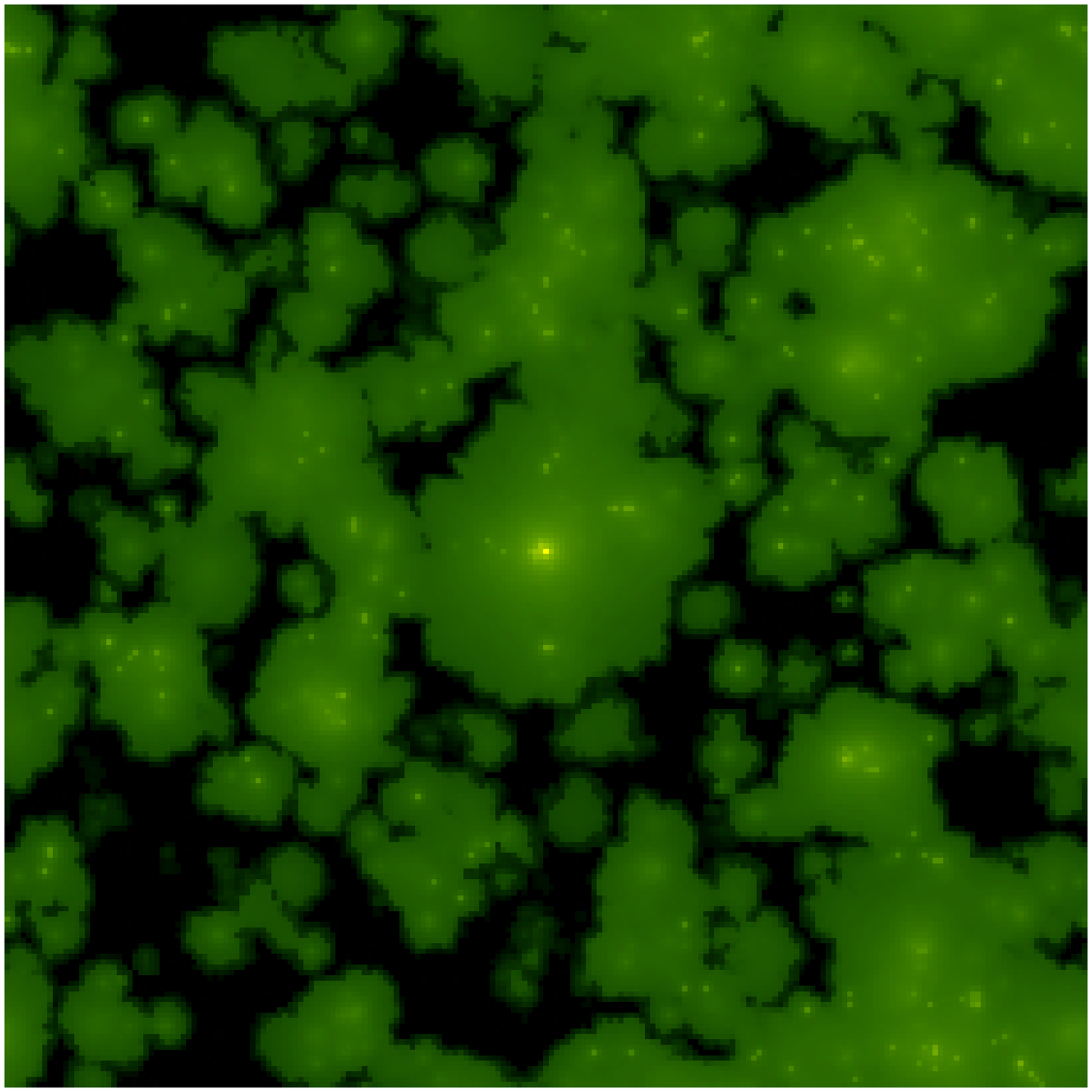}
  \includegraphics[width=2.in]{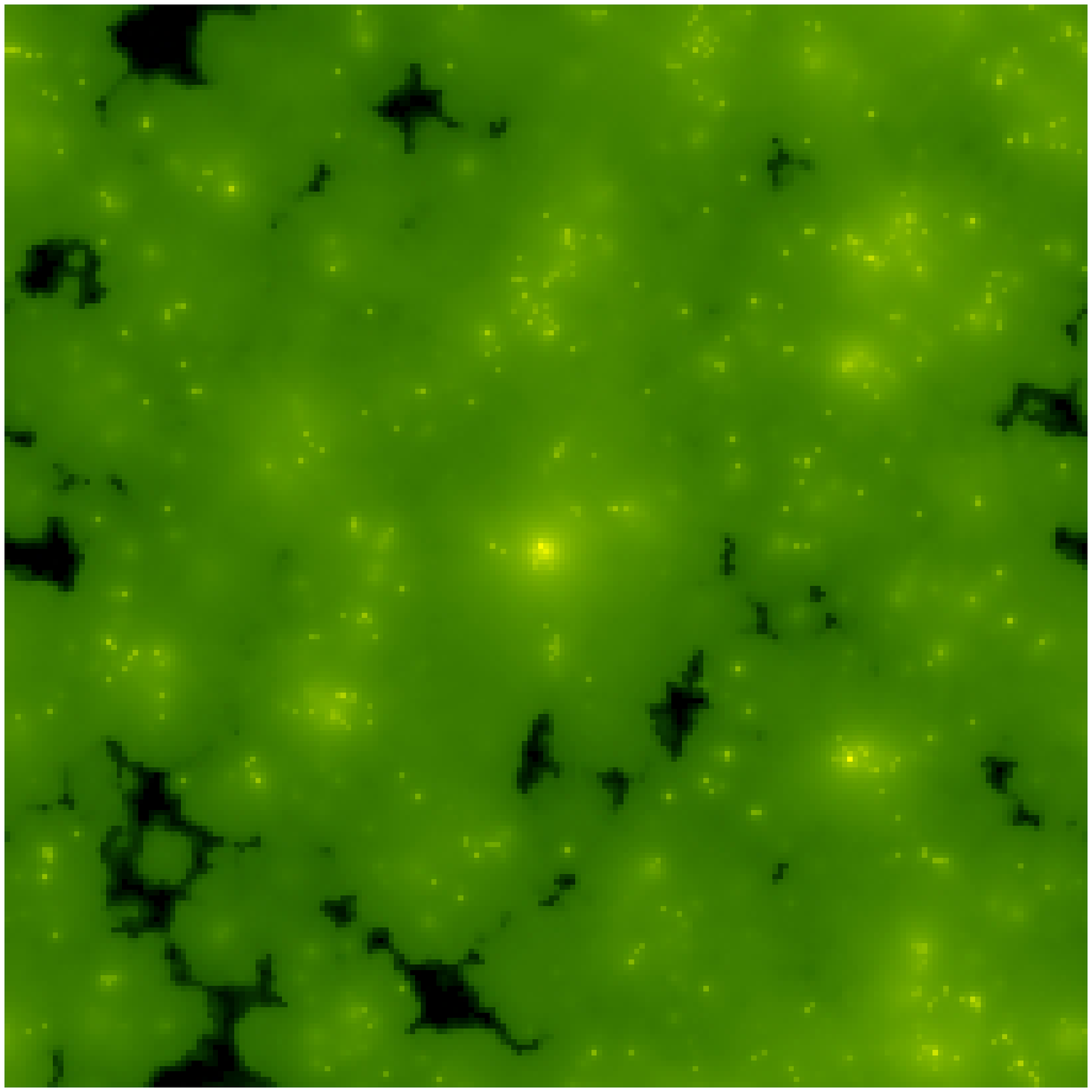}
  \includegraphics[width=2.in]{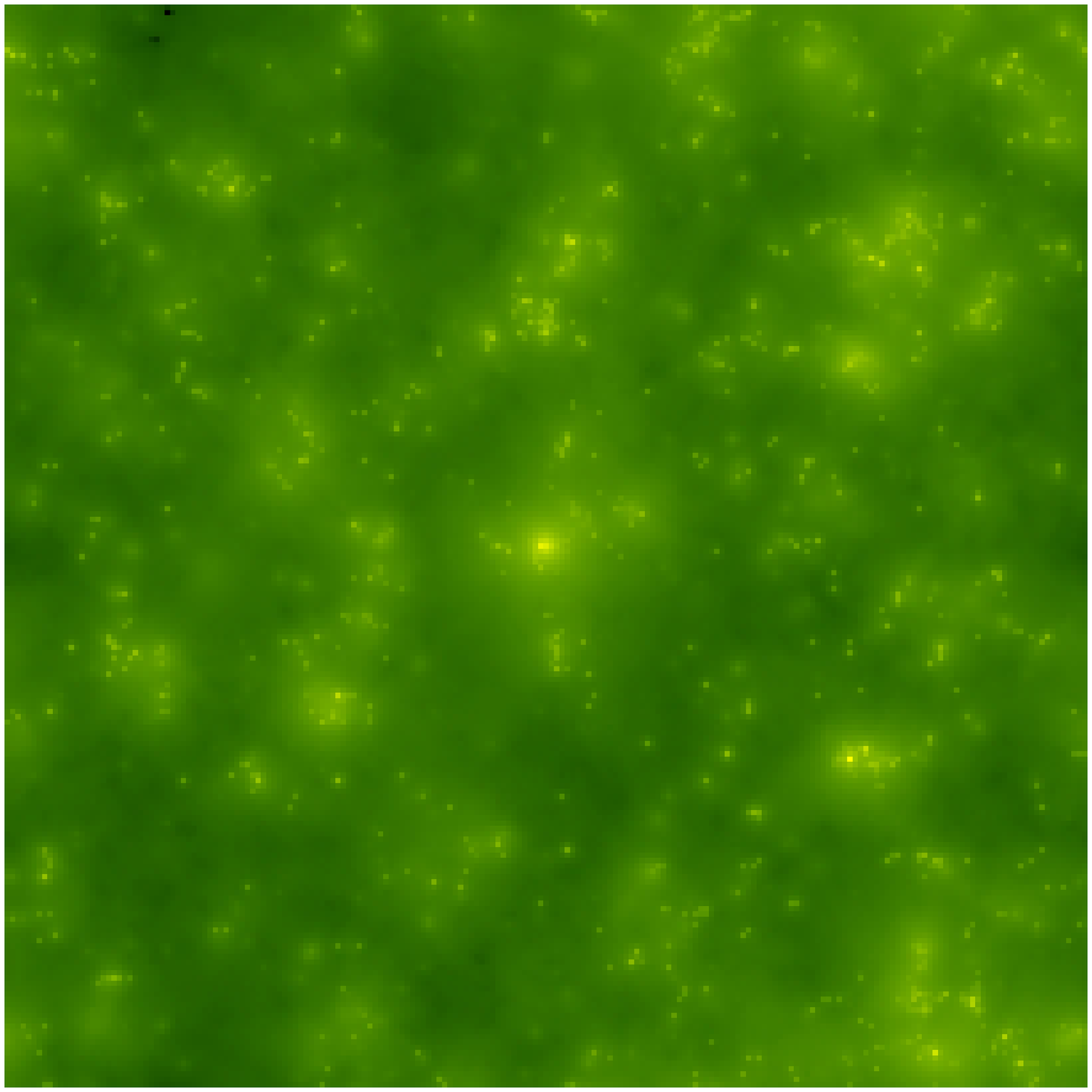}
\caption{The reionization history of a high density peak. 
  The images are centered on the most massive (at $z=6$) halo in our
  computational volume and are of size $100\,h^{-1}$Mpc to the side. The
  snapshots are at: (top row) $z=12.9$ (left; global mass-weighted ionized 
  fraction $x_m=0.001$), $z=10.1$ (middle; $x_m=0.10$), $z=9.0$ (right; 
  $x_m=0.28$), and (third row) $z=7.9$ (left; $x_m=0.66$), $z=7.0$ (middle; 
  $x_m=0.94$), and $z=6.0$ (right; $x_m=0.9999$). The underlying cosmological 
  density field (dark,green) is superimposed with the ionized fraction (light, 
  orange) and the cells containing ionizing sources (dark, blue dots), the 
  slices thickness of $0.5\,h^{-1}$~Mpc (1 cell) in the density and ionized 
  fraction fields and  $10\,h^{-1}$~Mpc in terms of sources. The corresponding 
  images of the (non-equilibrium) photoionization rates ($0.5\,h^{-1}$~Mpc 
  thickness) are shown in the second and bottom rows.
\label{peak_evol}}
\end{center}
\end{figure*}

\subsection{Reionization history}

In Figure~\ref{peak_evol} (top and third row) we illustrate the main stages of 
the reionization history of a high density peak and its intergalactic
environment. The particular peak we show here is the one surrounding the
largest-mass halo found in our computational volume at redshift $z=6$. For
clarity and convenience for the reader we have shifted the peak to the box
center using  
the periodicity of our computational volume. The first resolved source in 
this region forms at $z\sim16$. By redshift $z=12.9$ (top left; mass-weighted 
ionized fraction is $x_m=0.001$ at this time) the central halo has already 
undergone several major mergers with nearby halos and its total mass has grown 
to $1.5\times10^{10}M_\odot$. The combined effect of the central halo and the
nearby members of the same source cluster is to create a substantial ionized 
region (of size $R\sim1\,\rm h^{-1}Mpc$, defined as the radius at which the 
integrated continuum optical depth from the source at the Lyman limit reaches 
unity). At this time there are only a few ionized regions in our computational 
volume (and just two intersect the plane shown). By $z=10.1$ (top middle; 
$x_m=0.10$) many more H~II regions appear and both the sources and the 
ionizing regions are strongly clustered in space. The central region still 
remains the largest one. By redshift $z=9.0$ (top right; $x_m=0.28$) many more 
haloes have formed, most of them in large clustered groups. The H~II region 
surrounding the central peak remains among the largest 
($R\sim6\,\rm h^{-1}Mpc$), but several other ionized bubbles reach comparable 
sizes. 

At redshift $z=7.9$ (bottom left; $x_m=0.66$) some quite sizable regions, more 
than ten Mpc across, have percolated, reaching local overlap. The central 
bubble has grown to a size of $R\sim8\,\rm h^{-1}Mpc$. The reionization 
geometry becomes quite complex, with most ionized bubbles becoming 
interconnected, while leaving large neutral patches in-between. By $z=7.0$ 
(bottom middle; $x_m=0.94$) the notion of isolated, quasi-spherical H~II 
regions becomes rather meaningless, since all ionized regions have already 
merged into one topologically-connected region, although substantial neutral 
patches still remain interspersed throughout our volume. The volume remains 
on average quite optically-thick to ionizing continuum radiation. Finally, at 
$z=6.0$ (bottom right; $x_m=0.9999$; $R=18.7\,\rm h^{-1}Mpc$) our volume 
is well beyond overlap (which we define by $x_m>99\%$). Only by that time the 
volume becomes on average optically-thin to ionizing radiation.

\begin{figure}
  \includegraphics[width=3in]{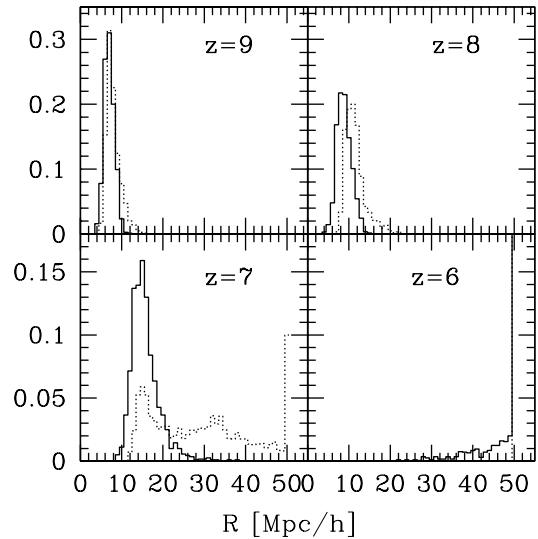}
\caption{Histogram of the bubble size distributions along LOS. Shown are the
  distances from the source (the most massive halo in our volume at $z=6$) 
  at which continuum optical depth (at the hydrogen ionization threshold) 
  surpasses $\tau=1$ (solid lines), or $\tau=4.6$ (dotted lines), at several 
  illustrative redshifts, as labelled.
\label{hist_fig}}
\end{figure}

The corresponding panels in the second and fourth row of
Figure~\ref{peak_evol} show images of the (non-equilibrium) ionization rate
distribution at the same redshifts. The distribution is highly inhomogeneous,
following the patchiness and peaking strongly in the vicinity of large source
clusters. The volume-averaged photoionization rates at redshifts
$z=(12.9;10.1; 9.0; 7.9; 7.0; 6.0)$ in units of $10^{-12}\,s^{-1}$ are 
$\Gamma_{-12}=(7.0\times10^{-2};1.3\times10^{-1};2.3\times10^{-1};
7.6\times10^{-1};1.3;4.9)$, growing strongly with time as larger fraction of 
the volume becomes ionized and ever more sources form. Except for the earliest
times, the peak photoionization rate values on the grid (corresponding to the
brightest points of the images) remain fairly constant with time, at the same
redshifts they are $\Gamma_{-12}=(1.0\times10^{2};2.0\times10^{3};
2.0\times10^{3};2.5\times10^{3}; 2.8\times10^{3};5.1\times10^{3})$. 
The photoionization rate distributions and evolution are discussed further in 
\S~\ref{photoion_rates_sect}.   

\begin{figure*}
  \includegraphics[width=3in]{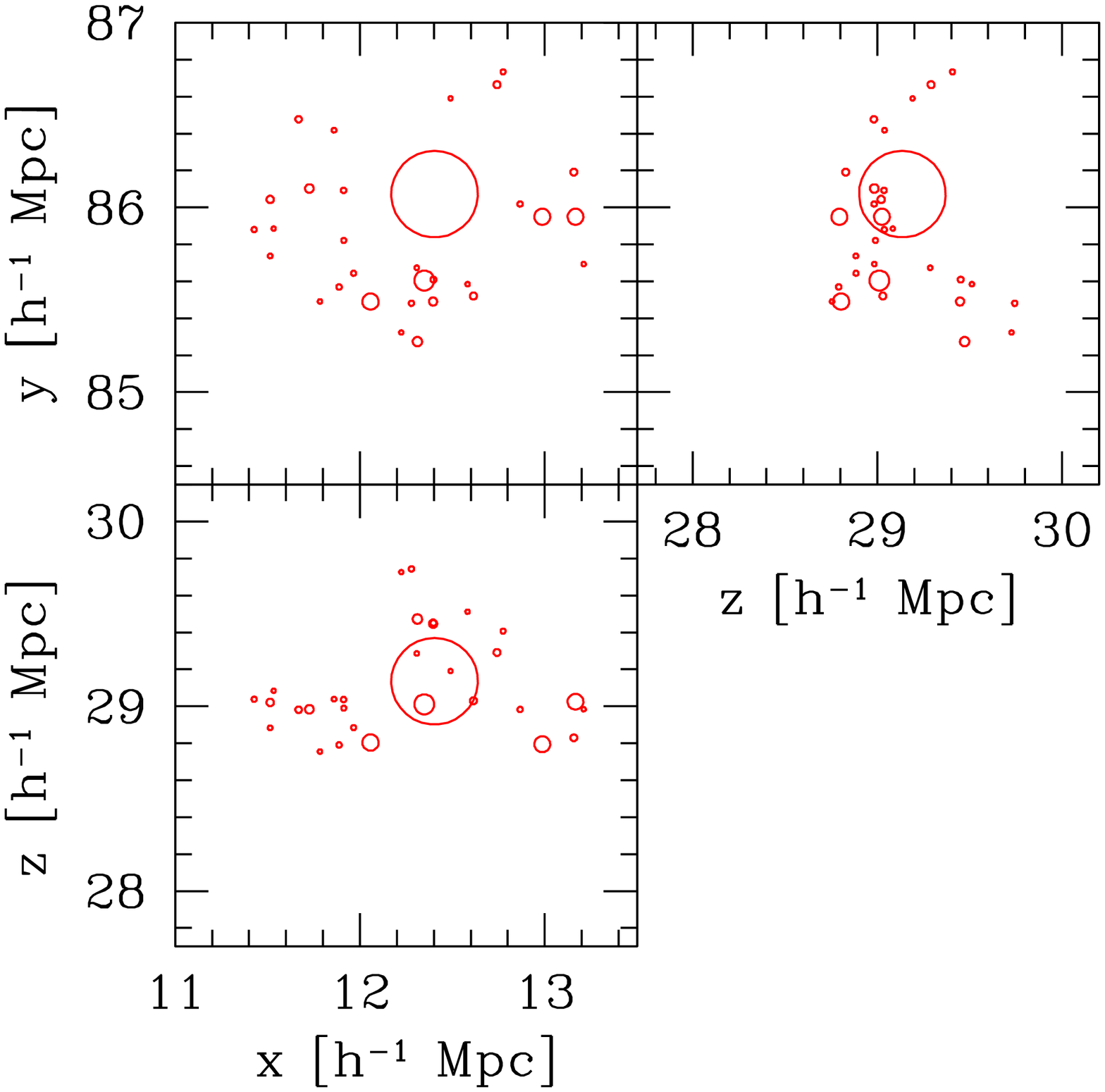}
  \includegraphics[width=3in]{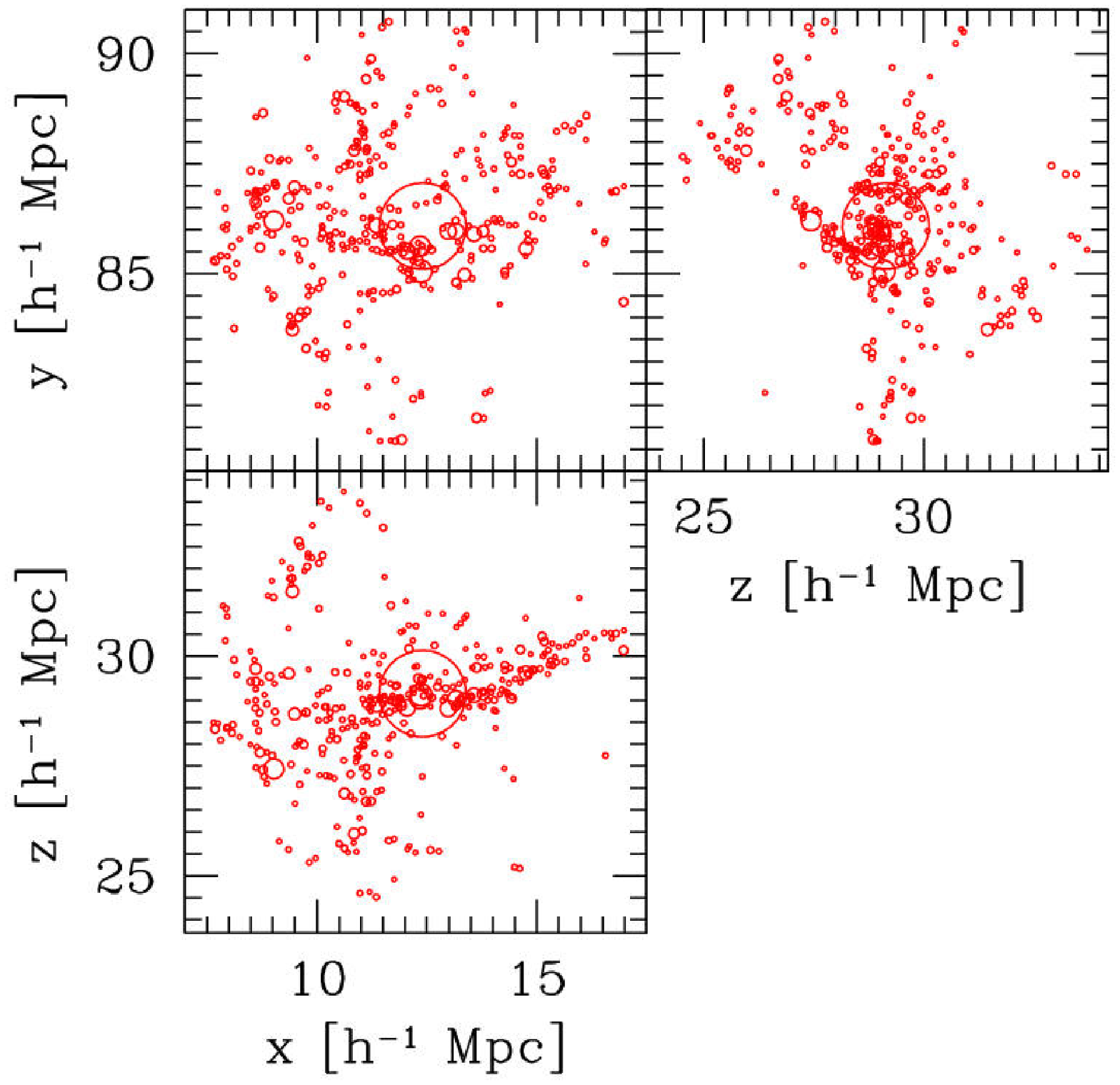}
\caption{Projected distributions  at $z=7$ of halos within 1~$\rm h^{-1}Mpc$ 
  comoving (left) and within 5~$\rm h^{-1}Mpc$ comoving from the most massive 
  object in the box. Circle areas are proportional to the mass of the 
  corresponding halo. 
\label{halo_dist_fig}}
\end{figure*}

To further characterize the shape and boundary sharpness of the central H~II 
region in Figure~\ref{hist_fig} we show a histogram of the radial distance $R$ 
from the central source at which the cumulative continuum optical depth at the
ionizing threshold of hydrogen along each LOS reaches unity (black) and 4.6
(red; this optical depth value corresponds to 1\% transmission). A
spherically-symmetric H~II region would correspond to a single value for the 
radius for a given optical depth, regardless of the directionality, while any
scatter around the peak would be a measure of the non-sphericity of the
ionized region. Furthermore, a comparison between the distributions for the
two optical depths measures the sharpness of the H~II region boundary, as 
follows.
A sharp transition from an ionized to a neutral gas along a given LOS results
in both optical depth values being surpassed simultaneously (since the neutral
gas is extremely optically-thick) and in such situation the histograms would
coincide. On the other hand, if the boundary of the ionized region is not
clearly defined due to local percolation, gas is highly ionized and the
optical depth along the LOS increases only slowly, thus the values of 1 and
4.6 are reached at very different distances from the source. 

At redshift $z=9$ ($x_m=0.28$)the H~II region surrounding the central source 
is largely spherical and its boundary is well-defined, albeit with some modest 
spread around the peak, indicating some departures from sphericity, in 
agreement with what was seen in Fig.~\ref{peak_evol}. At redshift $z=8$ 
($x_m=0.62$) the H~II region
remains fairly spherical, slightly more so for $\tau=1$. The $\tau=4.6$
histogram has a long tail at large values of $R$, up to $\sim25 h^{-1}Mpc$,
meaning that  a small percentage of the LOS reach 99\% opacity only at such 
fairly large distances. Both distributions are somewhat wider than before and 
start to depart from each other, reflecting the increasing ``fuzziness'' of
the bubble boundary as it merges with other nearby bubbles due to local
percolation.  

At $z=7$ ($x_m=0.94$) the $\tau=1$ distribution still retains a well-defined peak, albeit 
one accompanied by a long high-$R$ tail reaching out to tens of Mpc. However,
the $\tau=4.6$ histogram changes its character completely, becoming very
broad, with only a low peak close to the $\tau=1$ peak and several  
secondary peaks. A significant number ($\sim14\%$) of the LOS do not reach 
$\tau=4.6$ within $50\rm\,h^{-1}Mpc$ of the central source (these are collected 
in the last, $R=50\rm\,h^{-1}Mpc$ bin of the histogram). Finally, at $z=6$
($x_m=0.9999$), well beyond the overlap epoch ($z\sim6.6$, $x_m=0.99$) there 
is no indication of a defined 
ionized bubble. Both distributions are very broad, with no clear peak. At 
this time the optical depth is determined by the tiny remaining neutral 
fraction, which in its turn is dictated by the density variations of the 
Cosmic Web. Either value of $\tau$ is reached at a wide range of distances in
different directions, from $\sim10\rm\,h^{-1}Mpc$ to over $50\rm\,h^{-1}Mpc$.
The IGM becomes largely optically-thin to ionizing radiation and thus for most 
LOS neither $\tau=1$ nor $\tau=4.6$ are reached within 50 $\rm\, h^{-1} Mpc$ 
from the source. 

We also note that these results on the H~II region boundaries are valid for
soft, stellar (either Pop. II or Pop. III) spectra. Should hard sources, e.g.
QSOs, contribute significantly to reionization the ionized region boundaries 
will inevitably be thicker and the transition from ionized to neutral smoother,
due to the longer mean free paths of the hard photons. However, most current
observational indications are that stellar sources dominate reionization, in 
which case the main effect of the relatively few hard photons present will be
to heat the neutral IGM, but not to ionize it significantly.    

\begin{figure}
  \includegraphics[width=3in]{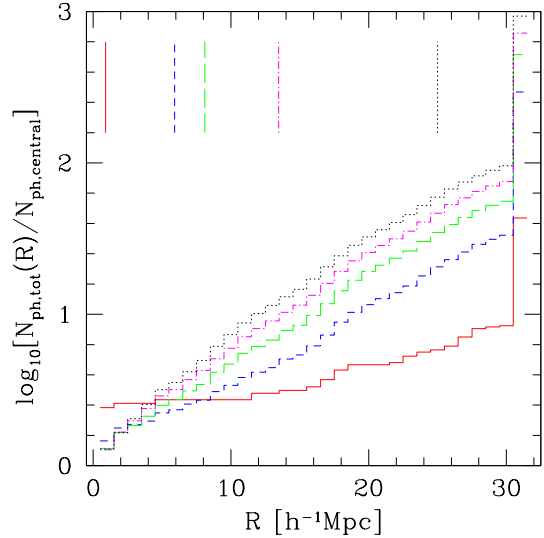}
\caption{Histograms in radial bins of the total cumulative emissivity from all
  halos within distance $R$, $N_{\rm ph, tot}(R)$, measured from the central
  object, here the most massive object in the computational volume, in units
  of the central object's emissivity, $N_{\rm ph,central}$. Shown are (bottom
  to top on the right) $z=12.9,9,8,7$ and 6.25 ($x_m=0.001,0.28,0.62,0.94$ and 
  0.998). The last radial bin contains the total emissivity of the 
  computational box at that time. The vertical lines on top (corresponding 
  line types and colors) indicate the average size of the H~II region at the 
  corresponding redshift. 
\label{emiss_fig}}
\end{figure}

\subsection{Halo clustering in the vicinity of a luminous source}

Taking a closer look at the halo clustering in the immediate vicinity 
of a high density peak in Fig.~\ref{halo_dist_fig}  we show the projected
spatial distribution of halos around the most massive halo at $z=7$. There 
are 31 resolved halos within 1~$h^{-1}$Mpc from the most massive halo 
and 360 resolved halos within 5~$h^{-1}$Mpc (both including the largest halo 
itself). The area of each circle is proportional to the mass (and thus, in 
accordance to our source model, to the luminosity) of the corresponding halo. 
Halos are distributed very anisotropically, concentrating preferentially 
along the density filaments and sheets of the local Cosmic Web. 

The mass of the most massive halo is $9\times10^{11}M_\odot$ at that time, 
well above any other halo in its vicinity. Nonetheless, the low-mass, but 
numerous halos surrounding the peak contribute significantly to the total 
ionizing emissivity. In order to quantify this point further, in 
Figure~\ref{emiss_fig} we show the cumulative emissivity vs. radial distance 
from the central halo for several redshifts spanning the whole range of 
interest here. At all redshifts only within $\sim2\,\rm h^{-1}Mpc$ from 
itself does the most massive halo dominate the total photoionizing emission 
(i.e. it contributes more than 50\% of the cumulative total emission coming 
from that region). The exception is the highest redshift ($z=12.9$), in which 
case the surrounding cluster of sources, rather than the central source, 
dominate the total flux even within $1\,\rm h^{-1}Mpc$ of the peak. The reason 
for this is the presence nearby of another halo of almost the same mass as the 
central one, which shortly thereafter merges with it. The cumulative emission 
is dominated by the small halo contribution beyond that distance. Within 
$10\,\rm h^{-1}Mpc$ the central source contributes only 10-30\% of the total 
emission. As more and more halos form and the lowest mass ones become 
relatively common the fractional contribution of the most luminous source 
to the total emissivity gradually decreases in time for all radii larger 
than a few comoving Mpc. At $30\,\rm h^{-1}Mpc$ the central source is dominated 
by the rest by 1.5-2 orders of magnitude. As fraction of the total emissivity 
of all sources in our computational volume (shown in the last bin of the 
histogram) the most luminous one contributes only $\sim2\%$ of the total at 
$z=12.9$, decreasing to $\sim0.1\%$ at $z=6.25$. 

\begin{figure}
  \includegraphics[width=3in]{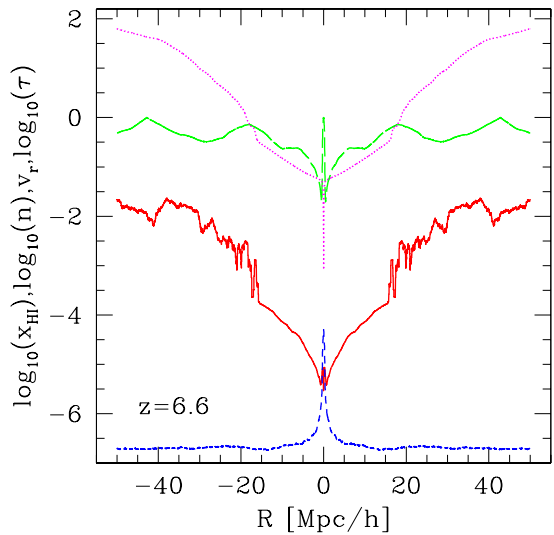}
\caption{Spherically-averaged profiles of (bottom to top on the left/right): 
  the comoving number density of hydrogen (in $cm^{-3}$; short-dashed, blue), the
  neutral hydrogen fraction (solid, red), radial velocity, $v_r$ (in $\rm 100\,
  km\,s^{-1}$; long-dashed, green) and continuum optical depth at $h\nu=13.6$~eV
  integrated from the source ($R=0$) outward, all at redshift $z=6.6$ 
  ($x_m=0.99$) vs. $R$, the comoving radial distance from the most massive 
  galaxy.
\label{spher_aver}}
\end{figure}

In Figure~\ref{emiss_fig} we also indicate the current H~II region mean radius 
(vertical lines of corresponding line types and colors). Within its own bubble 
the central source contributes $\sim50\%$ of the emission at $z=12.9$, 
decreasing to $\sim10\%$ at $z=7$ and $\sim1\%$ at $z=6.25$.

Recently, \citet{2005ApJ...625....1W} presented a semi-analytical model of the 
source clustering at high redshift. They found that a central galaxy at $z=7$ 
and with a velocity dispersion similar to our most massive source, 
$\sigma_V\approx200\rm\,km\,s^{-1}$) contributes $\sim40-70\%$ of the H~II 
region radius, or $\sim10-35\%$ of its volume (see their Fig. 2, right panel), 
a factor of a few larger compared to our results where we find that the central 
galaxy contribution to the total flux is $\sim1-10\%$. This discrepancy is most 
probably due to underestimate of source clustering in their bias model, and 
possibly also to the differences in the assumed source efficiencies. Furthermore,
we have only considered a single massive source, which of course does not account
for random statistical variations from source to source. Nevertheless, this result
underlines the importance of considering the luminous sources individually (as 
opposed to considering an averaged ``mean'' source) and using simulations in order
to account for nonlinear bias effects.

\subsection{IGM Environment of luminous sources at high redshift} 

\begin{figure}
  \includegraphics[width=3.2in]{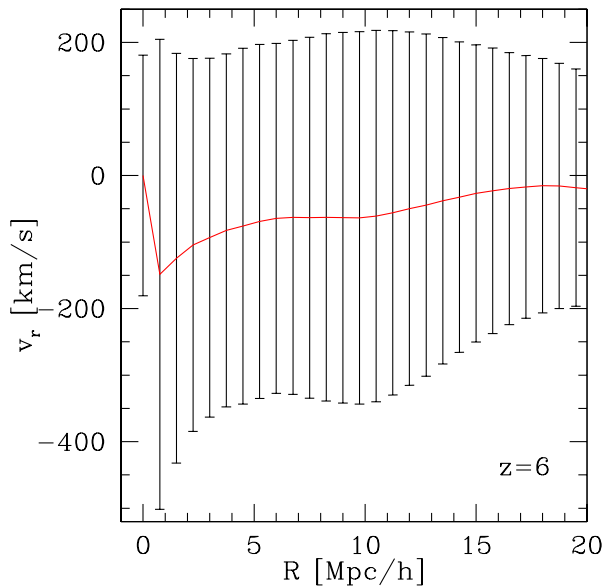}
\caption{Mean radial velocity at $z=6$ of the IGM with respect to the source 
(red solid line, negative means towards the halo, positive - away) and its rms 
variation (error bars), both plotted vs. (comoving) radius from the source.
\label{vel_r_fig}}
\end{figure}

The Ly-$\alpha$ absorption is strongly influenced by the local IGM environment -- 
density, velocity and ionization structure -- around the the source. As an 
illustrative example, let us consider a particular redshift, $z=6.6$, similar to 
the highest redshifts at which currently there are Ly-$\alpha$ source detections.
At this time the universe in our simulation is already highly-ionized, with a
mean mass-weighted ionized fraction of $x_m=0.939$, and a volume ionized fraction 
of $x_v=0.925$. In Figure~\ref{spher_aver} we show the spherically-averaged 
profiles around the central source of the gas number density, $n$, neutral 
fraction, $x_{\rm HI}$, integrated continuum optical depth ($\tau$, at 
$h\nu=13.6$~eV) and radial velocity, $v_r$. The radiative transfer cell which 
contains the central source is highly-overdense, with $\delta=n/\bar{n}-1=256$, 
which indicates that this cell roughly coincides with the source halo itself.
The radial density profile declines steeply away from the source. The overdensity 
is  $\delta=3.3$ at $R\sim1 \rm \,h^{-1}Mpc$, decreasing to $\delta=1$ at 
$R\sim2.5 \rm \,h^{-1}Mpc$, and approaching the mean density at distances beyond 
$R\sim10\rm \,h^{-1}Mpc$. 

The radial velocity profile shows an extended ($R\sim20\,\rm h^{-1}Mpc$) infall 
region, with the mean radial velocity peaking at $\sim150\,\rm km\,s^{-1}$ before 
dropping to zero inside the source halo itself. The proximity region of the central 
source ($R<15\rm\,h^{-1}Mpc$) is highly-ionized, with neutral fraction 
$x_{\rm HI}<10^{-4}$. The rest of the volume still has an appreciable neutral 
fraction ($\sim0.1-1\%$), however, and is thus on average still optically-thick, 
with the mean optical depth reaching $\tau=63$ at $R=50\rm\,h^{-1}Mpc$. 

However, the spherically-averaged quantities provide only a limited information 
about the state of the IGM surrounding each source. All quantities are distributed 
highly-anisotropically, and thus affect the Ly-$\alpha$ emission differently along 
each LOS. In particularly, the effect of the relative velocities of the IGM and 
source is relatively poorly studied at present. An optically-thick medium at rest 
with respect to a  Ly-$\alpha$ source would absorb the blue wing of the line and 
transmit its red wing, at longer wavelengths than the line centre at 
$\lambda_0=1215\,$\AA\, \citep[e.g.][]{1998ApJ...501...15M}. A relative motion of 
the IGM gas and the source along the LOS would result in either more or less 
transmission, depending on the motion direction. E.g. gas infall towards the source 
along the LOS from the observer would redshift it, resulting in some absorption of 
the red wing of the line. We note that in order to evaluate these velocity effects
with any precision much higher resolution simulations (and ones including outflows)
will be required. Our radiative transfer grid has cell size of $\sim0.5/h$~Mpc 
comoving, or $\sim70/h$~kpc physical at $z=6$, which roughly corresponds to the size
of the largest halos found in our box. The velocity and density fields used are at 
twice higher resolution ($\sim0.25/h$~Mpc comoving). Therefore, our results here 
should be considered as a guidance, illustrating that the velocity effects are quite
important and should not be ignored.

In Figure~\ref{vel_r_fig} we show the average IGM gas 
velocity relative to the source (line) vs. distance from it and the variance of that 
average velocity (error bars). As noted above, in the vicinity of the source the 
IGM on average infalls towards the halo. However there are large variations, of order 
hundreds of $km\,s^{-1}$ around this mean. E.g. a velocity offset of 200 $km\,s^{-1}$
at $z=6$ corresponds to $\Delta\lambda\sim6\,$\AA, of the same order as the typical 
observed line widths, and thus a relative motion of the IGM and source of this order
could have a very significant effect on the observed line. In the next section we 
would quantify the effect of peculiar velocities on the Ly-$\alpha$ line.

\begin{figure*}
  \includegraphics[width=3.2in]{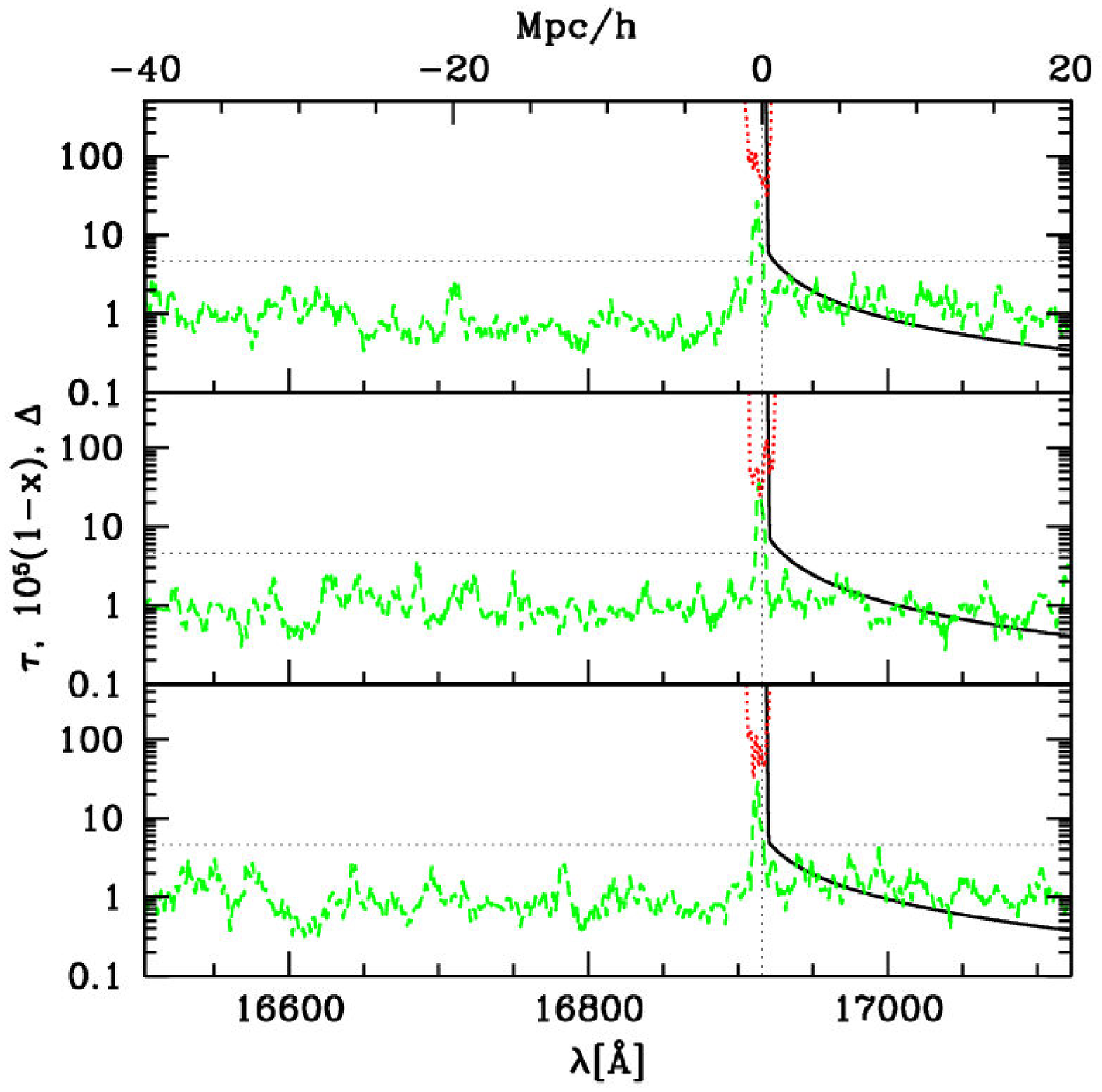}
  \includegraphics[width=3.2in]{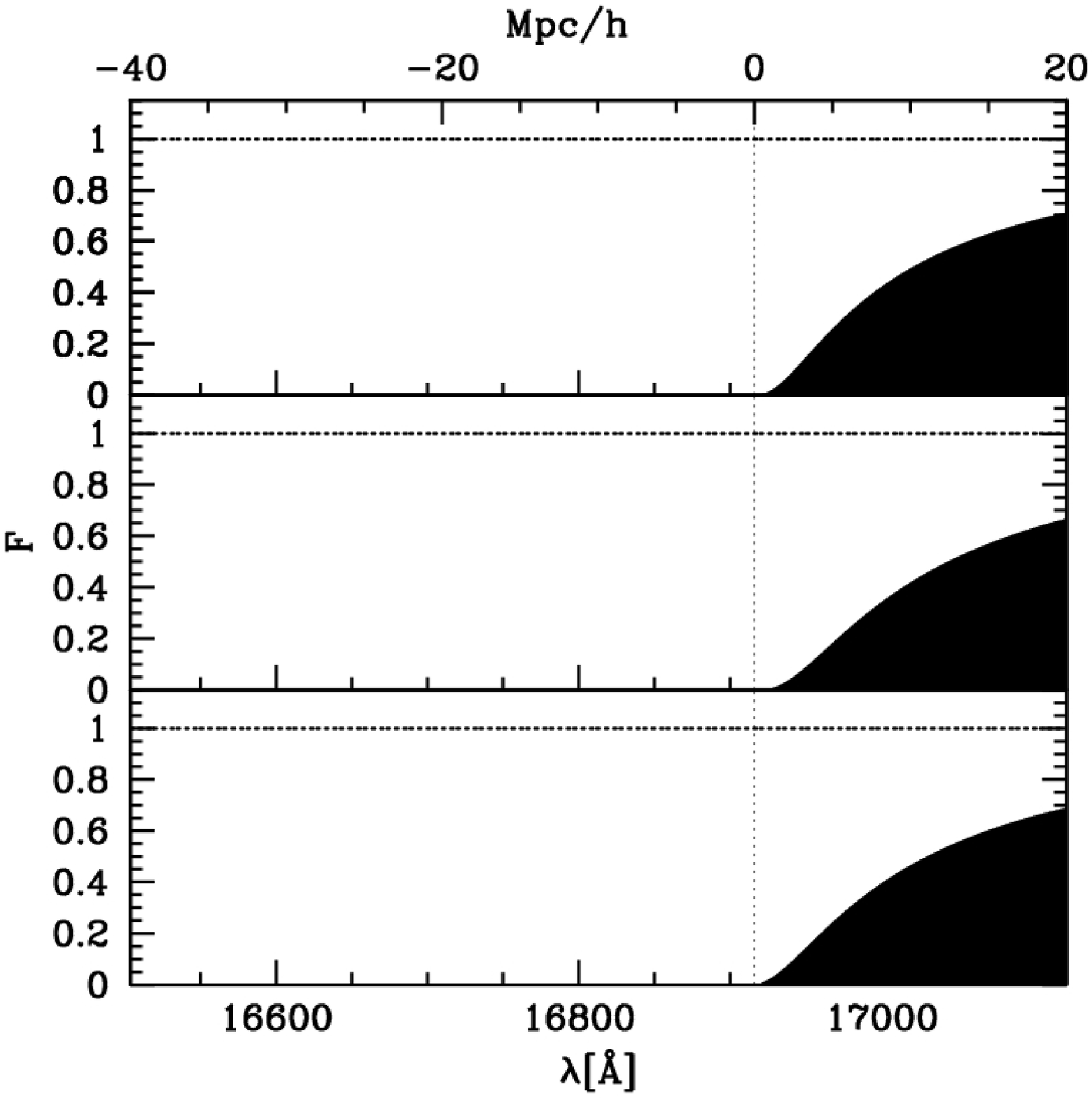}
  \includegraphics[width=3.2in]{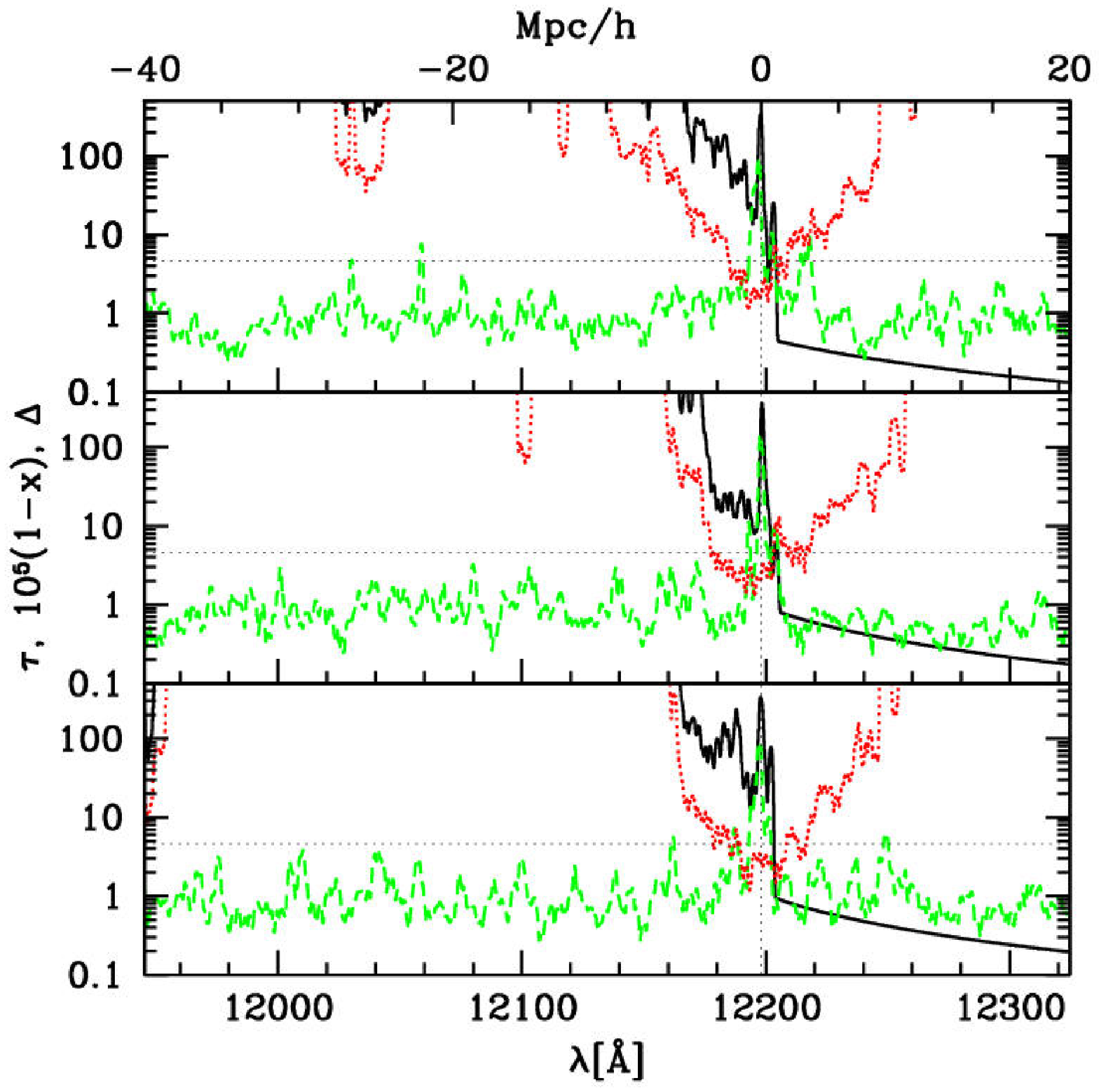}
  \includegraphics[width=3.2in]{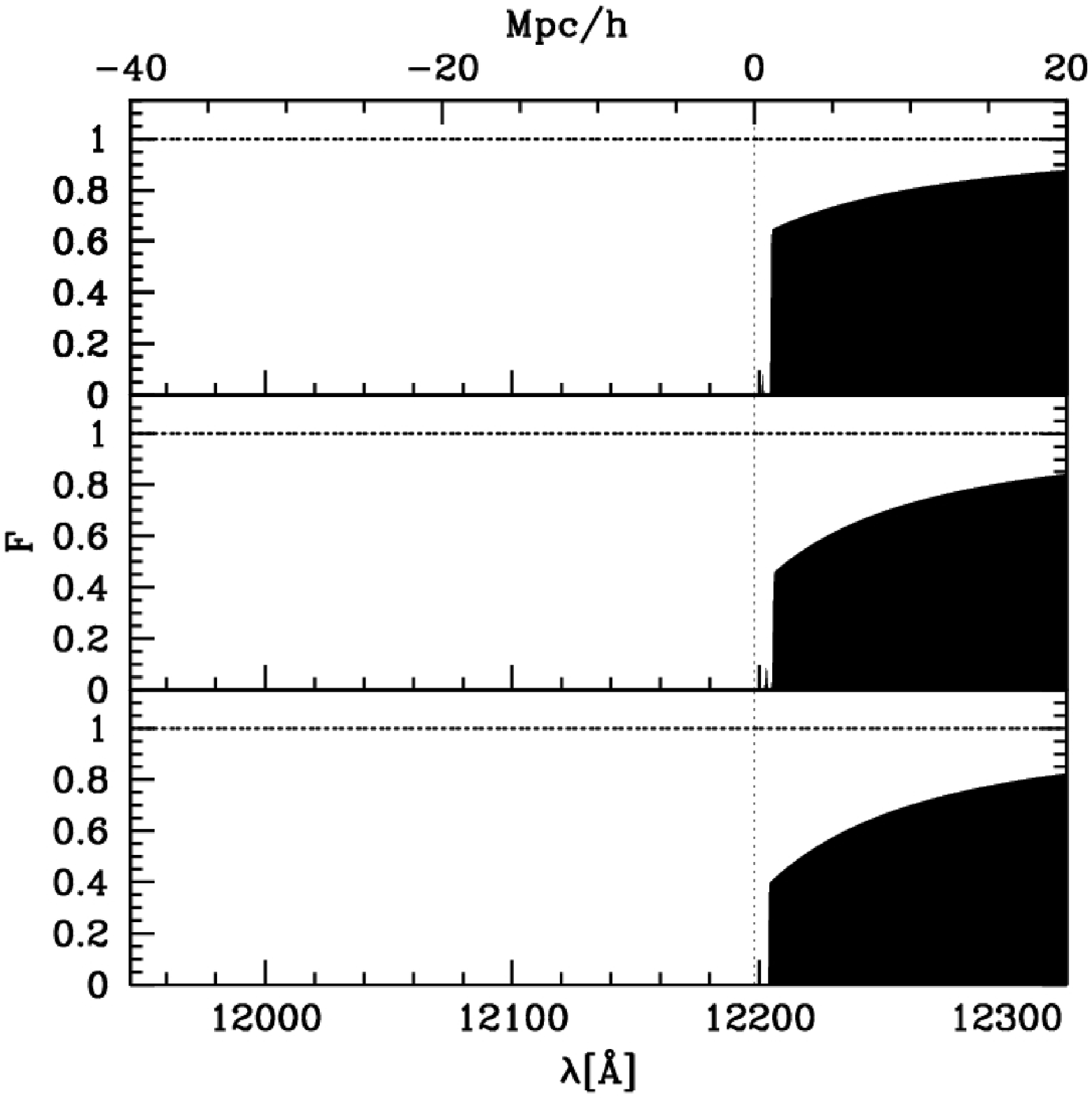}
  \caption{Sample LOS at redshifts $z=12.9$ (top; $x_m=0.001$) and 9.0 (bottom;    $x_m=0.28$) vs. 
    $\lambda$/comoving distance from the most massive galaxy. Shown are (left
    panels) the optical depth (solid), neutral fraction $x_{\rm HI}=1-x$
    ($\times10^5$; dotted) and density in units of the mean (dashed), and
    (right panels) the corresponding transmission. The vertical lines show the
    position of the central source (in redshift space, i.e. accounting for its
    peculiar velocity along the LOS). The horizontal lines on the left
    indicate the optical depth equivalent to 1\% transmission. On the right,
    the shaded region is the transmission in the case where the unabsorbed
    spectrum is flat (the horizontal dotted line).
\label{spectra}}
\end{figure*}

\section{Observability of high-$z$ Ly-$\alpha$ sources}

\subsection{Absorption spectra of luminous sources}
\label{spectra:sect}

\begin{figure*}
  \includegraphics[width=3.2in]{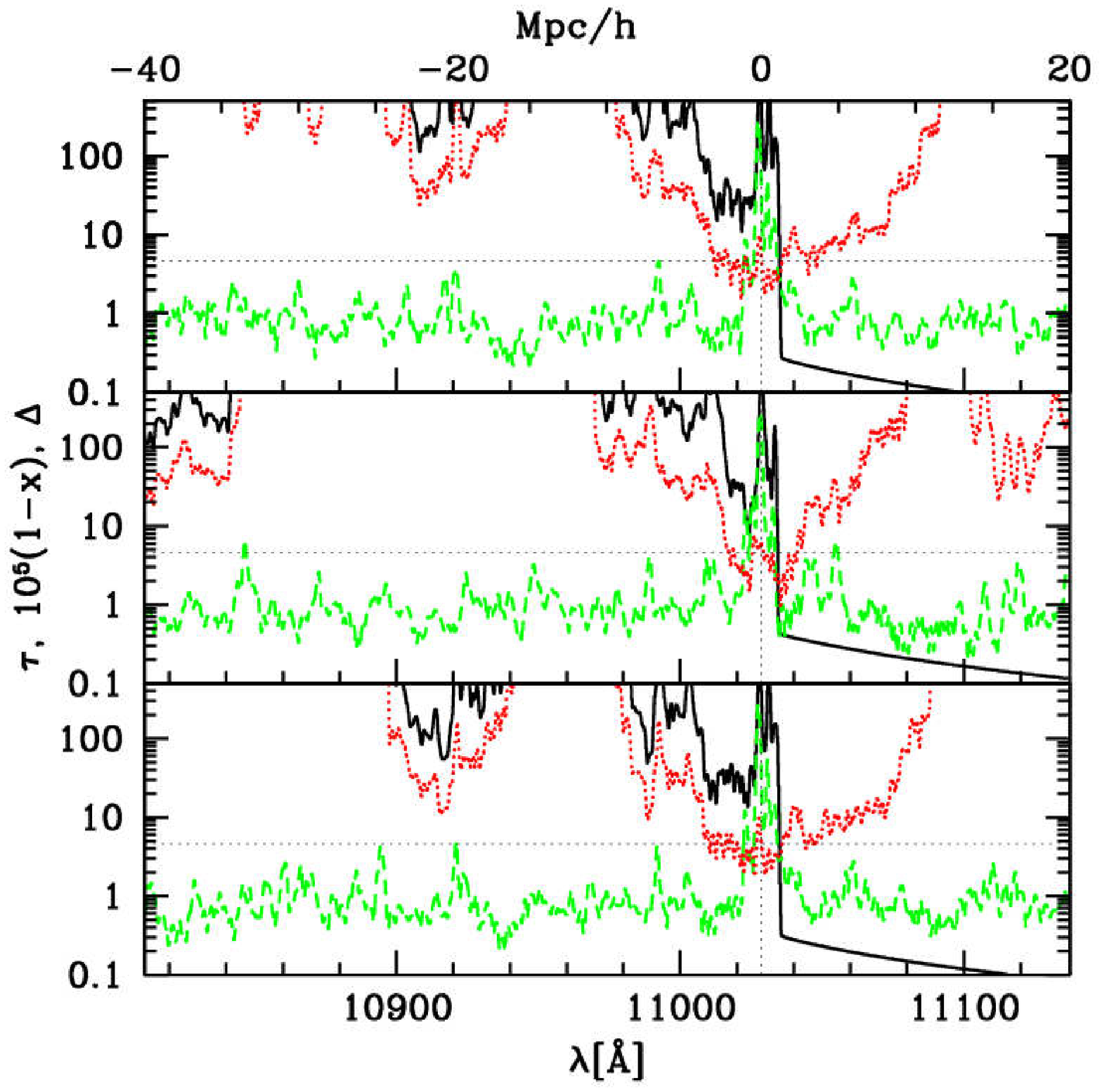}
  \includegraphics[width=3.2in]{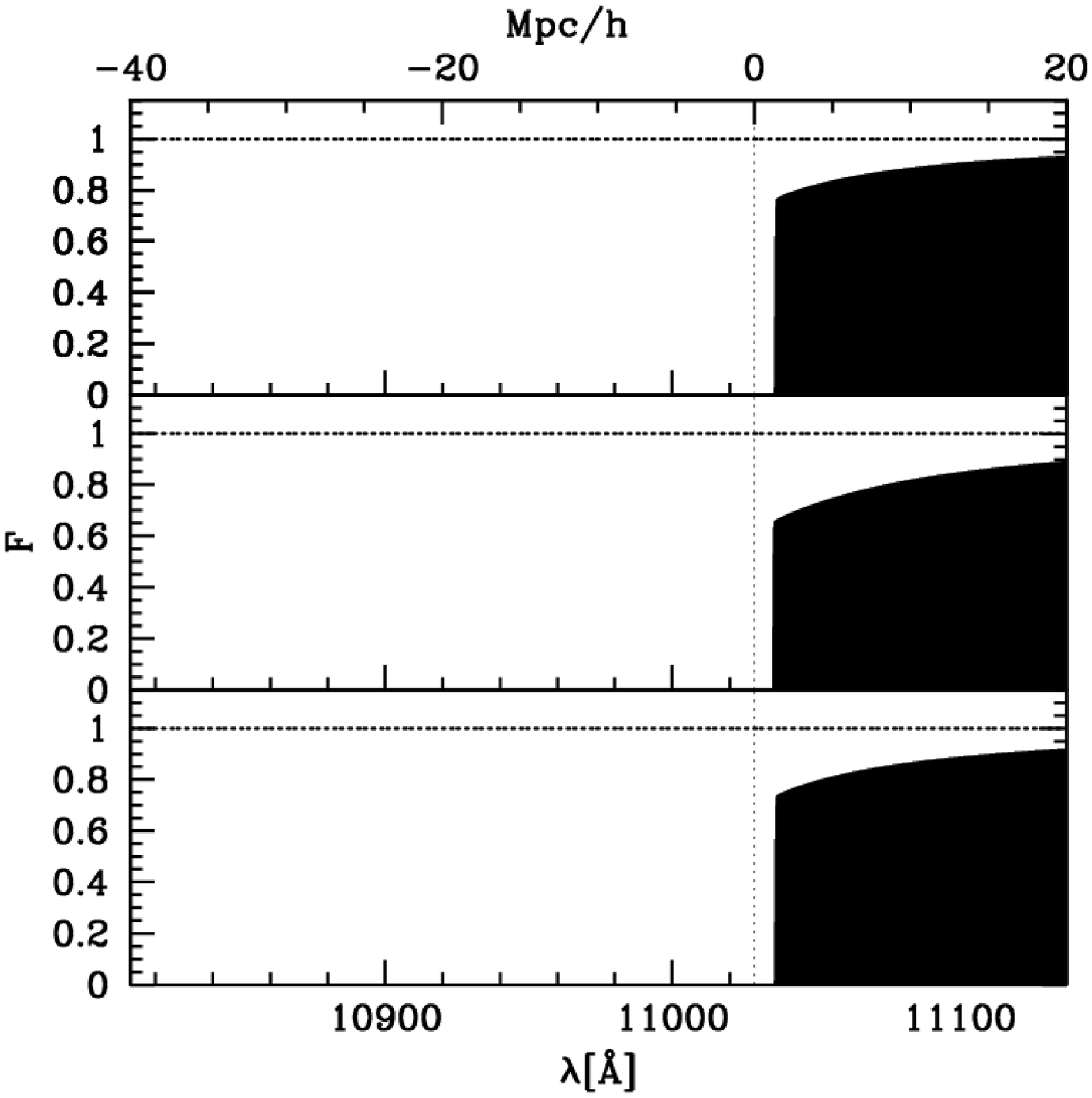}
  \includegraphics[width=3.2in]{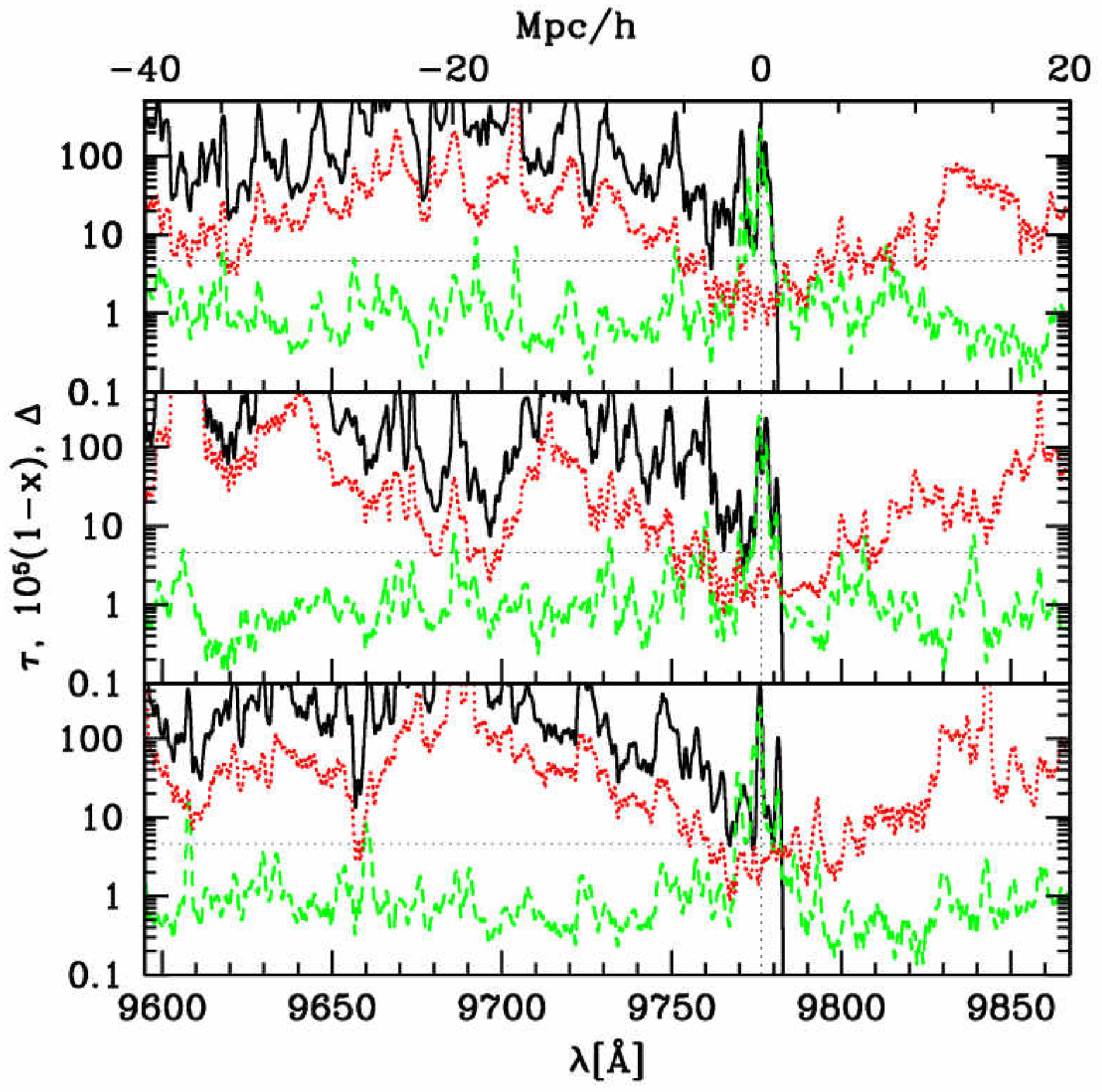}
  \includegraphics[width=3.2in]{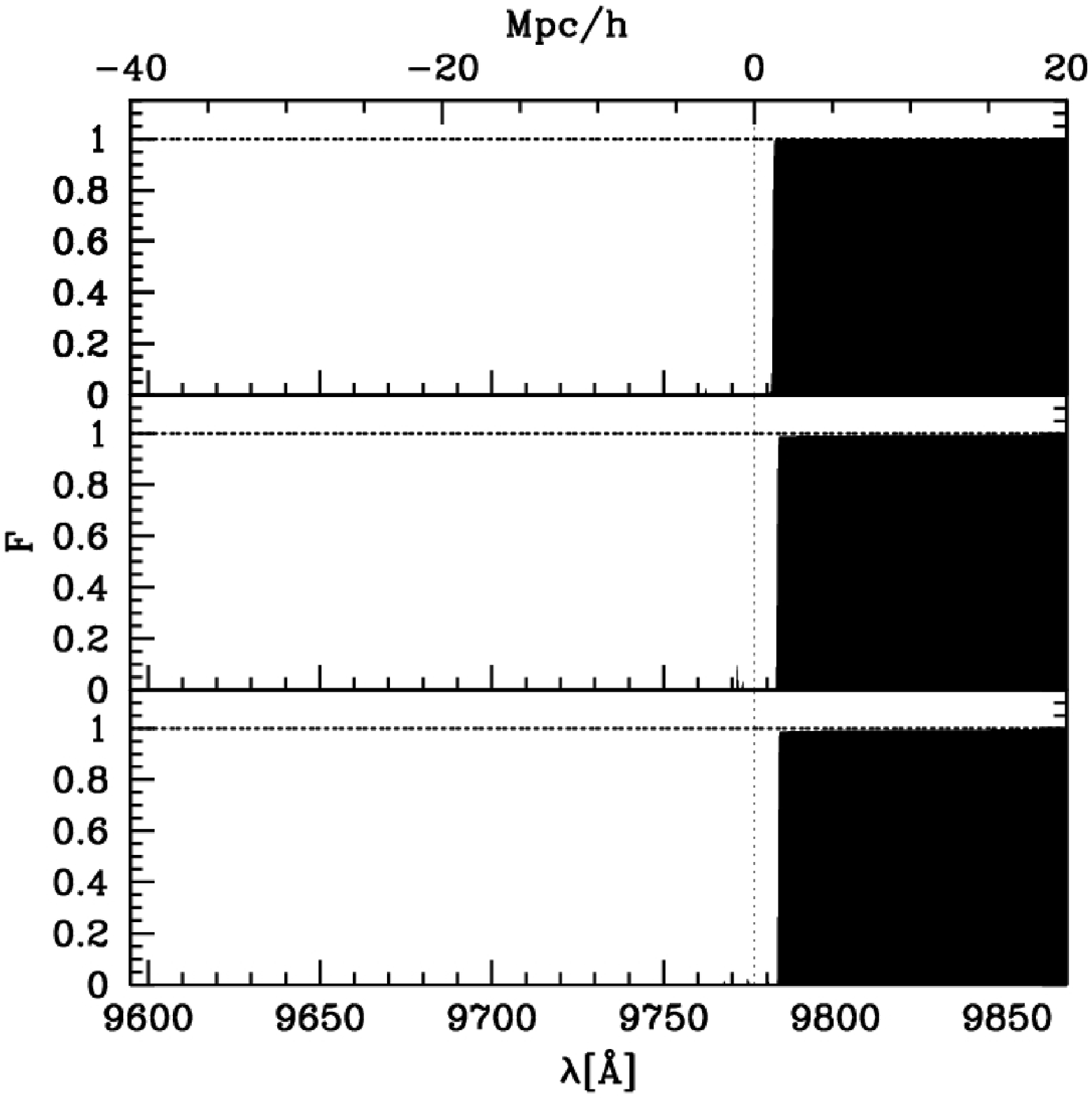}
\caption{Same as Fig.~\ref{spectra}, but at redshifts $z=8.1$ (top; $x_m=0.62$) 
and 7.0 (bottom; $x_m=0.94$).
\label{spectra2}}
\end{figure*}

\begin{figure*}
  \includegraphics[width=3.2in]{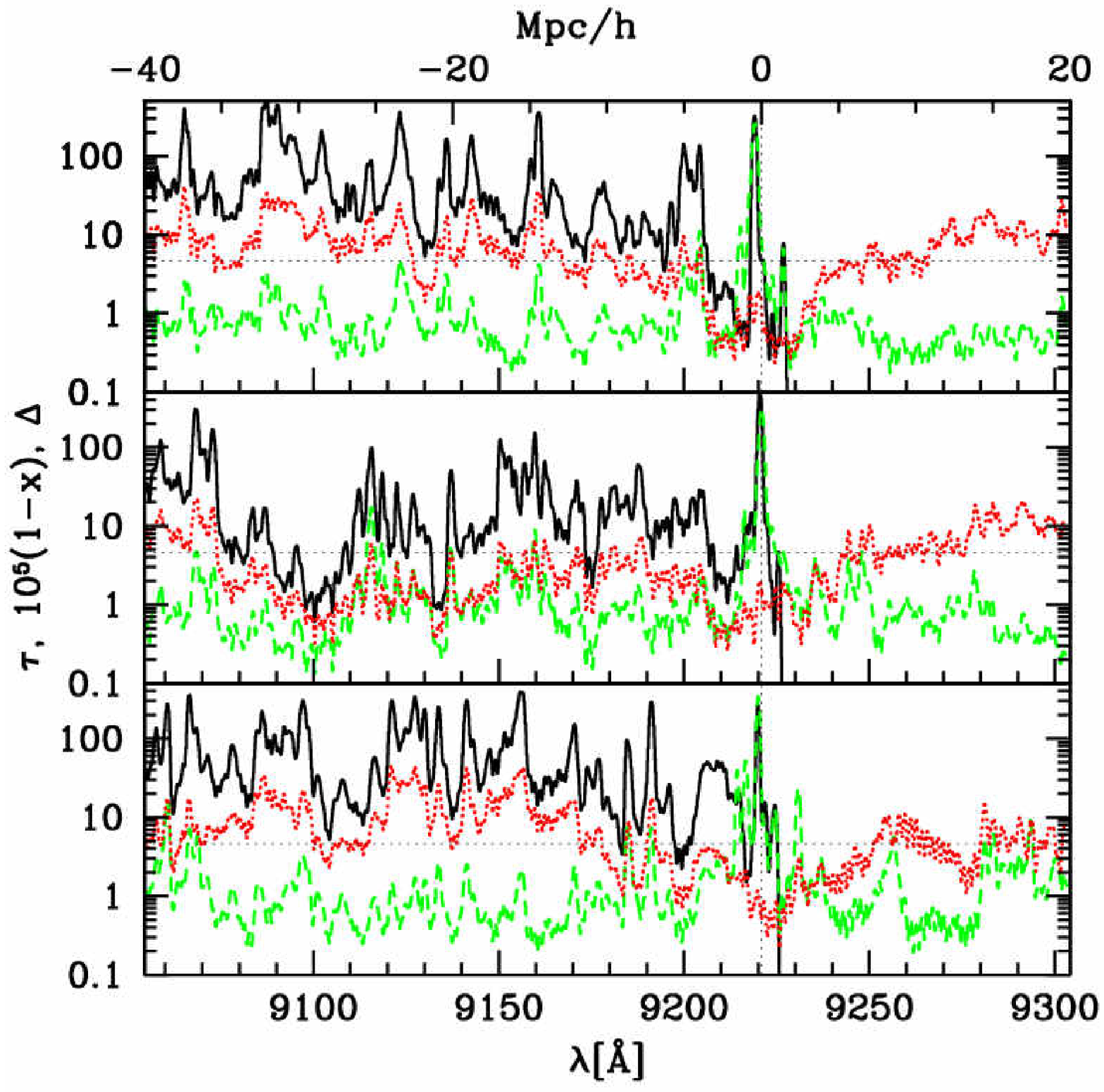}
  \includegraphics[width=3.2in]{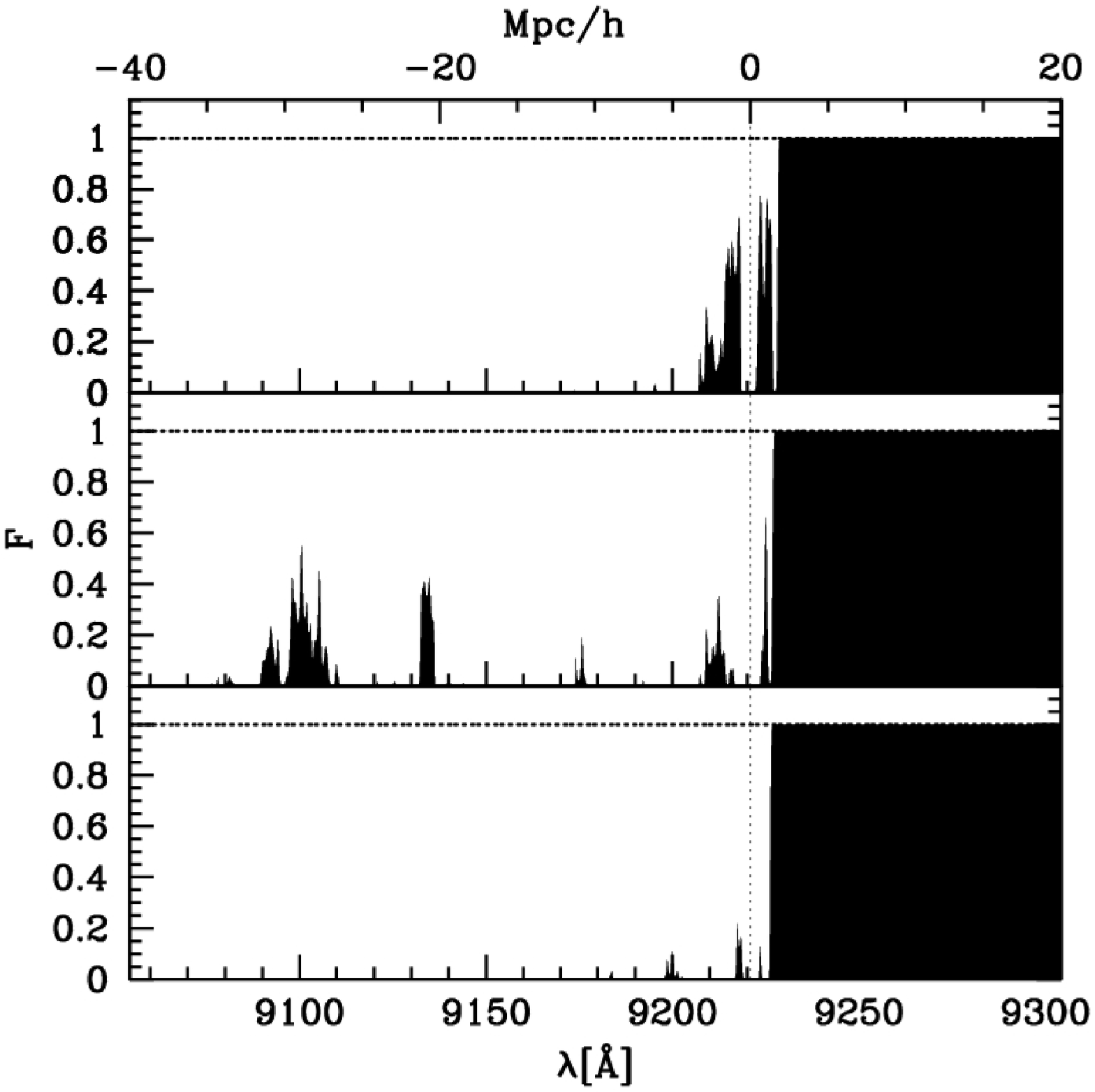}
  \includegraphics[width=3.2in]{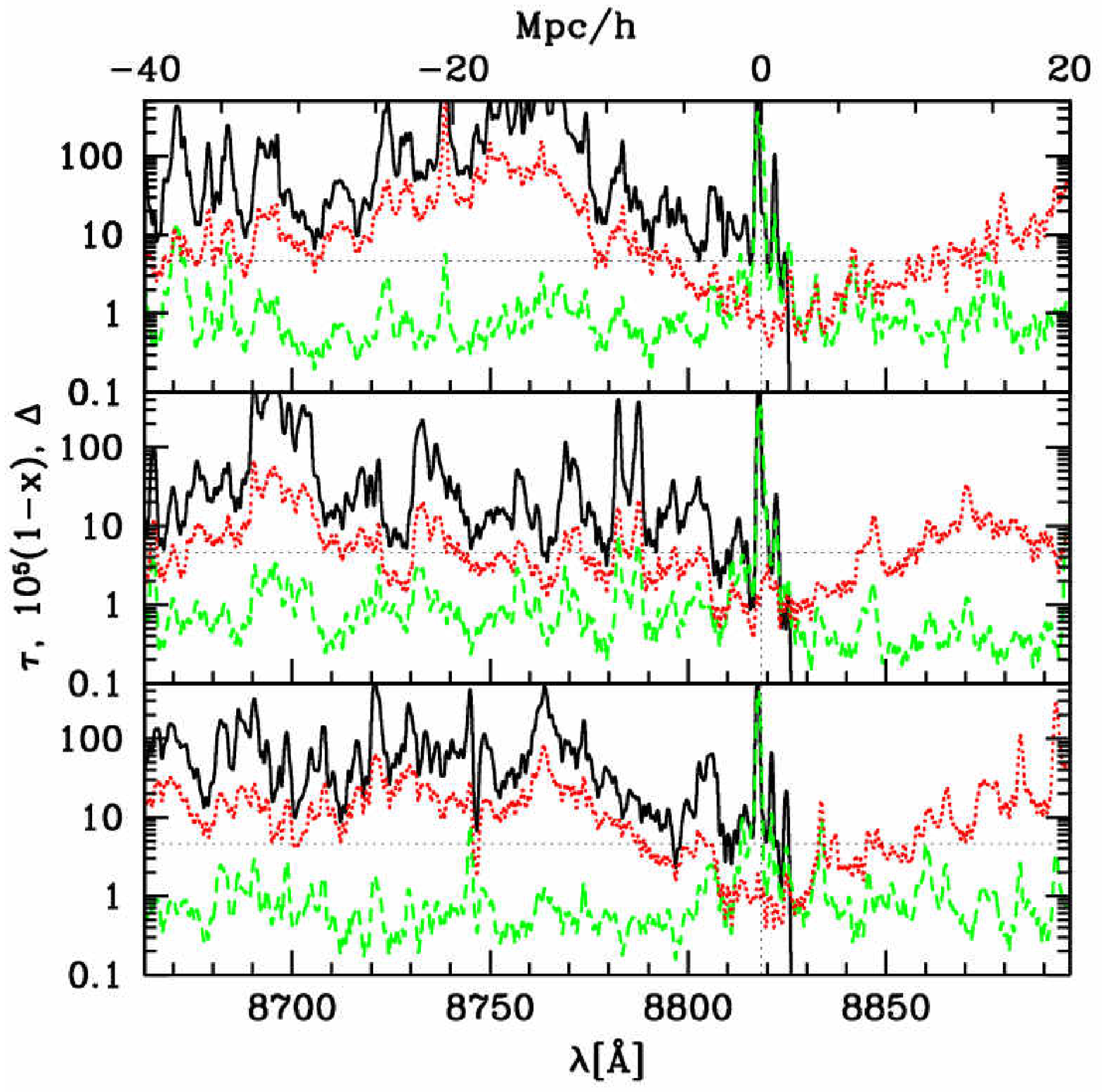}
  \includegraphics[width=3.2in]{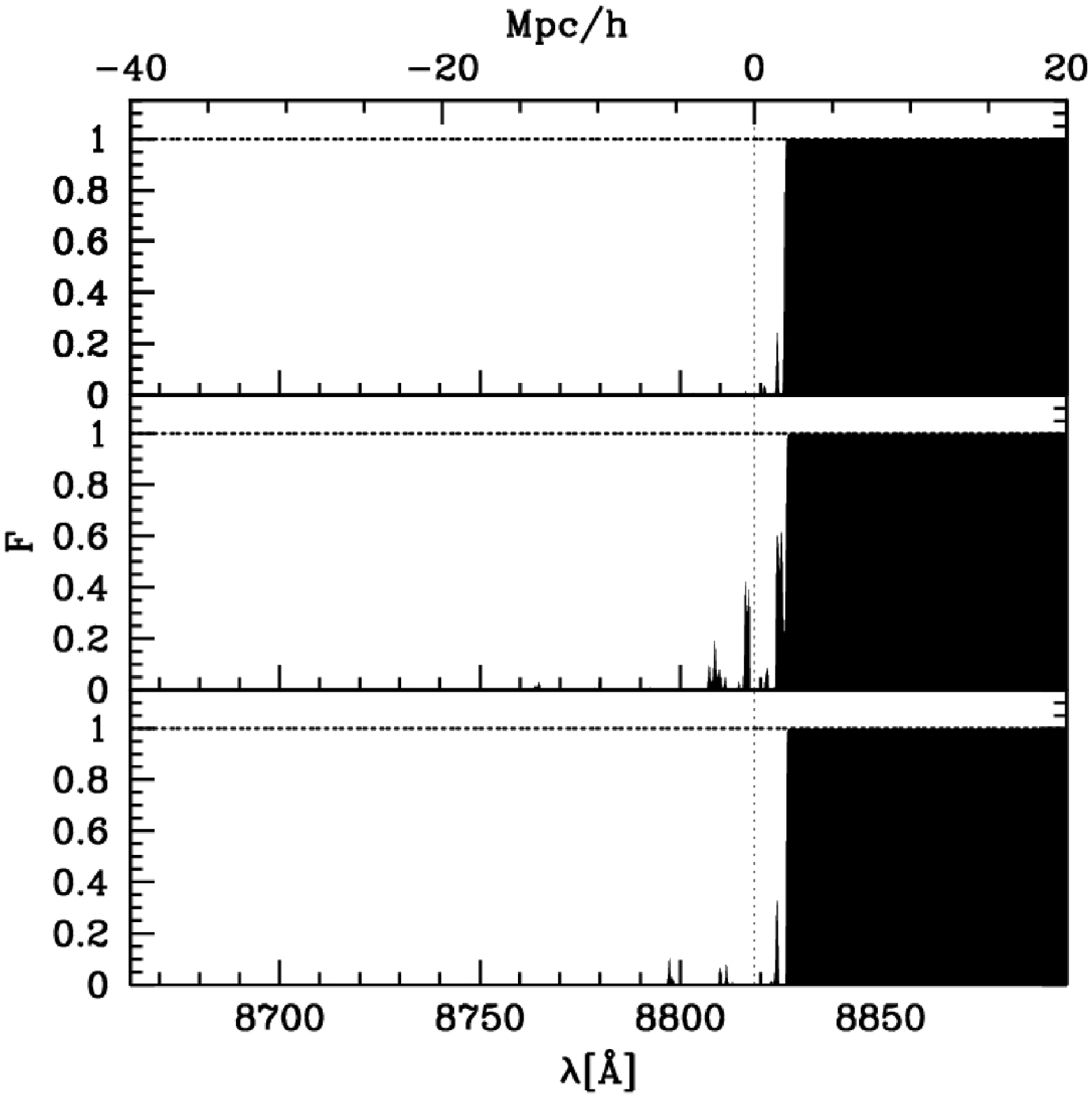}
\caption{Same as Fig.~\ref{spectra}, but at redshifts $z=6.6$ (top; $x_m=0.99$)
 and 6.25 (bottom; $x_m=0.998$).
\label{spectra3}}
\end{figure*}

\begin{figure*}
  \includegraphics[width=3.2in]{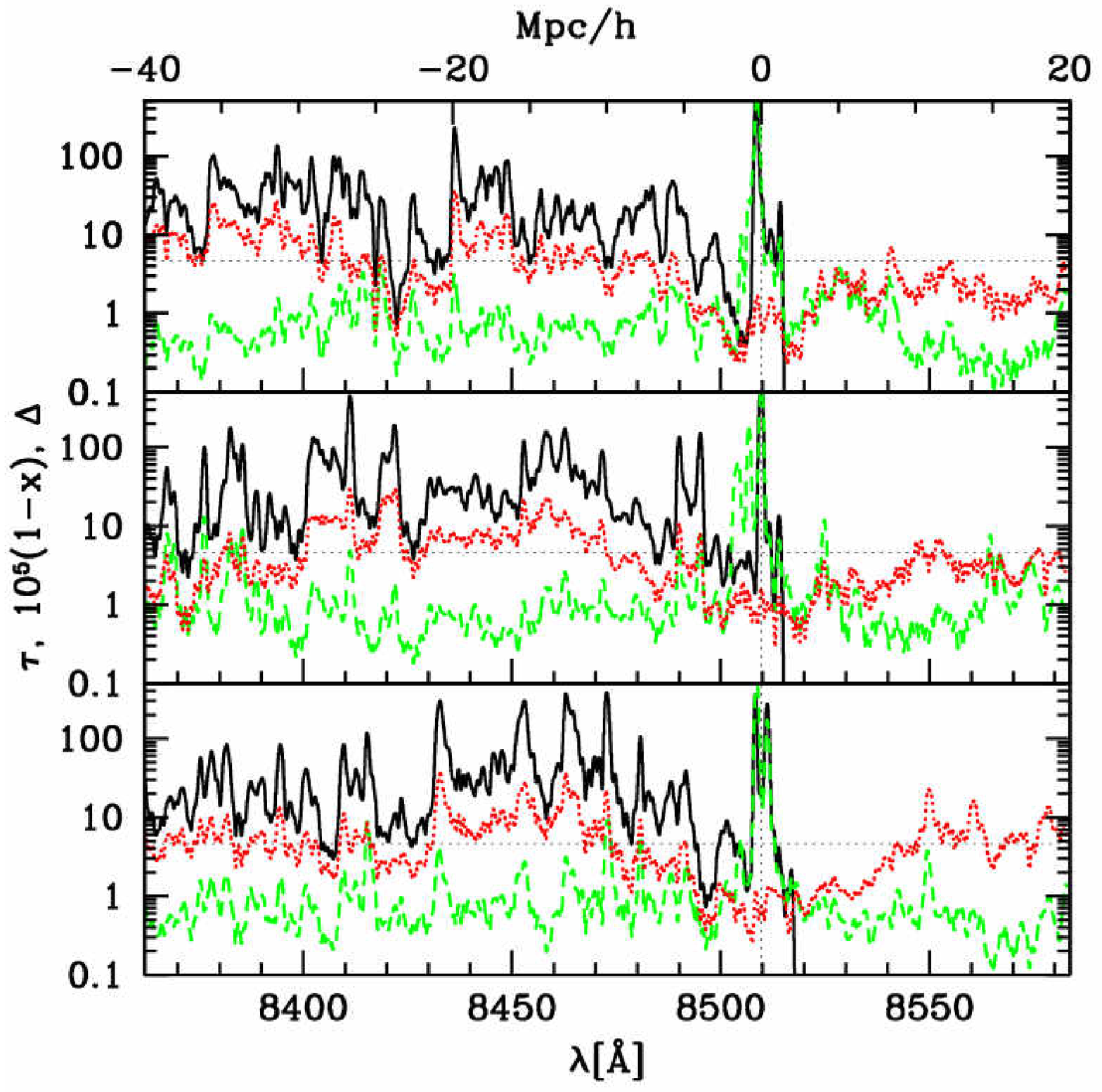}
  \includegraphics[width=3.2in]{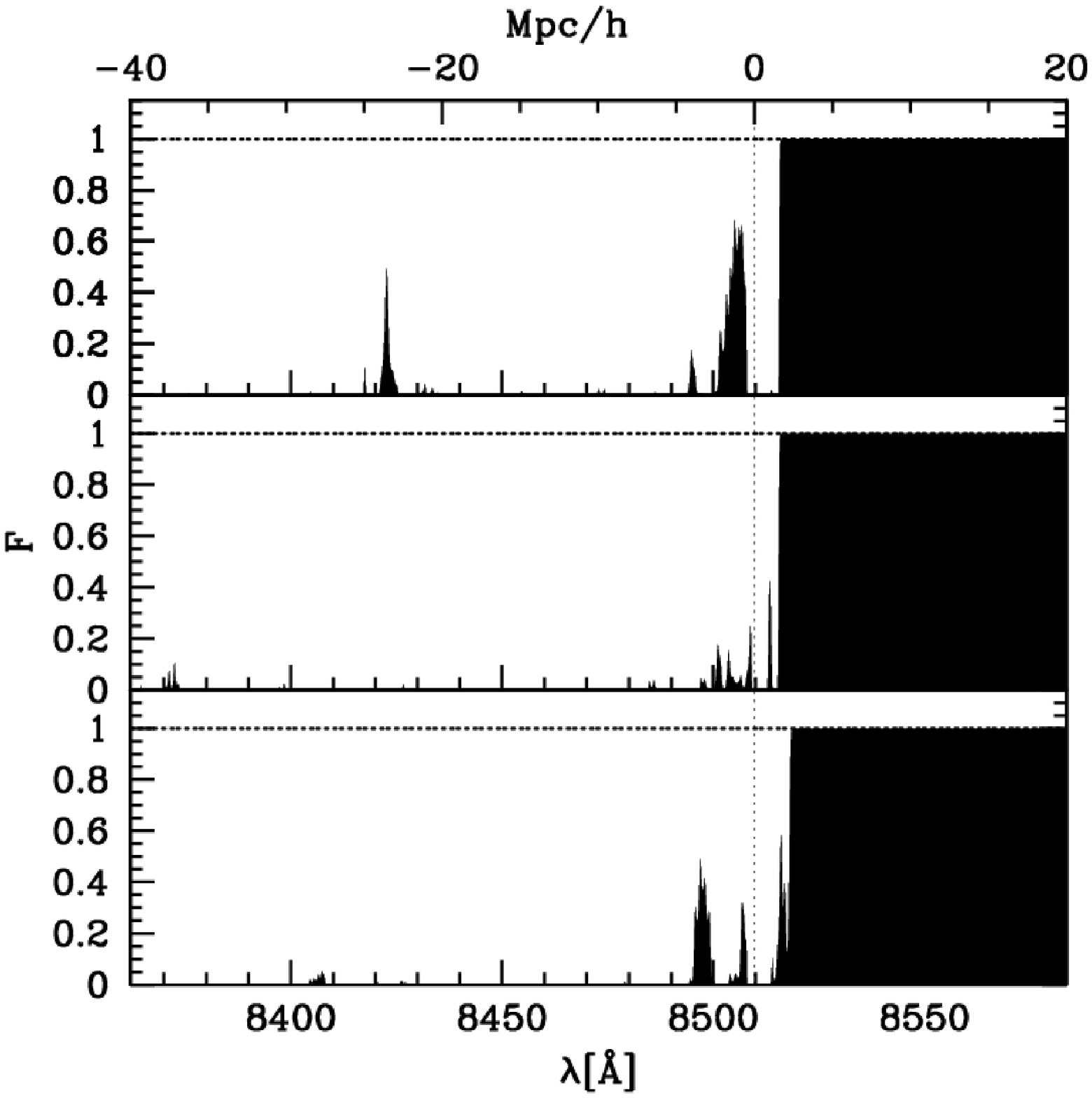}
\caption{Same as Fig.~\ref{spectra}, but at redshift $z=6.0$ ($x_m=0.9999$).
\label{spectra4}}
\end{figure*}

\begin{figure*}
  \includegraphics[width=3.2in]{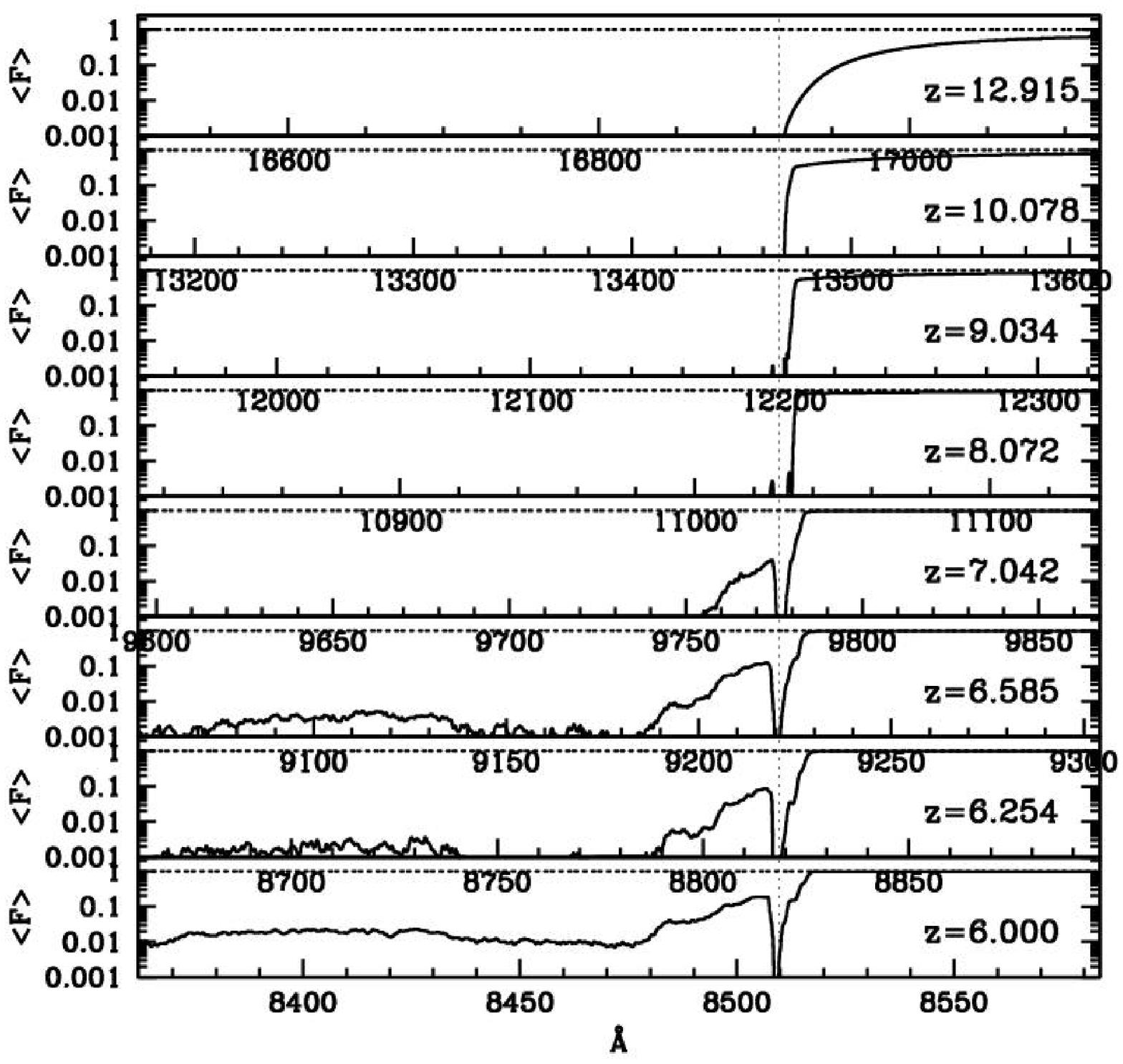}
  \includegraphics[width=3.2in]{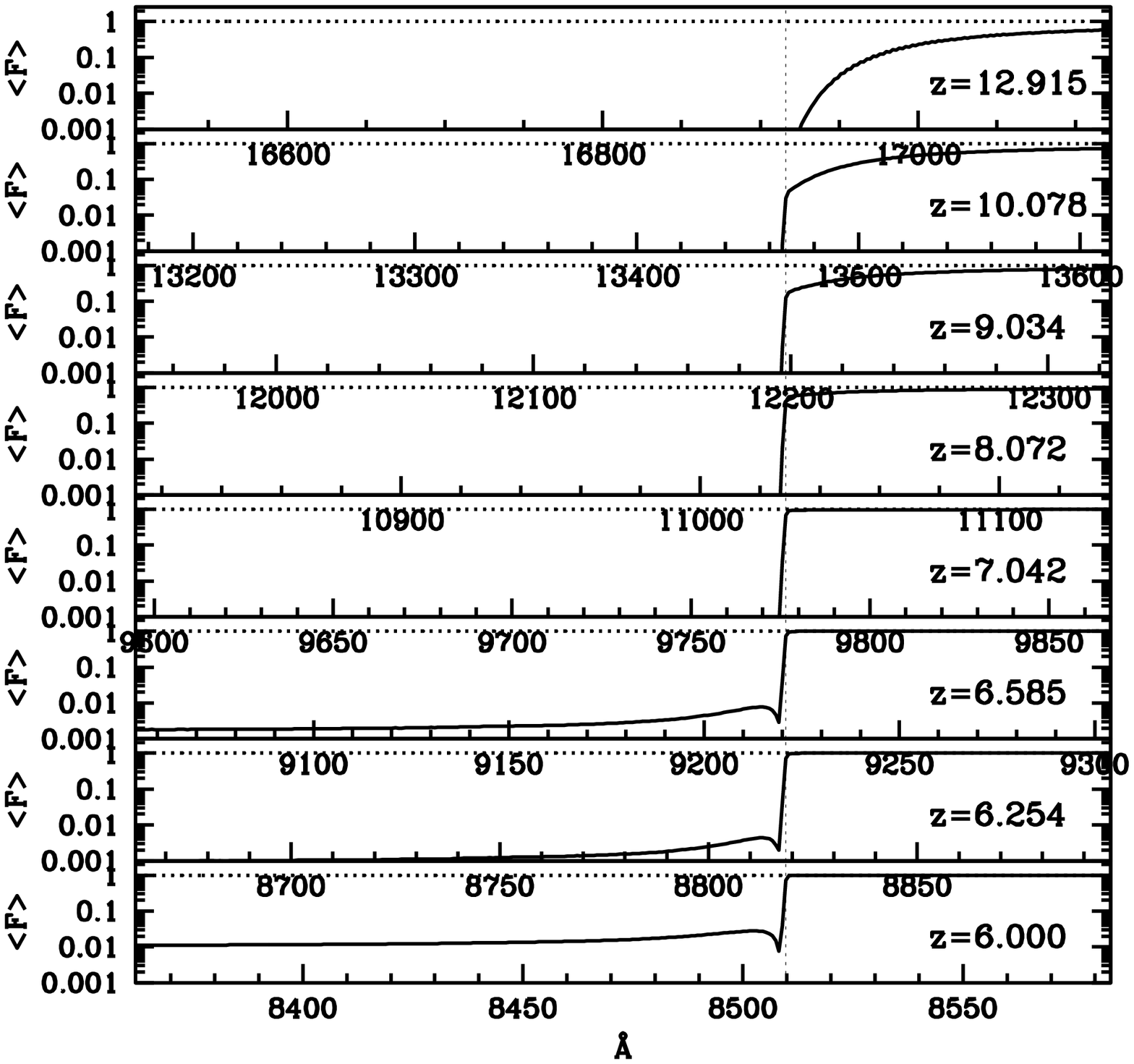}
\caption{Mean radially-averaged transmission around the most massive source
  (left) and average over all sources (right) at several representative 
  redshifts, as labelled. The corresponding ionized fractions by mass are
  $x_m=0.001, 0.105, 0.279, 0.618, 0.939, 0.992, 0.9986$ and 0.9999. 
\label{meanF_fig}}
\end{figure*}

Much of the information about high-redshift Ly-$\alpha$ sources and IGM is
based on absorption spectra. We thus start our discussion of the observable
signatures by presenting and discussing some sample spectra intersecting the 
position of the most luminous source in our computational volume. 
Figures~\ref{spectra}-\ref{spectra4} we show sample spectra along three 
random lines-of-sight  at a few selected redshifts spanning the complete range 
of interest here. On the left panels we show the distributions of 
Ly-$\alpha$ Gunn-Peterson optical depth, $\tau_{\rm GP}$, neutral fraction, 
$x_{\rm HI}=1-x$ (multiplied by $10^5$ for clarity), and gas density in units 
of the mean, $\Delta=n/\bar{n}$. On the right panels we show the corresponding 
Gunn-Peterson transmission spectra for flux level of unity, $\exp(-\tau_{\rm GP})$.
The horizontal lines on the left panels indicate the optical depth of 
$\tau=4.6$, which corresponds to 1\% transmission. For reference, this value is 
roughly equal to the optical depth of a hydrogen gas with neutral fraction of 
$x_{\rm HI}=10^{-5}$ at the mean density at redshift $z=6.6$. All quantities shown 
are in redshift/wavelength space and in the observer ($z=0$) frame. For 
reference, on the top axis of each figure we show the approximate corresponding 
distances in real space. Finally, in Figure~\ref{meanF_fig} we show the mean, 
averaged over all random LOS, transmission spectra at the same redshifts, for 
the most massive source (left) and mean over all sources (right).

The Lyman-$\alpha$ absorption as a function of wavelength is computed using 
the standard procedure (e.g., \cite{1998MNRAS.301..478T}). The optical 
depth and transmission results include the redshift-space distortions due 
to the local peculiar velocities, relative to the peculiar velocity of the 
source (i.e., after applying peculiar velocity distortions, the whole spectrum 
has been shifted slightly so the source is returned to it's real space 
position). The temperature of the gas is assumed to be $10^4$ K when computing
thermal broadening, consistent with the assumption adopted for the
simulations.  

The nominal resolution of our spectra is $R\sim6000-12000$ at $z=6$ (higher
at higher redshifts), based on our grid resolution of $203^3$ (radiative 
transfer) and ($406^3$ density and velocity fields). This resolution roughly 
corresponding to the one for medium-resolution observed spectra.  In reality 
the situation is more complicated. Due to the non-linear 
transformations between our raw simulation data and the final spectra the 
limited simulation resolution can affect the results even if it were better 
than the observational resolution. A separate issue pointing in the same
direction is the fact that the real data has effectively infinite resolution 
in the transverse direction (i.e., the width of the light beam), and thus in
order for us to be accurate we have to ensure we are resolving essentially 
all the transverse structure, which is not the case for the current 
simulations. As a result, our spectra should not be considered completely 
realistic predictions, but rather as a guidance showing some important 
features to be expected from real spectra, as discussed below. Future
higher-resolution, more detailed simulations will be better suited to make 
realistic predictions of the actual detailed spectral properties. 

At early times ($z=12.9$; Figure~\ref{spectra}, top) the H~II region surrounding 
the most massive source is still quite small (see also Figs.~\ref{peak_evol}, 
\ref{hist_fig} and \ref{emiss_fig}) and the Ly-$\alpha$ emission of the source is 
completely suppressed by the damping wing of the Ly-$\alpha$ line profile, 
rendering it unobservable. The ionized region grows quickly after that and by $z=9$ 
reaches size of $\sim6\rm\,h^{-1}Mpc$ (Figure~\ref{spectra}, bottom). Regardless, 
the Ly-$\alpha$ optical depth even within the source proximity region remains quite 
high, at few up to $\sim10$, which allows through only a very weak transmission. 
The damping wing slightly weakens compared to the higher redshifts, but is still 
quite substantial, and still depresses most of the red wing of the emission line. 

Some of the continuum immediately behind the source in redshift space (within a 
few \AA) is absorbed due to gas infall towards the density peak. Because of that 
additional velocity towards the source the gas in front of it is slightly 
redshifted and thus absorbs at wavelengths red-ward of the source position.
The line center at the luminous source's position is completely dark due to the
high density in the middle of the density peak, regardless of the very low neutral 
fraction there. These features persist throughout the evolution and are very 
characteristic for all luminous sources, since these always associated with high 
density peaks and surrounded by infall.   

At redshift  $z=8.1$ (Figure~\ref{spectra2}, top) a weak damping wing is still 
present and essentially no transmission occurs on the blue side of the line. 
The sources at this time may be potentially visible with very deep observations.
Only by redshift $z=7$ (Figures~\ref{meanF_fig}) the ionized region is sufficiently 
large for the damping wing to effectively disappear. However, the gas in the ionized 
region still has a significant neutral fraction, and is sufficiently dense to absorb
all photons on the blue side of the Ly-$\alpha$ line along most LOS. On 
average a weak transmission at a few percent level starts coming through in the 
proximity region of the source (Figure\ref{meanF_fig}). As more and more sources 
form and the ionizing flux rises the first transmission gaps start to appear at 
$z<7$, both in the mean IGM away from the peak and in the source proximity 
(Figure~\ref{spectra3}). At redshift $z=6.6$, our nominal overlap time 
(Figures~\ref{spectra3} and \ref{meanF_fig}), the proximity region extends for 
$\sim30$~\AA\ and has become fairly optically-thin, allowing up to 30-40\% 
transmission. Most of the volume still remains optically-thick, but some substantial 
transmission regions appear in the IGM away from the peak. We also note that
there are significant variations between the different LOS, with some allowing
for much more transmission than others. The size and properties of the
proximity region also vary due to its asymmetry and the anisotropy of nearby
structures. Finally, during the post-overlap epoch (Figures~\ref{spectra3},
bottom, \ref{spectra4} and \ref{meanF_fig}) the IGM slowly becomes more
optically-thin to Ly-$\alpha$ and gradually approaches the state of the
Ly-$\alpha$ forest. There are no more clearly-defined H~II bubbles. Only a few
isolated low-density regions remain neutral. This is due to the inside-out  
character of the reionization process, whereby the high-density regions are 
preferentially ionized first, while the voids, where structure formation is delayed 
are ionized last.

Comparing the radially-averaged mean transmission around a luminous source (high 
density peak) and the average for all sources (i.e. the mean behaviour around a 
typical source; Figure~\ref{meanF_fig}) we see both some similarities and several 
notable differences. The mean damping wing has similar evolution with redshift in 
the two cases, strong at high redshift ($z>10$), gradually becoming weaker 
($z\sim8-9$) and finally disappearing at later times ($z<7$). Naiively, one might 
expect that compared to an average source the luminous ones would suffer from 
weaker damping since such sources typically reside in the middle of large H~II
regions, far from their boundaries. In fact, this expectation proves only partially 
correct. The damping is somewhat more pronounced around an average source than 
around a luminous one, but the differences we observe are rather modest, reflecting 
the fact that weaker sources are strongly clustered around the same density peaks 
that host the luminous ones, thus largely share the damping (or its lack) with the 
central source. For the same reason the damping becomes irrelevant at about the same 
time in both cases, which would not have been the case if the weak sources were 
residing in smaller, isolated bubbles. 

The infall around the high density peak leads to some redshifted absorption that 
appears behind the redshift-space position of the source. The same behaviour is not
seen for a typical source. Such smaller halos tend to move more in tandem with their
surrounding IGM, often towards the nearest high density peak. Some local infall 
should exist also for these halos, but this is at very small scales,
unresolved here. However, these scales are small compared to the typical
emission line width (see next section) and thus we do not expect that such
local infall has significant effect on the emission line.

One final important difference between a luminous and an average source is that the 
latter does not typically have a proximity transmission region on the blue side of 
the line. The spectra of the luminous sources, on the other hand exhibit extended 
high-transmission (10-60\% transmission) regions within 5~Mpc$\,\rm h^{-1}$ 
($\sim30$~\AA). This behaviour is again due to the high source clustering around 
the density peak. The combined effect of all sources is to achieve much lower
neutral gas fraction regardless of the higher gas density there. The line
center coinciding with a high density peak remains optically-thick, however,
unlike the line center of a typical source. Away from the proximity region the
absorption is largely saturated, but there are a number of transmission gaps
with up to a few per cent transmission. Future work would quantify the
statistics of these features and its evolution. 

\begin{figure*}
  \includegraphics[width=3.2in]{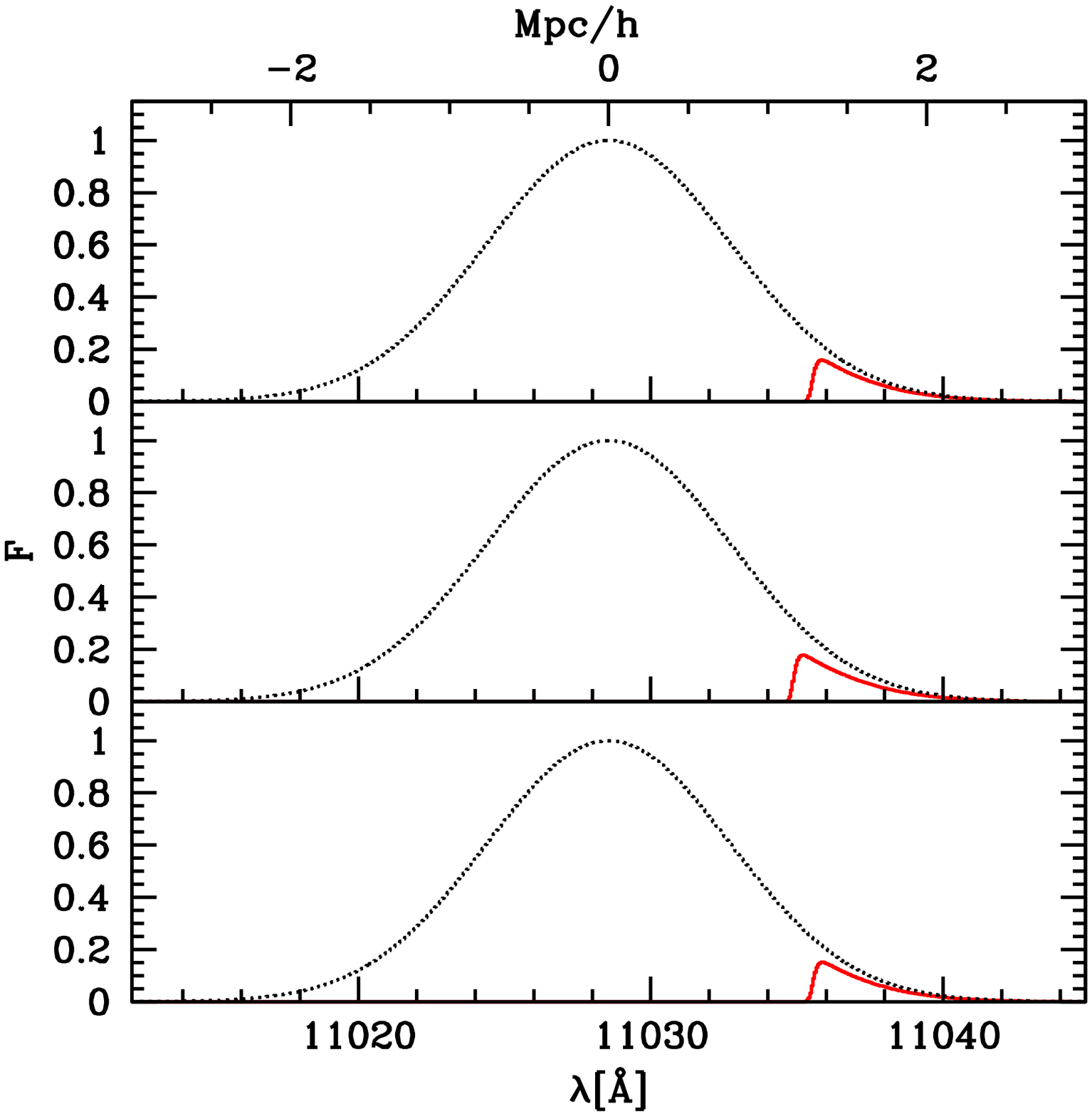}
  \includegraphics[width=3.2in]{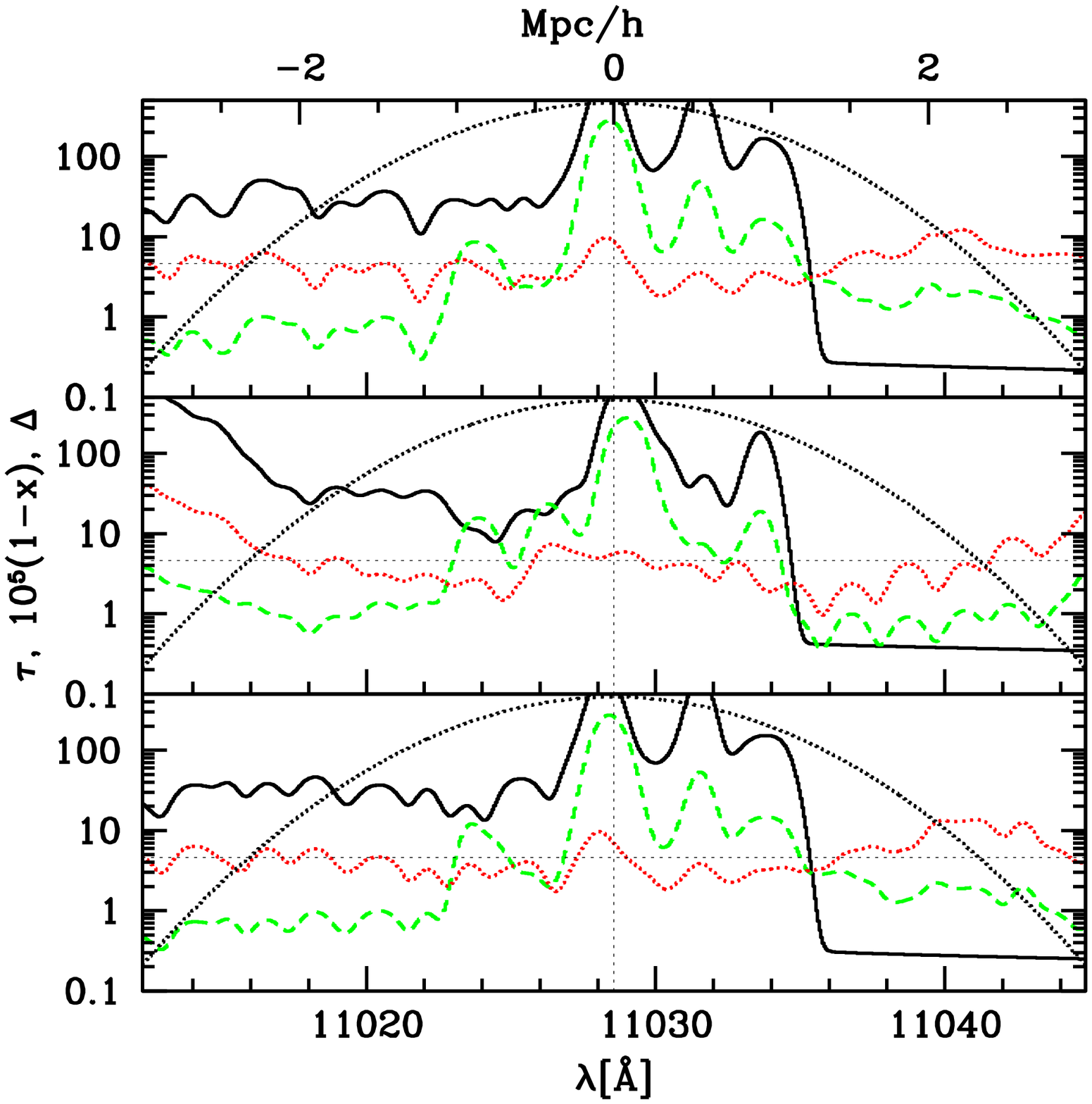}
  \caption{Emission lines: (left panels) intrinsic line, assumed a Gaussian
    with rms width of $160\,\rm km\,s^{-1}$ (black, top), and transmitted one
    (red, bottom) for three sample LOS at redshift $z=8.1$ ($x_m=0.62$), and 
    (right panels)
    the corresponding optical depth (solid, black), neutral fraction $x_{\rm
      HI}=1-x$ ($\times10^5$; dotted, red) and density in units of the mean
    (dashed, green), all in redshift space (i.e. accounting for the relative
    velocities). The intrinsic emission line is also shown for reference.   
\label{line8}}
\end{figure*}

\begin{figure*}
  \includegraphics[width=3.2in]{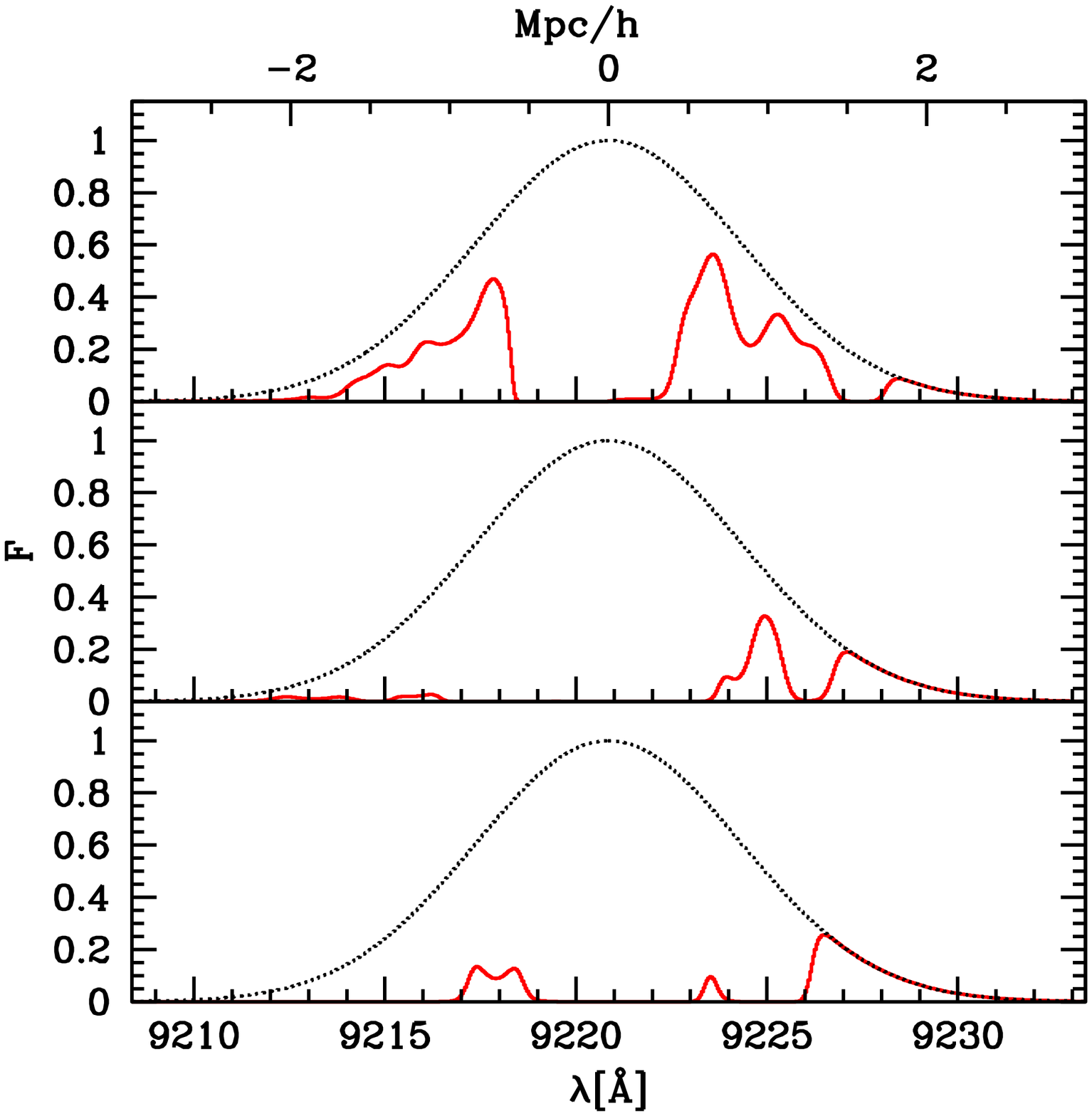}
  \includegraphics[width=3.2in]{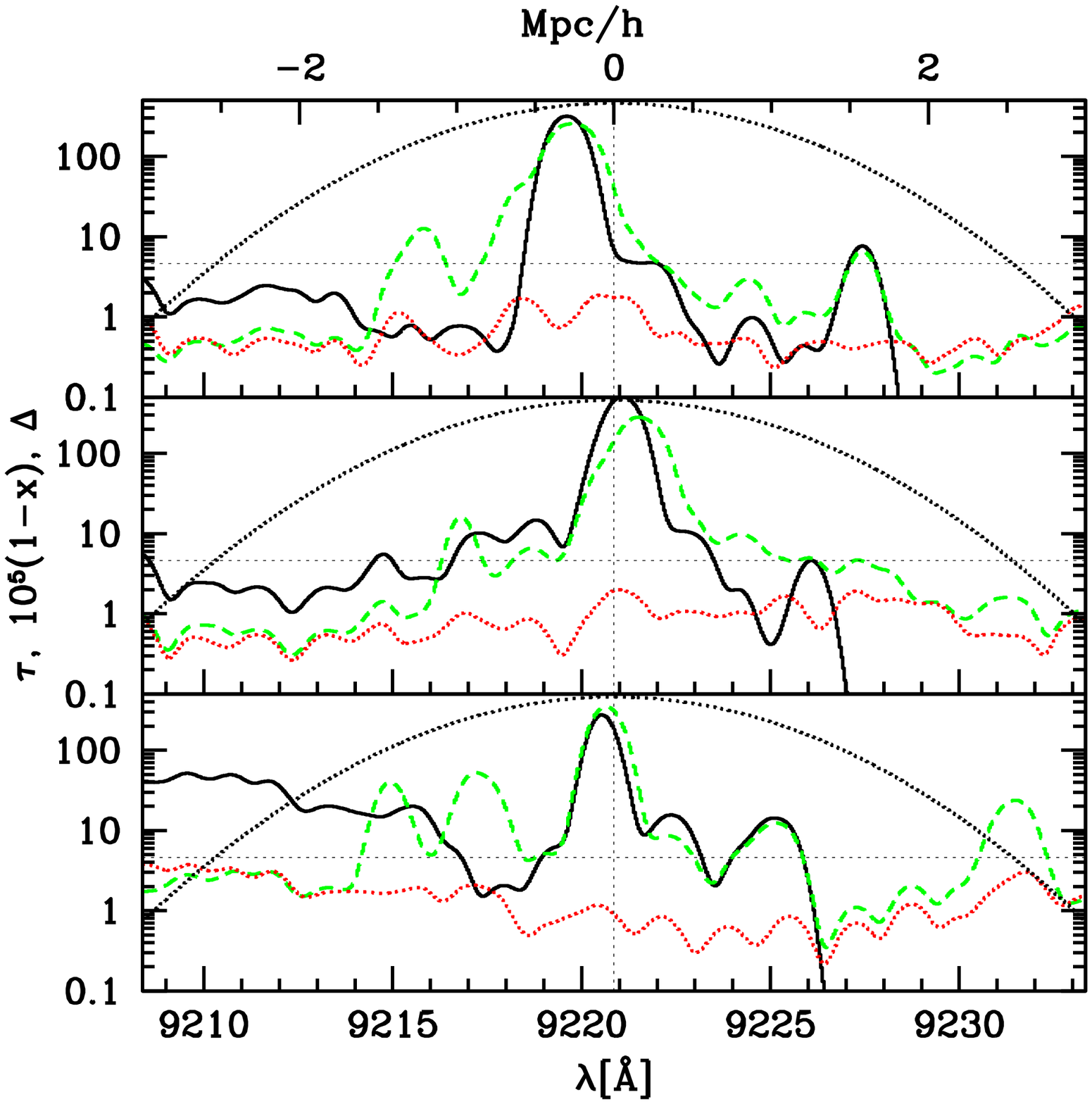}
\caption{Same as Fig.~\ref{line8}, but at redshift $z=6.6$ ($x_m=0.99$).
\label{line6.6}}
\end{figure*}

\subsection{Emission line shape and its evolution}

\begin{figure*}
  \includegraphics[width=3.2in]{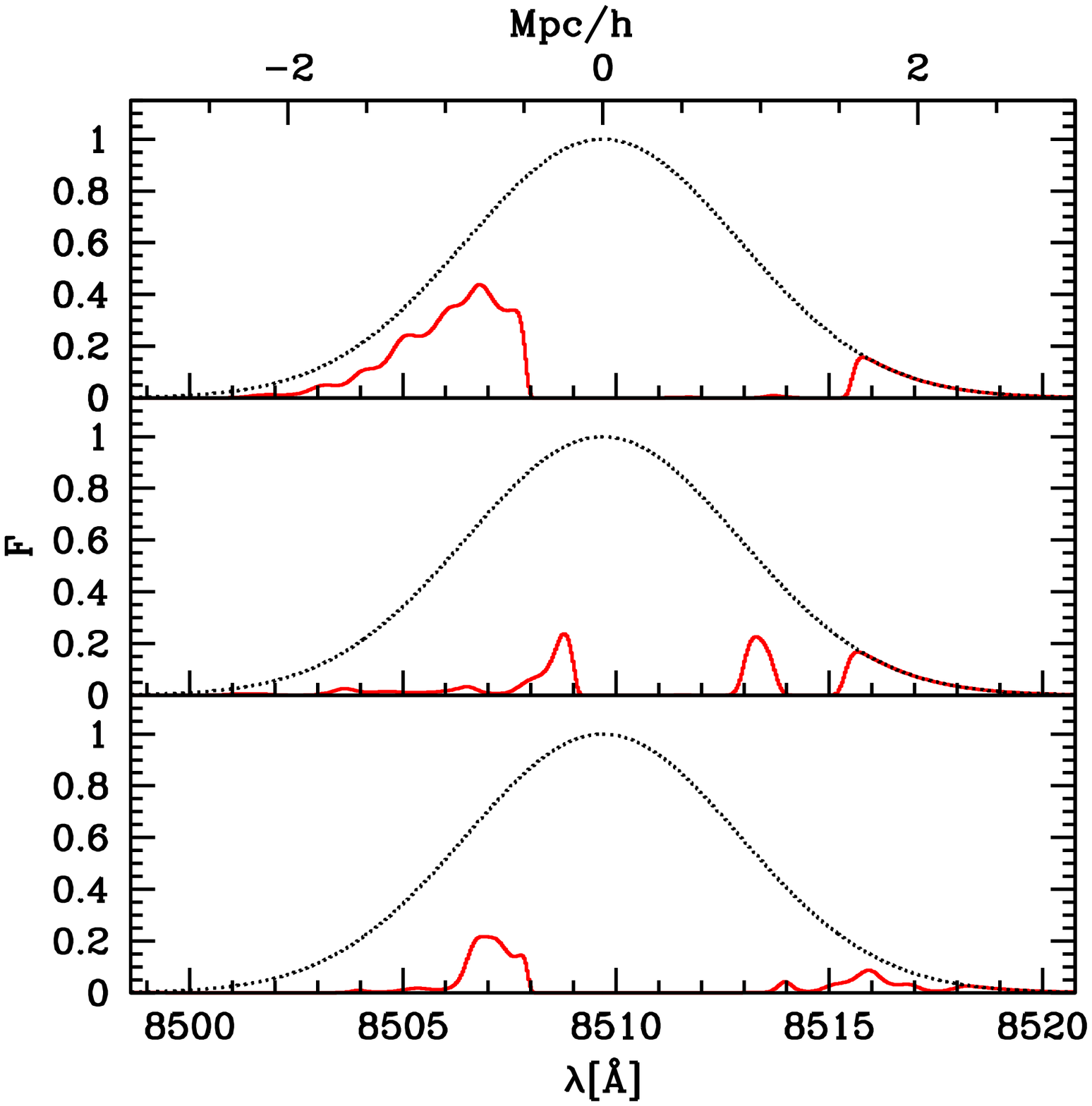}
  \includegraphics[width=3.2in]{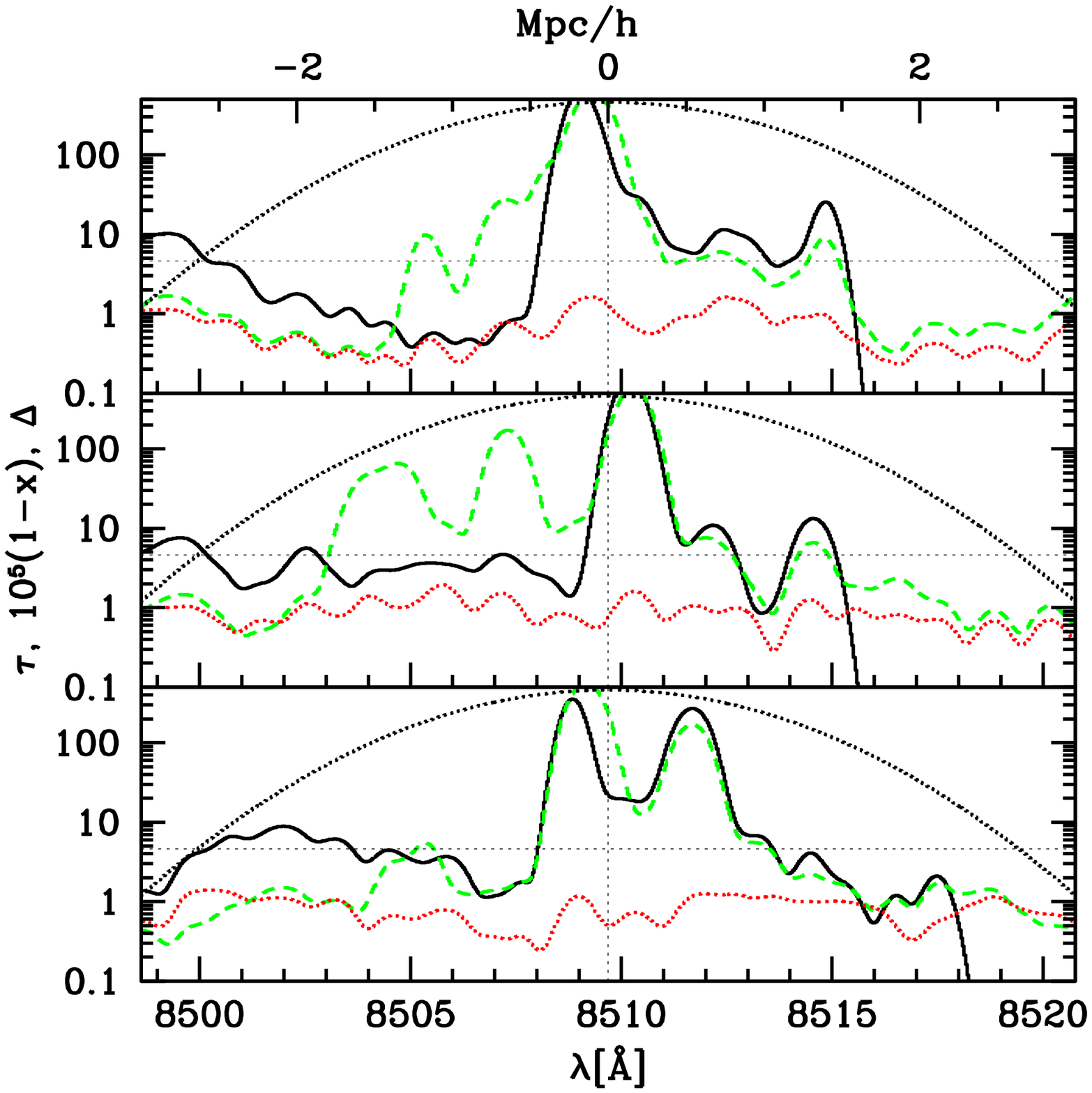}
\caption{Same as Fig.~\ref{line8}, but at redshift $z=6$ ($x_m=0.9999$).
\label{line6}}

  \includegraphics[width=3.2in]{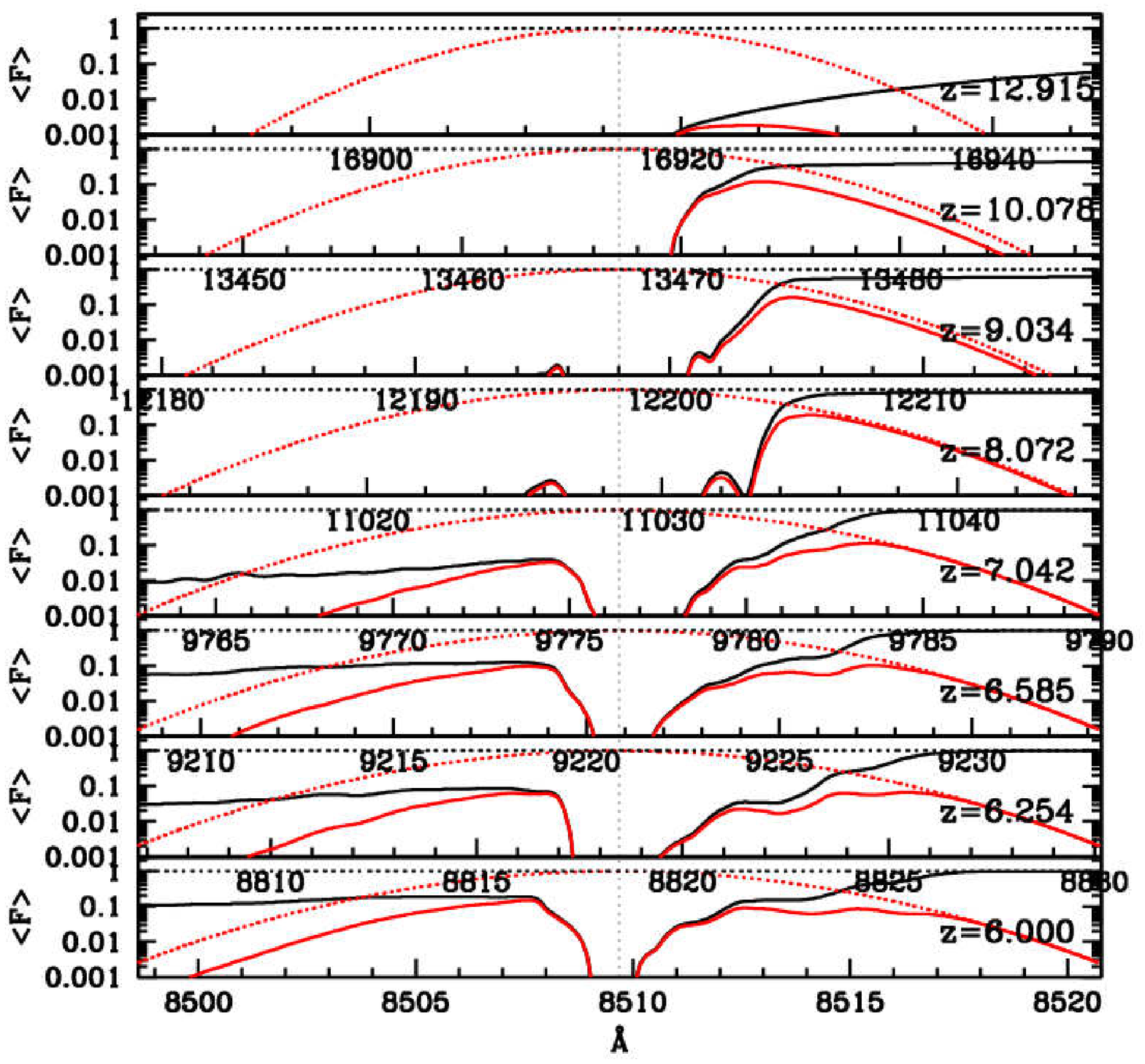}
  \includegraphics[width=3.2in]{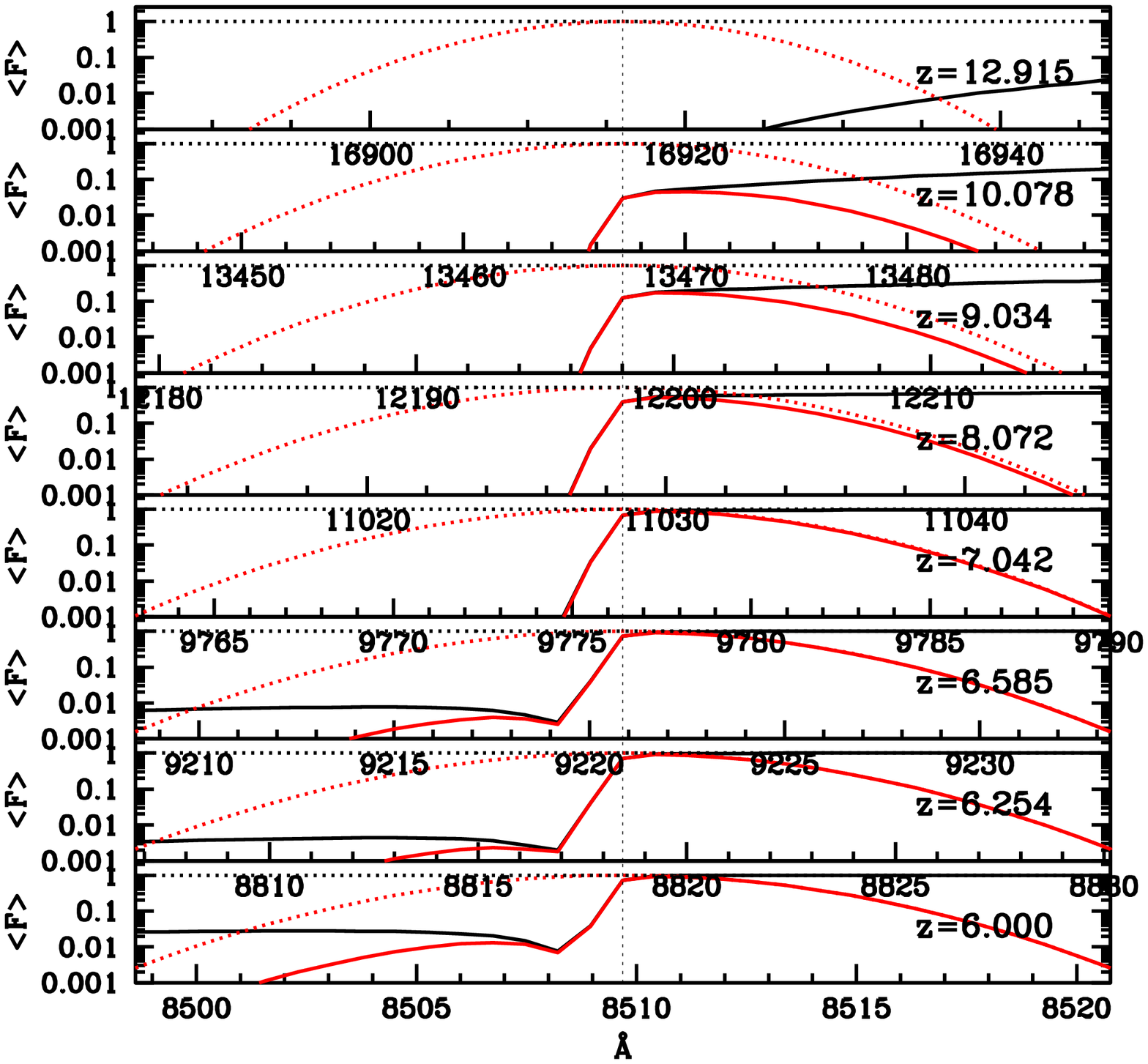}
\caption{Evolution of the mean emission lines for most massive source (left
  panels) and average over all sources (right panels). Shown are the intrinsic
  emission line (dotted, red; assumed a Gaussian with rms width of $160\,\rm
  km\,s^{-1}$, normalized to one at the peak), the transmitted line
  (solid, red), and mean absorption (solid, black) for three sample LOS
  several redshifts, as labelled. 
\label{meanFline_fig}}
\end{figure*}

In order to study the effect of IGM absorption on the line profile shape we model 
the intrinsic Ly-$\alpha$ line as a Gaussian with an rms width of $160\,\rm
km\,s^{-1}$ and peak amplitude normalized to unity. In Figures~\ref{line8},
\ref{line6.6} and \ref{line6} we show sample results for several LOS through
the most luminous source at redshifts $z=8$, 6.6 and 6, respectively. These
examples are picked to illustrate the typical cases for the observed line
shape. On the left panels we show the assumed intrinsic (black) transmitted
(red) emission line, while on the right panels we show the corresponding
distributions of Ly-$\alpha$ Gunn-Peterson optical depth, $\tau_{\rm GP}$,
neutral fraction, $x_{\rm HI}=1-x$ (multiplied by $10^5$ for clarity), and gas
density in units of the mean, $\Delta=n/\bar{n}$ and again (for reference) the
intrinsic emission line.

At redshift $z=8.1$ (Figure~\ref{line8}) the emission line shape is fairly
regular and does not vary strongly for different LOS. The blue wing is
generally highly absorbed and no appreciable flux comes through and much of the
red wing is absorbed as well. The reasons for this behaviour become clear from 
the right panels. The neutral fraction is fairly uniform throughout this
region, at $x_{\rm HI}\sim4\times10^{-5}$, thus the GP optical depth largely
follows the local density field. The high density of the peak and its
immediate vicinity results in optical depths of $\tau_{\rm GP}>10$ everywhere
and $\tau>100$ at the peak itself. The gas infall towards the peak leads to a
significant absorption of the red wing. The only transmitted flux comes from
the far red side of the line (slightly depressed by the weak remaining damping
wing). 

At overlap (1\% global neutral fraction by mass; $z=6.6$) the line shape
becomes much more irregular and varies significantly between the different LOS
(Figure~\ref{line6.6}). Significantly larger fraction of the flux is
transmitted, both on the red and on the blue side of the line. The neutral
fraction in the vicinity of the source is still fairly uniform, but much lower
than at the earlier time, at $x_{\rm HI}\lesssim\times10^{-5}$. Thus the GP 
optical depth again largely follows the density distribution. The effect of
the gas infall towards the peak is still present and some of the red wing is
absorbed, but much less so than at higher redshifts since the infalling gas
is more highly-ionized. The line center remains absorbed and all damping wing
effects have disappeared.   

After overlap (Figure~\ref{line6}) the neutral fraction gradually declines,
decreasing the optical depth and allowing ever more flux to be transmitted.
The neutral fraction is still fairly uniform and thus the optical depth mostly
follows the density fluctuations. The line shapes remain rather irregular,
with significant variations between the different LOS. There is transmission
in both red and blue wing of the line.
 
In Figure~\ref{meanFline_fig} we show the evolution of the mean (i.e. averaged
over all LOS) observed emission line shape for the most luminous source in our
volume (left) and average over all sources (right). In both cases the line
starts completely damped ($z=12.9$). The later evolution of the mean line
shape differs significantly, however. By redshifts $z=9-10$ a significant
fraction of the red wing of the line is transmitted for the luminous source,
except for the line center, while much less of the red wing is transmitted for
an average source. At later times ($z\sim7-8$) this situation is reversed -
practically all of the red wing of the line is transmitted for an average
source, but much of the flux is still absorbed for the luminous source due to
the high density peak in the middle and its surrounding infall. As a word of
caution we should note that some of this effect is in fact not physical but 
numerical, since our simulations do not resolve well the detailed structure
around the smaller halos. However, as we also mentioned above, these
resolution effects should be modest considering that the emission line is
fairly wide and thus reasonably well-resolved and any corrections due to
smaller-scale structures will not affect much of the line. 

On the blue side of the line there are further important differences between
the luminous and average sources. The strong clustering of sources around the
density peaks result in very high fluxes and thus a more pronounced
highly-ionized proximity region blue-ward of the line center. Thus,
significantly more of the blue wing of the luminous source line is
transmitted, up to 10\% on average at $z=6$, vs. only $\sim1-2\%$ for 
an average source.  

An interesting consequence of the very high absorption observed at 
the line center for massive sources and the redshift-space distortions 
due to the gas infall (the latter similar to the one studied 
theoretically in a more idealized setup by \citet{2004ApJ...601...64B}) 
is that the Ly-$\alpha$ line naturally takes a double-peaked (or even 
multiple-peaked in some cases) observed profile. This suggests that in
principle it might be possible to use the line profiles of bright 
Ly-$\alpha$ sources to study the infall surrounding their host halos. In 
practice this might be difficult due to a number of complications. The 
line structure is quite different along different LOS, partly (as we 
pointed above in \S~2.4) due to the very anisotropic velocity structure 
surrounding the source, as well as its own peculiar motion. Furthermore,
our analysis only takes into account the effects of the IGM on the line 
shape, while realistic ones will be affected also by the host galaxy's 
internal structure, outflows, etc. Modelling all those effects correctly
will require high-resolution radiative-hydrodynamic simulations, which
is well beyond the scope of this work.

\subsection{Evolution of the mean transmissivity}
\label{meanFabg_sect}

\begin{figure}
  \includegraphics[width=3.in]{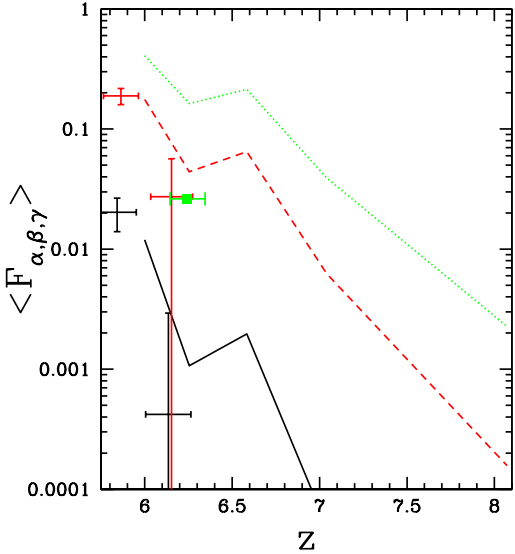}
\caption{Overall mean transmission fraction (not necessarily near a source) in
Lyman-$\alpha$ (solid, black), Lyman-$\beta$ (dashed, red), and 
Lyman-$\gamma$ (dotted, green). The points with horizontal error-bars show the
measurements of \citep{2006AJ....132..117F}.  The (black, red), (lower,
upper), points with vertical error bars show Ly-($\alpha$,$\beta$), while the
green dot shows Ly-$\gamma$ (see text). 
\label{meanFabg}}
\end{figure}

Figure~\ref{meanFabg} shows the mean transmission fraction as a function of
redshift, for the Lyman-$\alpha$, $\beta$, and $\gamma$ transitions. 
The latter two are, respectively, 6.2 and 17.9 times weaker than 
Lyman-$\alpha$, so transmission can be seen in cases where Lyman-alpha would
be opaque \citep{2006AJ....132..117F}.  At the low redshift end of our
simulation data these quantities have been observed in the highest redshift 
quasar spectra \citep{2006AJ....132..117F}. Fig.~\ref{meanFabg} shows the 
\citep{2006AJ....132..117F} (points with error bars).  At $z>6$, the 
Ly-$\alpha$ and Ly-$\beta$ transmission measurements are only upper 
limits.  It is unclear exactly what error bar should be assigned to the 
Ly-$\gamma$ point.  Fan et al. (2006) presented two upper limits and a 
detection in Ly-$\gamma$, which, taken together, support a detected 
mean transmission level (plotted in the figure) well below our 
prediction, with relatively small ($\sim 40$\%) errors;  however, with 
only these three points we can not be sure that the sample variance 
isn't substantially larger.
It appears that the reionization model in our simulations reproduces 
roughly the correct tail-end of reionization, but the data favor 
somewhat less transmission than is present in the model, i.e., a weaker 
radiation background and possibly slightly later reionization.

\begin{figure}
  \includegraphics[width=3.in]{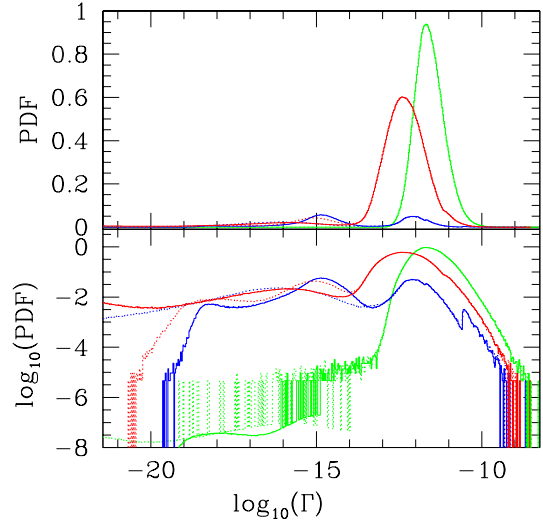}
\caption{PDF of the photoionization rate at  at $z=10.1$ (blue; $x_m=0.105$),
  $z=7.0$ (red; $x_m=0.94$), and $z=6.0$ (green; $x_m=0.9999$). We show the 
  actual, non-equilibrium
  rates (solid) and the corresponding equilibrium rates (dotted, same color at
  each redshift). All PDF's are normalized to have an area of unity below the 
  curve. 
\label{Gamma_pdf}}
\end{figure}

\subsection{Photoionization Rates}
\label{photoion_rates_sect}

In Figure~\ref{Gamma_pdf} we show the normalized probability density 
distributions (PDFs) of the nonequilibrium photoionization rates for 
all cells in our computational volume at three representative redshifts 
- $z=10.1$ ($x_m=0.105$), 7.0 ($x_m=0.94$) and 6.0 ($x_m=0.9999$)
(early times, late times and well after overlap) in linear (top) and log 
(bottom) scales. For comparison we also plot the photoionization rates 
if the corresponding cells were in ionization equilibrium.

\begin{figure*}
  \includegraphics[width=3.2in]{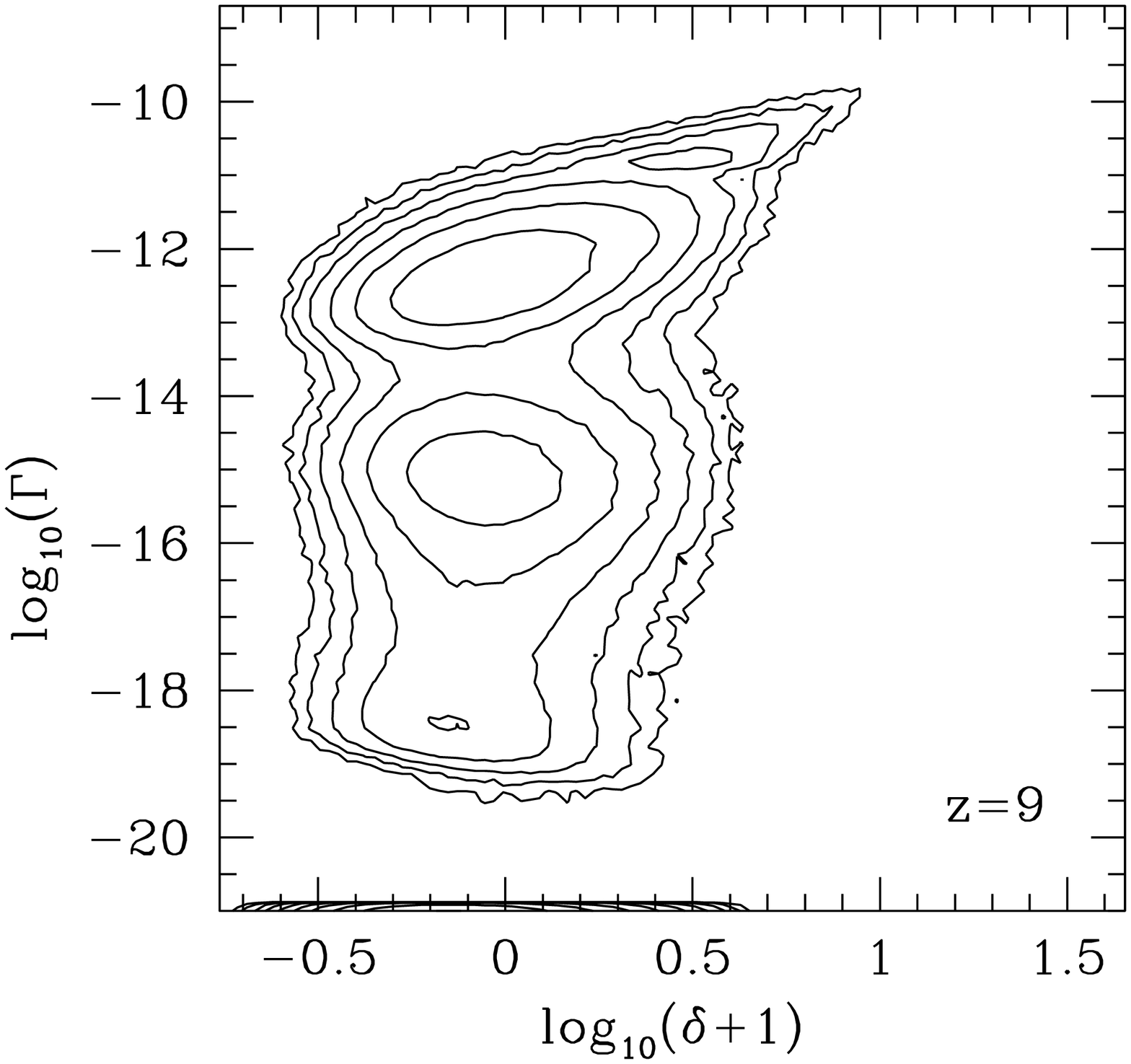}
  \includegraphics[width=3.2in]{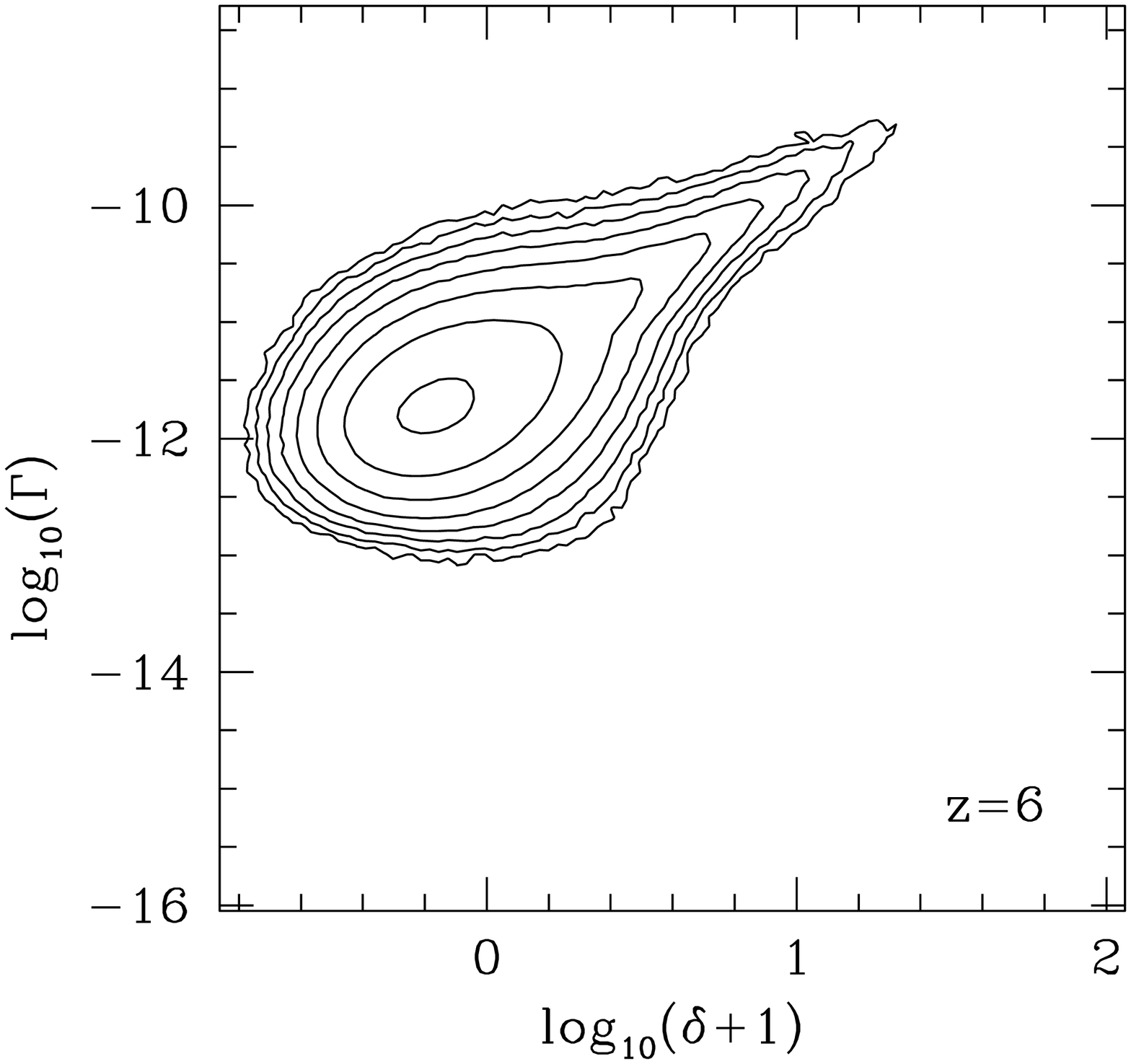}
\vspace{-0.5cm}
\caption{Photoionization rate - overdensity correlation  at $z=9$ (left; 
  $x_m=0.28$) and $z=6$ (right; $x_m=0.9999$). Contours are logarithmic, 
  from 10 cells up every 0.5 dex.
\label{Gamma_distr}}
  \includegraphics[width=2.3in]{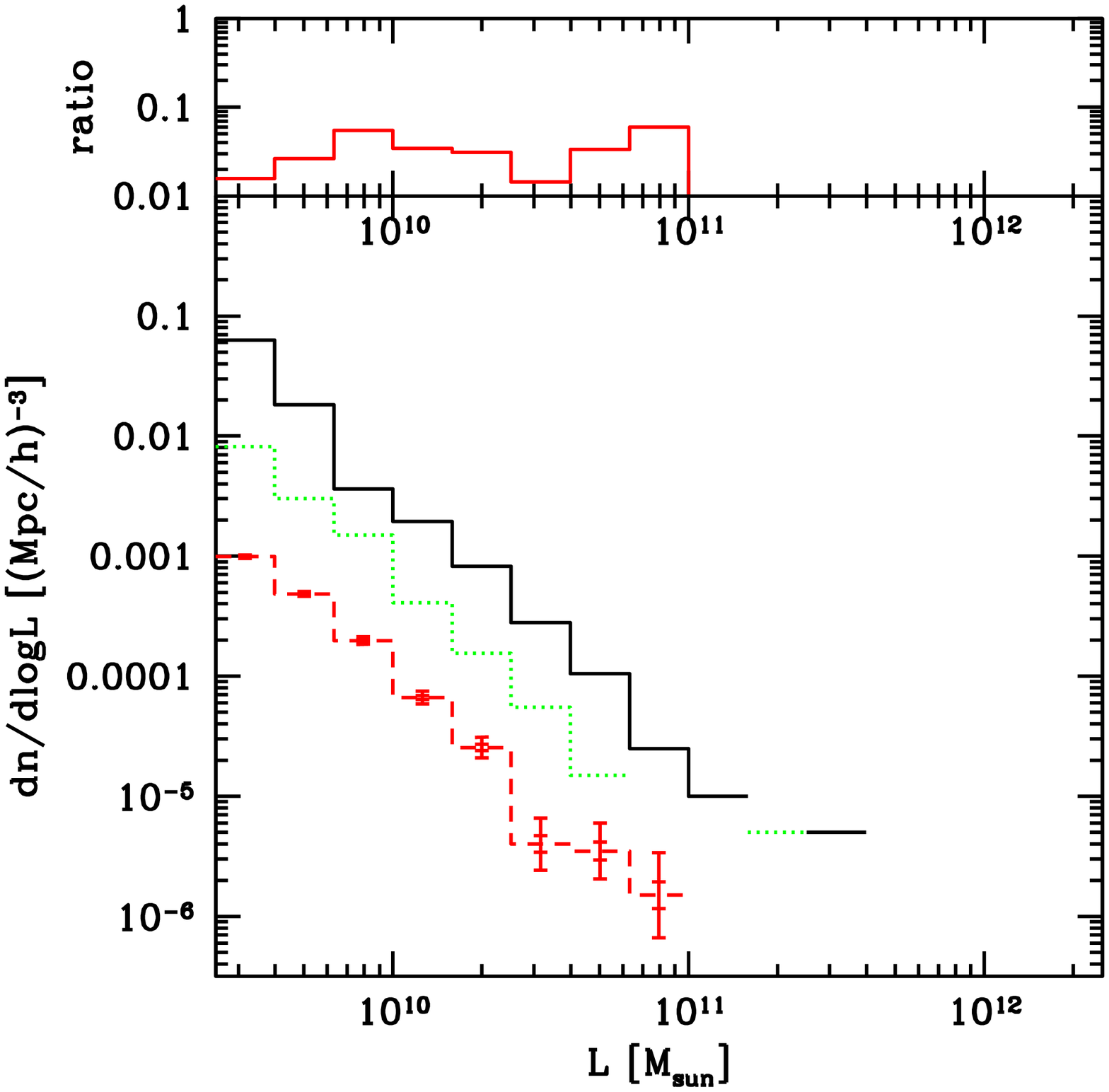}
  \includegraphics[width=2.3in]{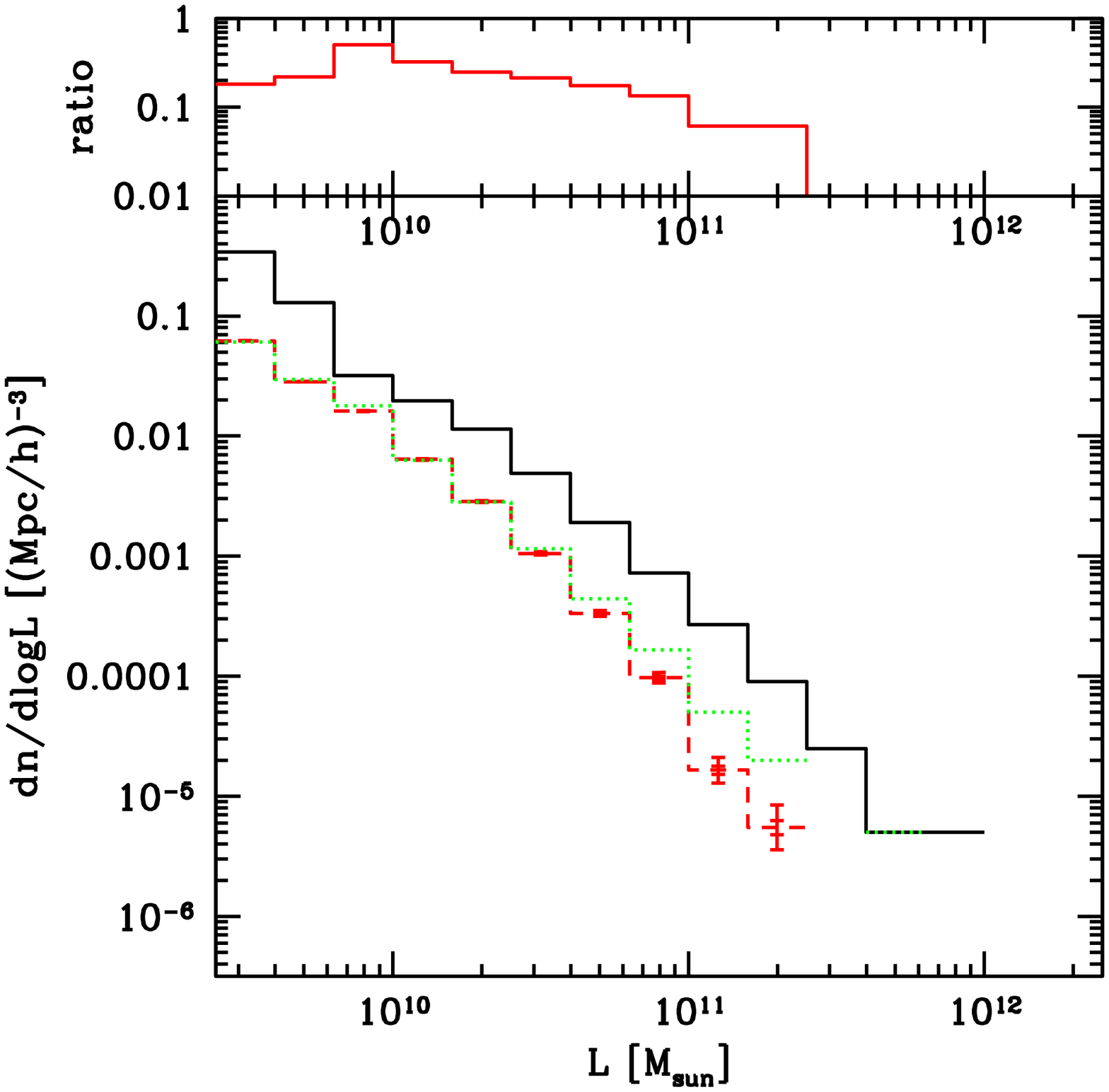}
  \includegraphics[width=2.3in]{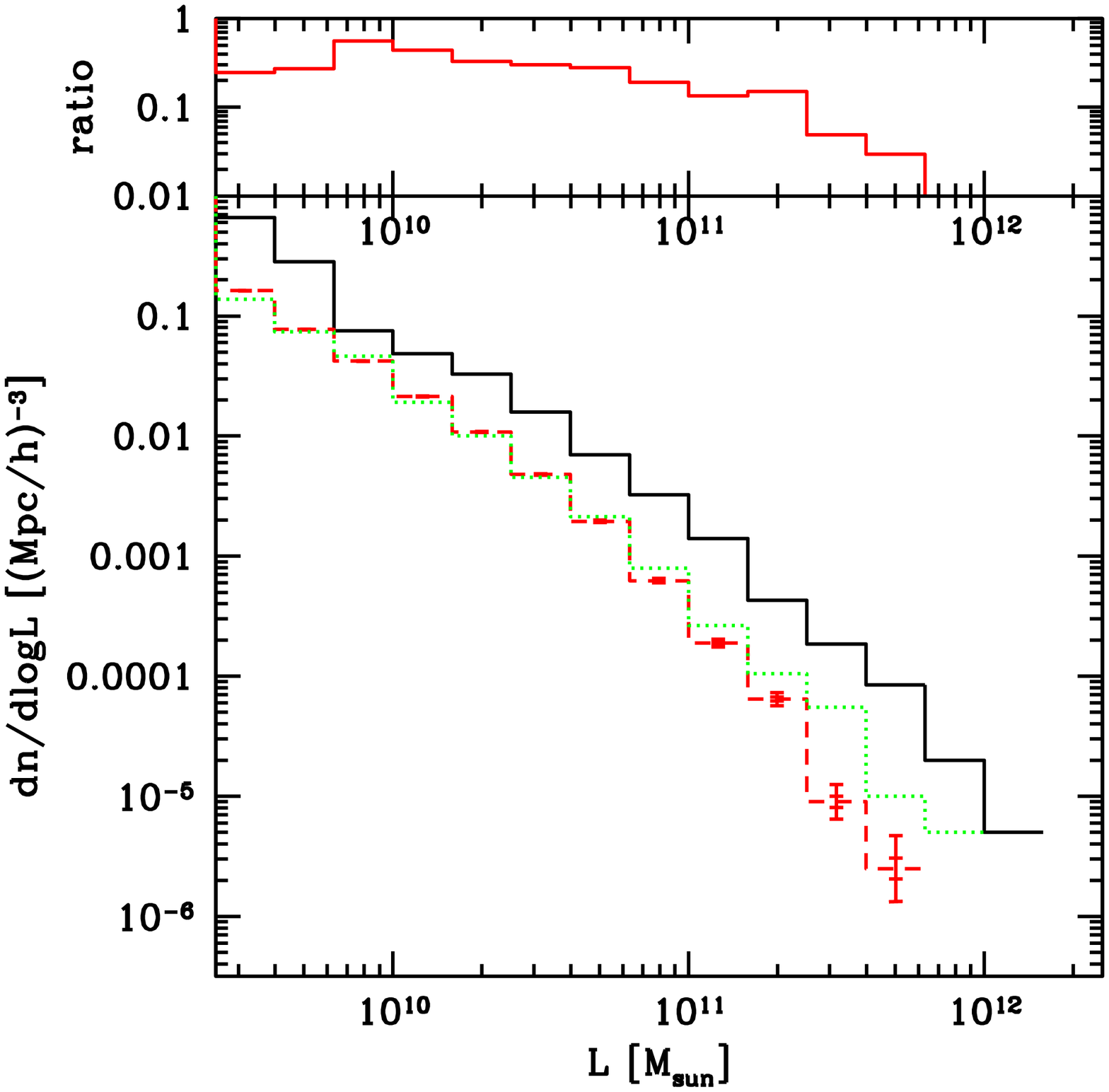}
\caption{ (bottom panels) Ly-$\alpha$ luminosity function of high-redshift
  sources without (black) and with absorption included (red) at redshifts
  $z=9$ (left; global mass-weighted ionized fraction $x_m=0.28$), $z=7$
  (middle; $x_m=0.94$) and $z=6.0$ (right; $x_m=0.9999$). For reference, the
  green, dotted line shows the result if each source is assumed 50\% absorbed,
  which would be the case if e.g. all of the blue wing of the emission line
  were absorbed, while all of the red wing were transmitted. The error bar in
  each bin reflects the number of sources in that bin found in our
  computational volume. (top panels) Bin-by-bin ratio of the observed to the
  intrinsic luminosity function.  
\label{lum_funct:fig}}
\end{figure*}

The right peak of each PDF distribution reflects the most common 
photoionization rate values in the ionized regions. The peak position 
remains at $\Gamma_{-12}\sim1$ throughout the evolution, with only 
slight shifts. The distributions are quasi-Gaussian, but with long
non-Gaussian tails at both high and (especially) low values of 
$\Gamma$. As could have been expected, the fraction of high-$\Gamma$ 
cells, all of which either contain sources or are in the immediate 
vicinity of a source, grows strongly with time, as ever . The highest 
photoionization rate values we find reach $\Gamma_{-12}\sim10^3-10^4$.
This peak value rises over time, as a consequence of the growth of
galaxies and the large number of sources forming in and around the 
density peaks. In the ionized regions the equilibration time is short
and photoionization equilibrium is 
generally a good approximation. The cells with lower values of $\Gamma$ 
(below $\Gamma_{-12}\sim0.01-0.1$ correspond to the ionization fronts or 
neutral regions. There are many more cells in I-fronts at $z=10.1$ than 
at later times, when most of the IGM is already ionized. The equilibrium 
and actual rates differ widely in those regions, indicating that the 
assumption of ionization equilibrium would be a very poor approximation 
there.

The photoionization rate-density correlations at $z=9$ and $z=6$ are shown in 
Figure~\ref{Gamma_distr} as contour plots. At high densities there is a clear, 
and fairly tight, correlation between the density and the photoionization rate.
This reflects the fact that many more sources form in overdense regions, 
resulting in higher photoionization rates. At densities just above the mean 
the correlation becomes much less tight and around the mean densities it 
becomes nonexistent. At high redshifts regions with $1+\delta\sim1$ can have 
any value of $\Gamma_{-12}$ from $10$ (for cells close to sources) down to 
essentially 0 (in neutral and shielded cells). There are three broad peaks of
the distribution, at $\Gamma_{-12}\sim1$ (the H~II regions), 
$\Gamma_{-12}\sim10^{-3}$ (cells at and around I-fronts), and 
$\Gamma_{-12}\sim0$ (neutral regions). By $z=6.0$, which is well after overlap
both the neutral and self-shielded regions and the I-fronts have mostly 
disappeared and $\Gamma_{-12}>0.1$ almost everywhere, rising with time as
more galaxies form. These values are fairly high compared to the
photoionization rate values found from the Lyman-$\alpha$ forest at
$z\sim2-4$ \citep{2002ApJ...570..457C,2004ApJ...617....1T,2006AJ....132..117F,
2007astro.ph..3306B}, in agreement with the high mean transmitted flux we
found in \S~\ref{meanFabg_sect}. Both point to somewhat lower ionizing source
efficiencies than the ones assumed here and to a correspondingly later end of 
reionization. 

\subsection{Luminosity function of high-z Ly-$\alpha$ sources}

The luminosity function is an important statistical measure of the
properties of high=redshift galaxies. It depends on both the intrinsic
luminosity of the galaxies and the absorption in the surrounding IGM. 
In Figure \ref{lum_funct:fig} we show our results of the size of the effect of
absorption on a luminosity function of high $z$ objects. For this fiducial
case we assume that the Ly-$\alpha$ luminosity is simply proportional to the
mass of our halos (similar to our model for the ionizing sources). We compute
the reduction in luminosity of each halo due to absorption
(Figures~\ref{spectra}-\ref{spectra4} show examples of this suppression). We 
assume that the intrinsic Lyman-$\alpha$ emission line is a Gaussian with 
an rms of $160\,\rm km\,s^{-1}$. Luminosity function of high-redshift sources
without (in this fiducial case this is just the halo mass function; solid
line) and with absorption included (dashed) at redshifts $z=9,7$ and 6. For
reference, the dotted line shows the result if each source is assumed
50\% absorbed, which would be the case if e.g. all of the blue wing of the
emission line were absorbed, while all of the red wing were transmitted.
Top panels show the bin-by-bin ratios of the observed to the intrinsic
luminosity function. Note that due to the binning at fixed luminosity
(intrinsic or observed) this ratio is not the same as the average suppression
per source of a given mass (i.e. the absorption shifts the curve both down and
to the left).

\begin{figure*}
  \includegraphics[width=3.2in]{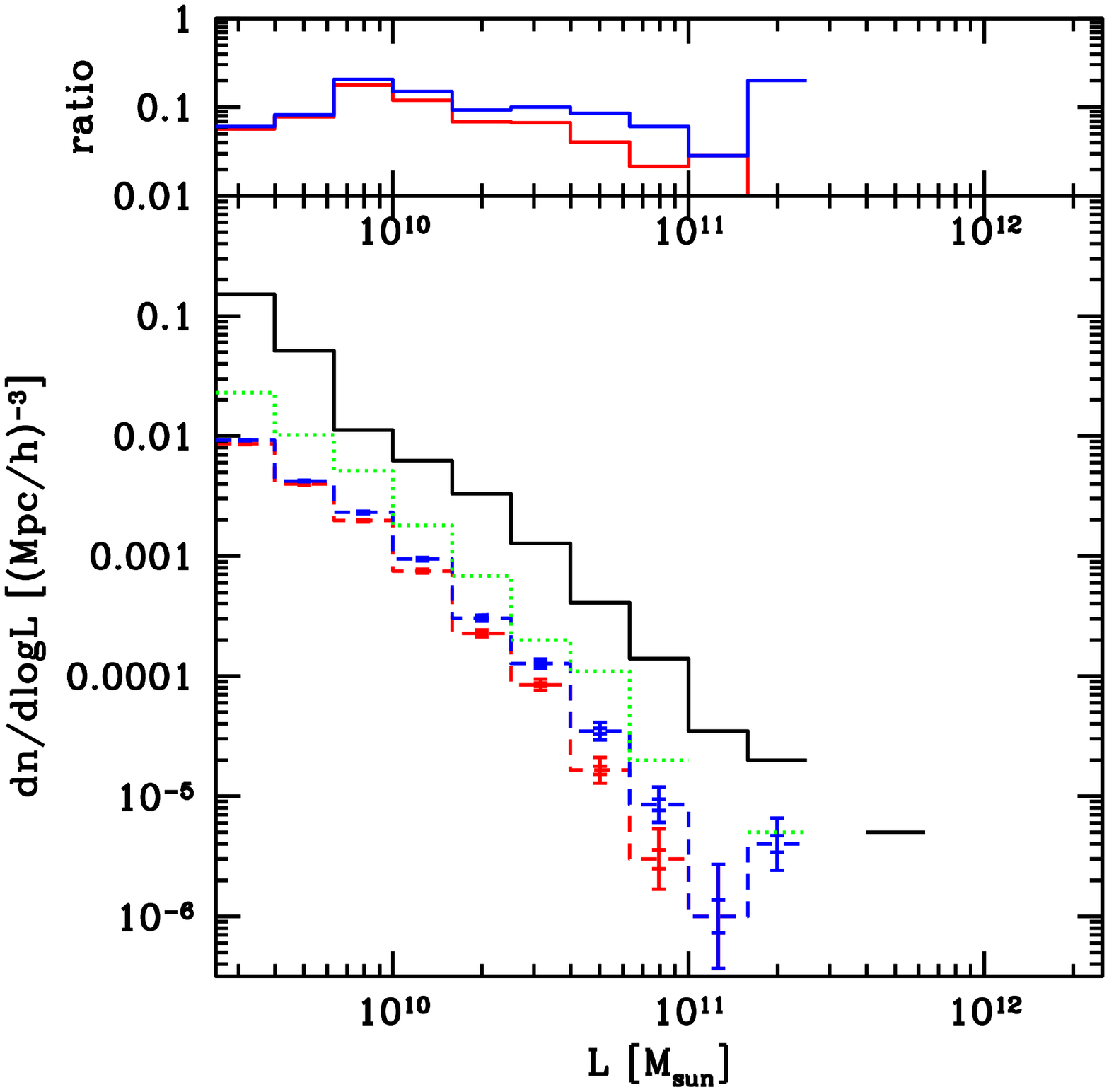}
  \includegraphics[width=3.2in]{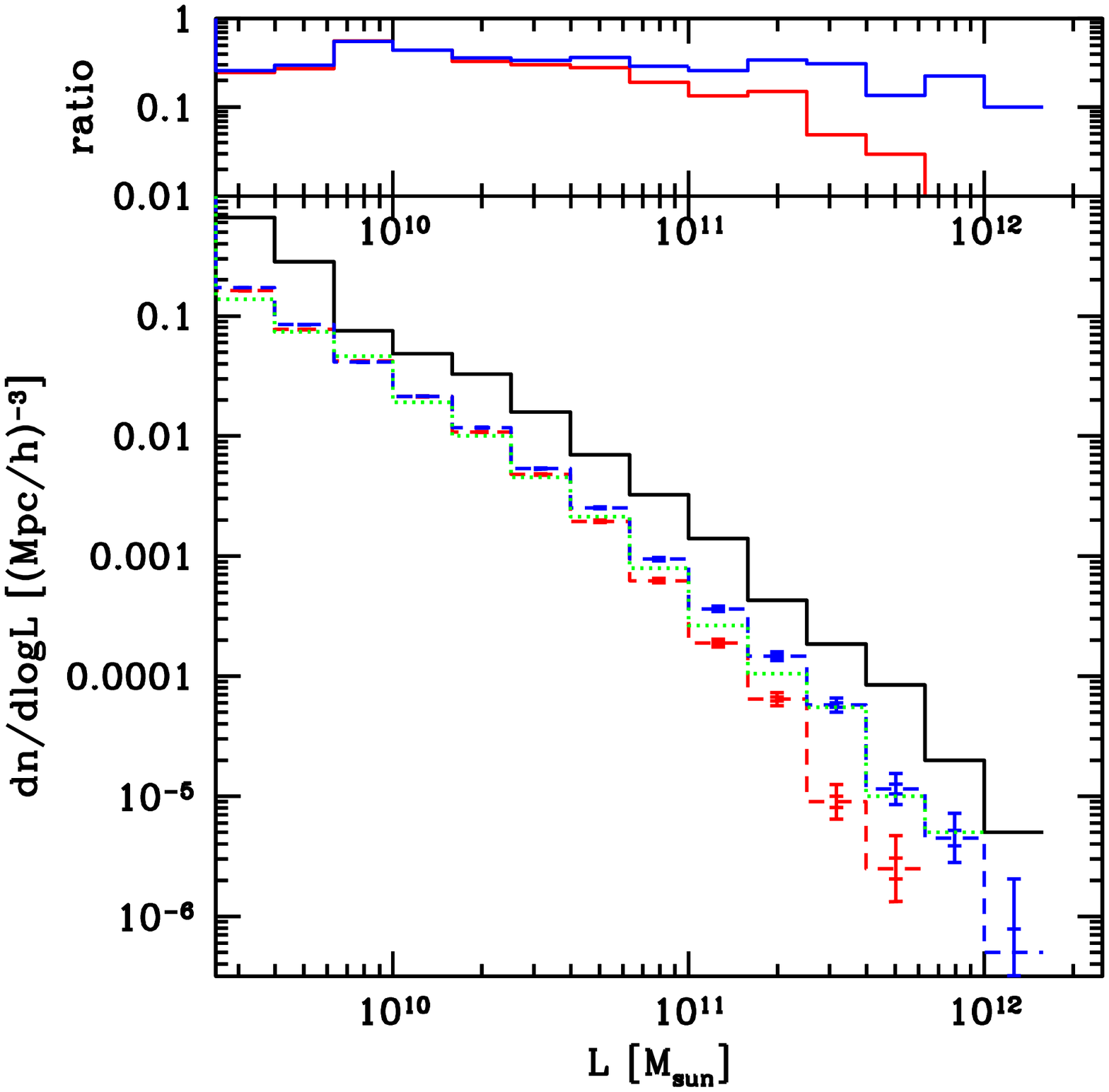}
\caption{Luminosity function at redshifts z=8.1 (left; $x_m=0.62$) 
 and z=6 (right; $x_m=0.9999$) if peculiar velocities are ignored 
 (blue, long-dashed). For reference, we also show the data from 
 Fig.~\ref{lum_funct:fig}, same notation. 
\label{lum_funct_nopecv:fig}}
\end{figure*}

\begin{figure*}
  \includegraphics[width=3.2in]{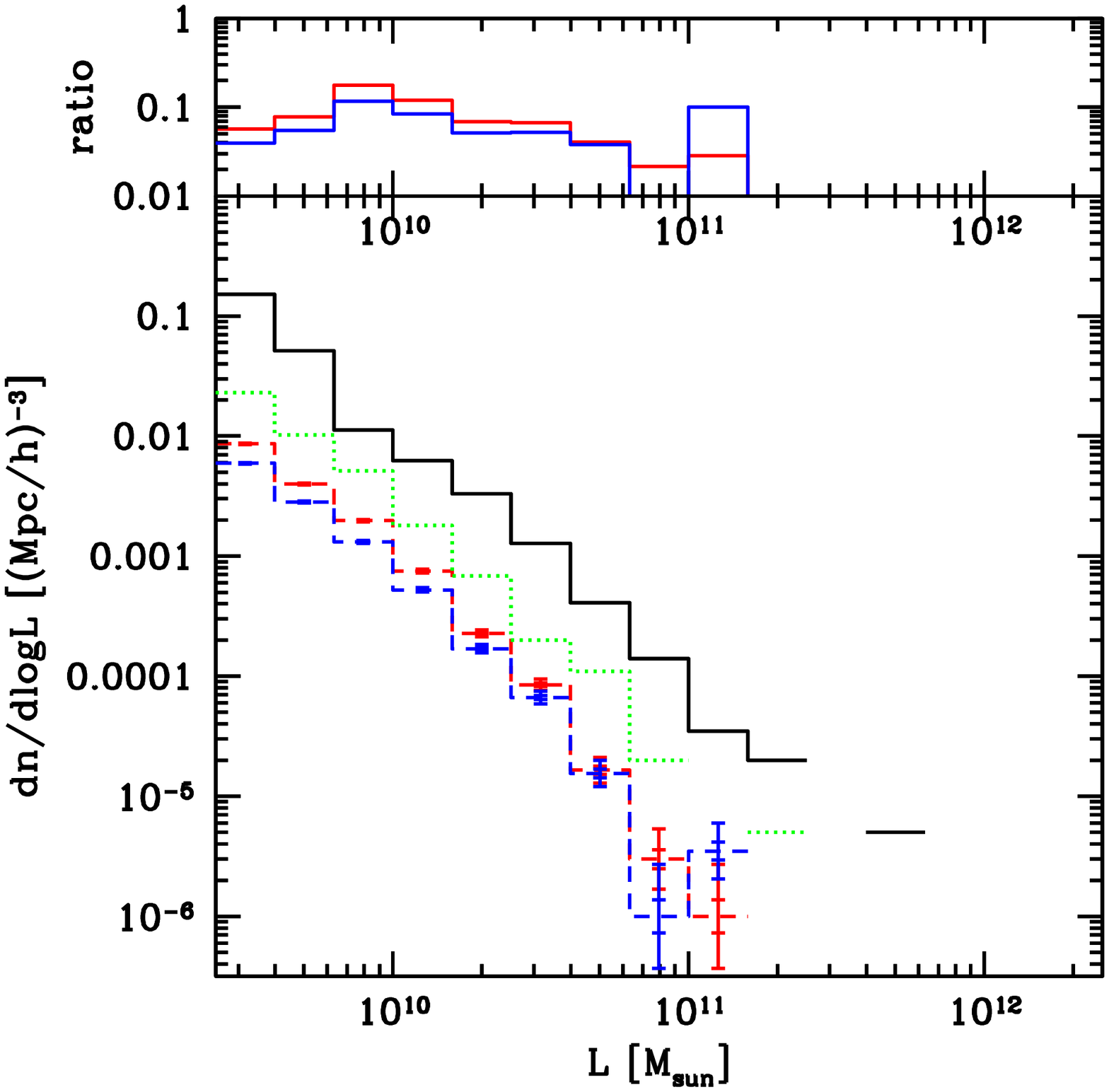}
  \includegraphics[width=3.2in]{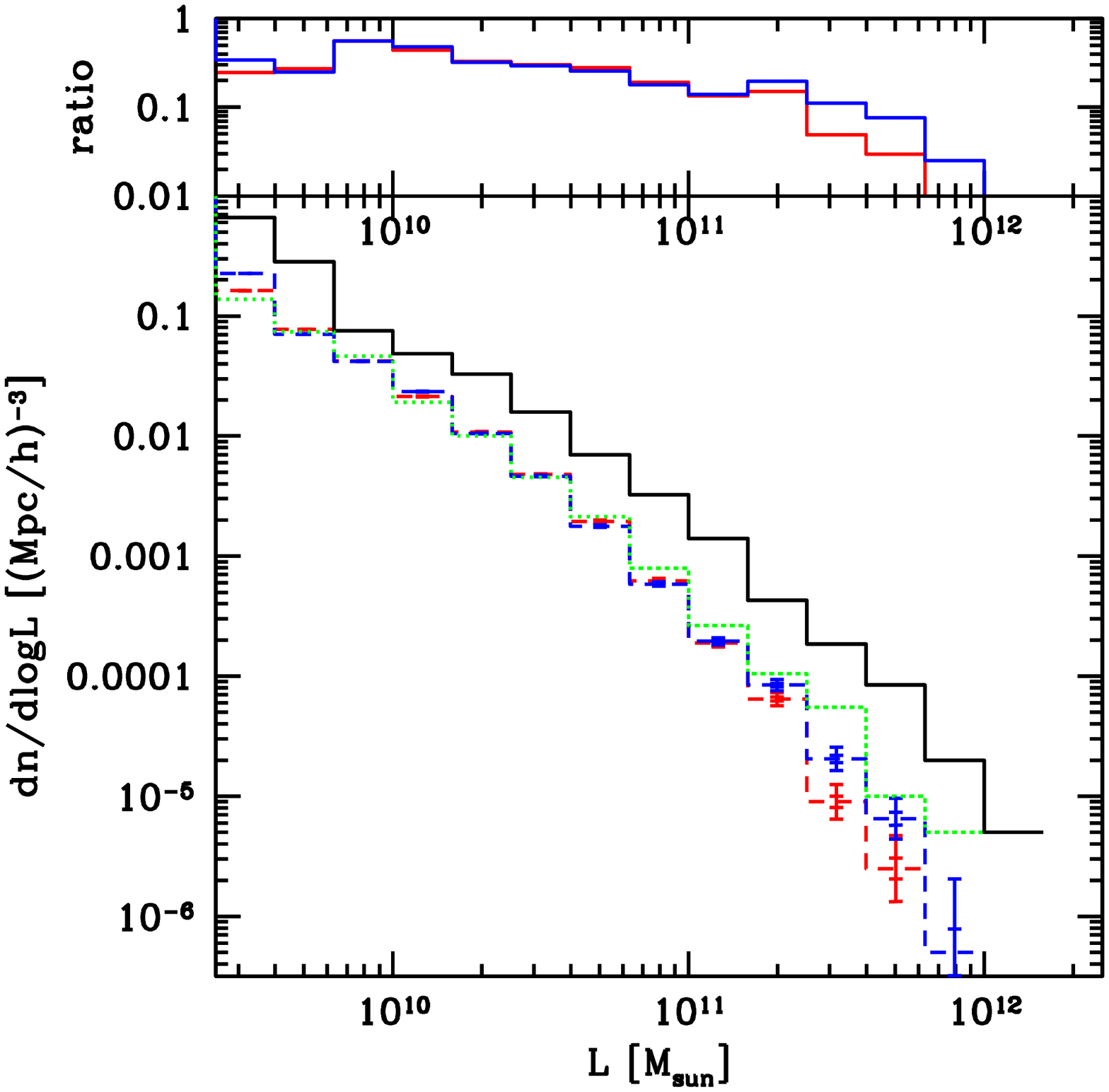}
\caption{Luminosity function at redshifts z=8.1 (left; $x_m=0.62$) 
 and z=6 (right; $x_m=0.9999$) for variable emission line width
  (blue, long-dashed; see text for details). For reference, we also show the
  data from Fig.~\ref{lum_funct:fig}, same notation. 
\label{lum_funct_varwid:fig}}
\end{figure*}

The luminosity function exhibits clear evolution from high to low redshift. 
As we discussed in \S~\ref{spectra:sect}, at high redshift the damping wings
are strong and thus not only the blue side, but also significant part of the
red side of the line is absorbed, as evidenced by the significant difference
between the dashed and dotted lines. As a result of this absorption, the
number of sources per luminosity bin drops by one to two orders of magnitude. 
During the later stages of reionization the damping wing effectively
disappears. As a consequence, the change in the faint end of the luminosity 
function due to IGM absorption is on average well-represented by simply 
reducing each source luminosity by 50\%, which would be the case if the blue 
half of the line were absorbed and the red half were not. The situation is 
different at the bright end of the luminosity function, however, where on 
average significantly more than half of the intrinsic flux is absorbed at 
both $z=7$ and $z=6$. The shape of the luminosity function shows some 
evolution, as well, which is in part due to an evolution in the shape of the 
halo mass function and in part to the higher mean absorption for the more
massive sources. The higher average absorption levels for the luminous 
sources is consequence of the infall which surrounds the high density peaks 
they are in. In order to demonstrate this, we re-calculated the luminosity 
function at $z=8.1$ and $z=6$ using exactly the same data, but setting all 
peculiar velocities to zero. The results are shown in 
Figure~\ref{lum_funct_nopecv:fig}. With no peculiar velocities present the 
intrinsic emission of all sources is absorbed on average at roughly the 
same level, by factor of $\sim 10$ at $z=8.1$ and by factor of 2 at $z=6$.
The resulting luminosity function at late times agrees well with the one 
where we simply assumed 50\% absorption. Early-on this is not the case as 
a consequence of the still-present damping wing.

\begin{figure*}
  \includegraphics[width=2.3in]{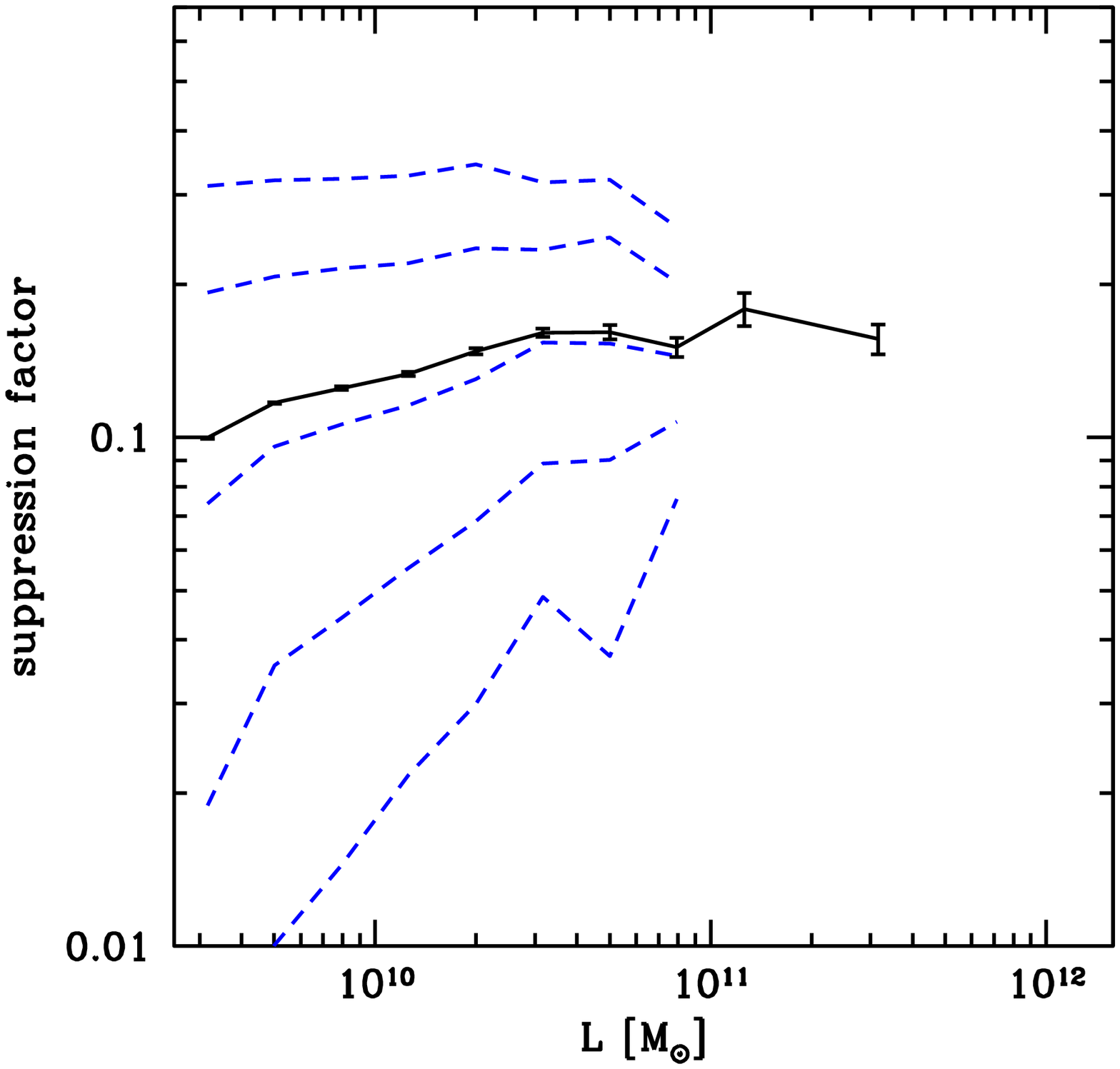}
  \includegraphics[width=2.3in]{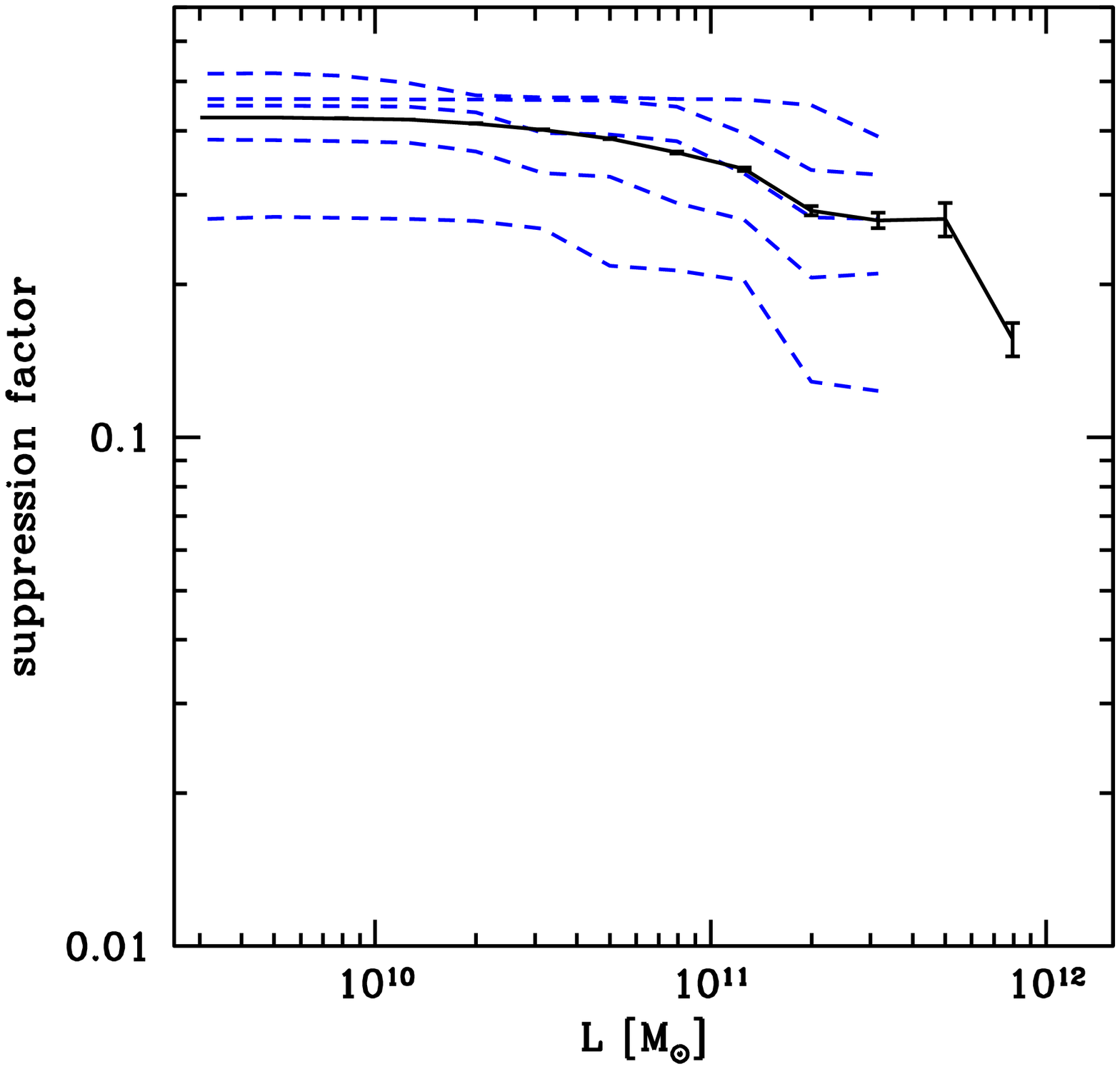}
  \includegraphics[width=2.3in]{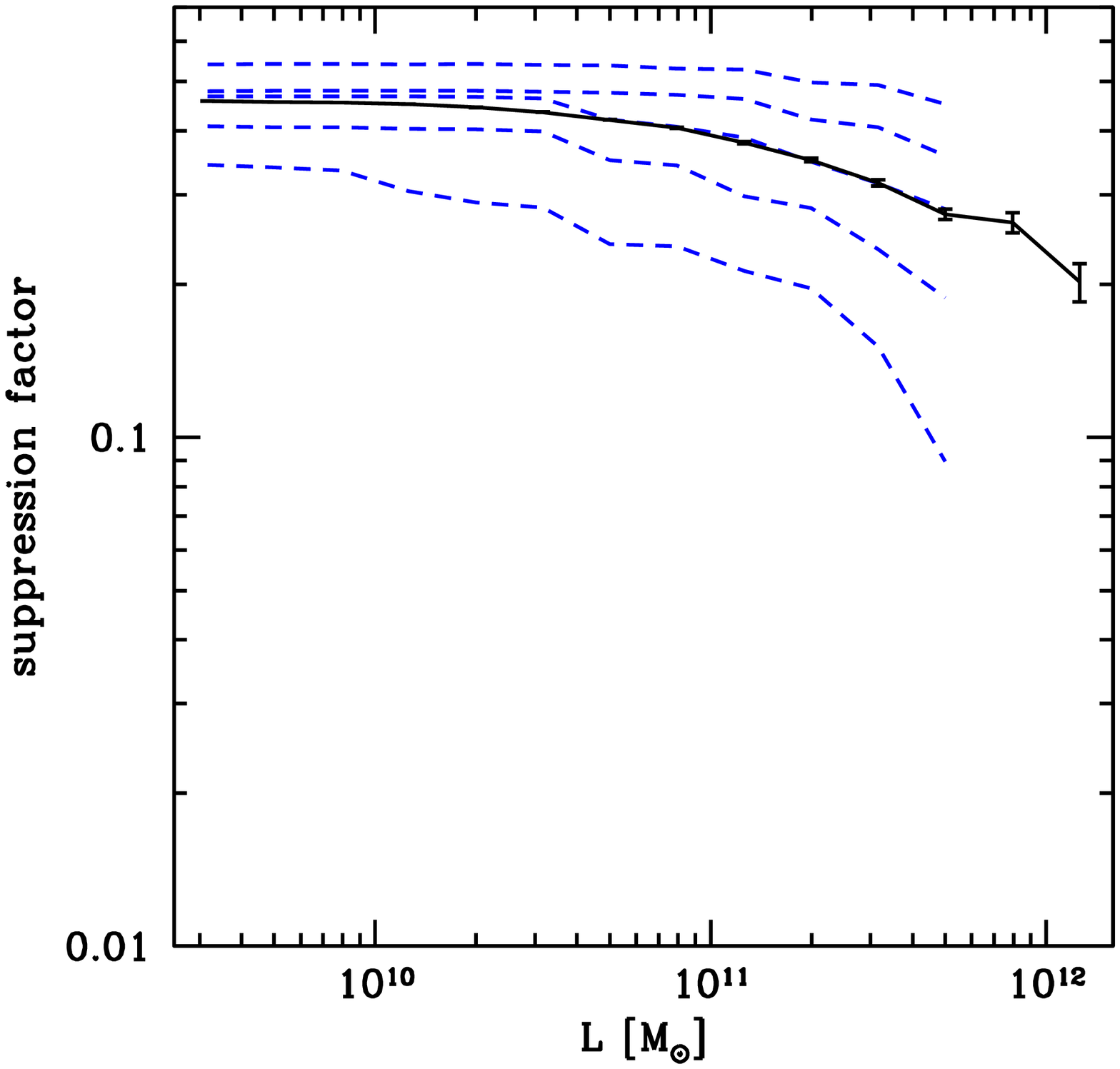}
\caption{Transmission fraction as a function of luminosity at redshifts 
  $z=9.0$ ($x_m=0.28$; left), $z=7.0$ ($x_m=0.94$; center) and $z=6.0$ 
  ($x_m=0.9999$; right). Shown are (dashed lines, top to bottom) 0.023, 
  0.16, 0.5, 0.84, and 0.977 percentiles (e.g., 2.3\% of points have 
  suppression less than the uppermost line). The center line is the 
  median of all the LOS. We required at least 50 LOS in each bin
  for sampling the distribution properly. There are 10 random LOS per
  source. We also plot the mean (solid line) for all LOS in each bin.
\label{sup_dist:fig}}
\end{figure*}
\begin{figure*}
  \vspace{-0.3cm}
  \includegraphics[width=2.3in]{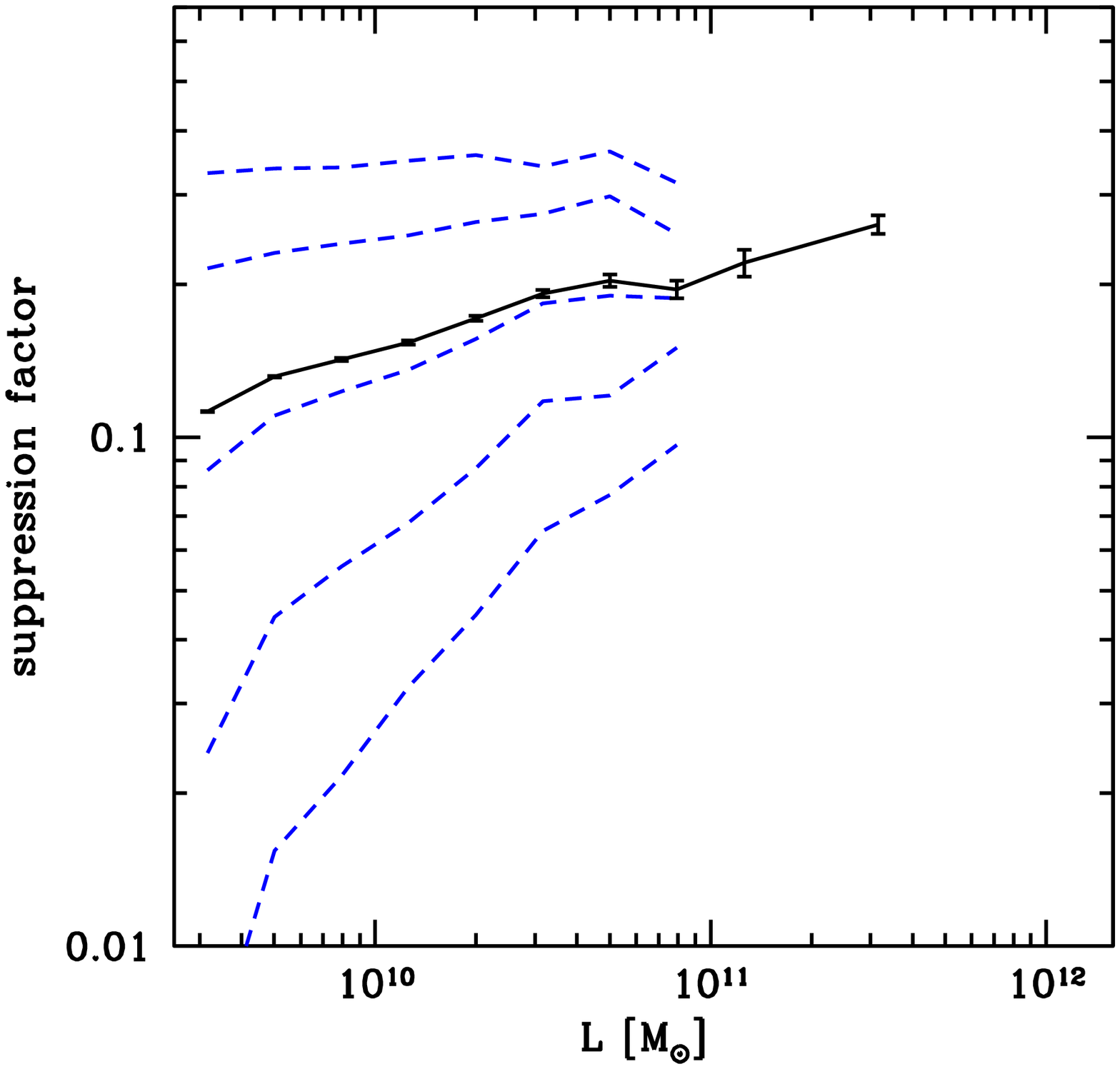}
  \includegraphics[width=2.3in]{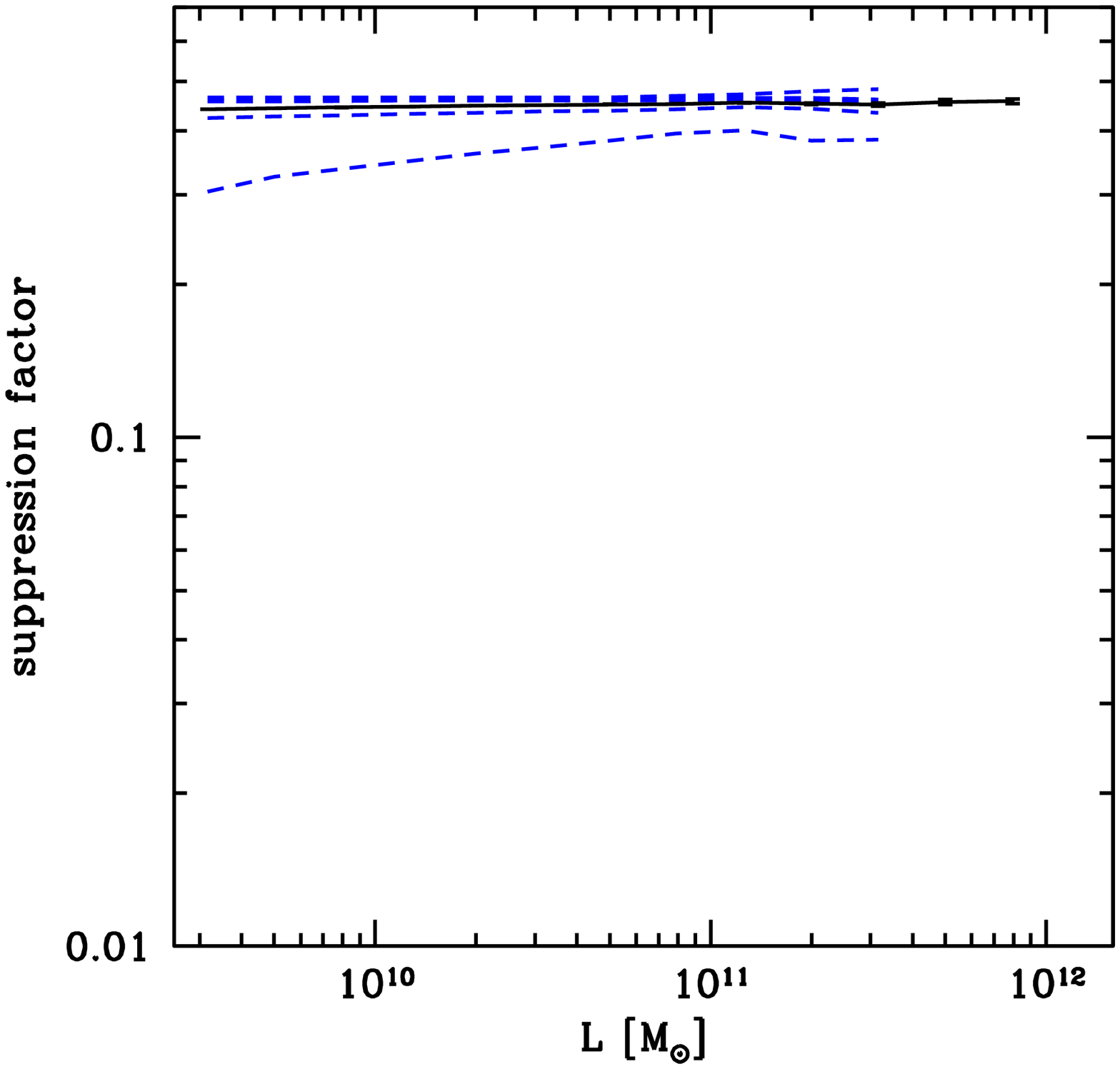}
  \includegraphics[width=2.3in]{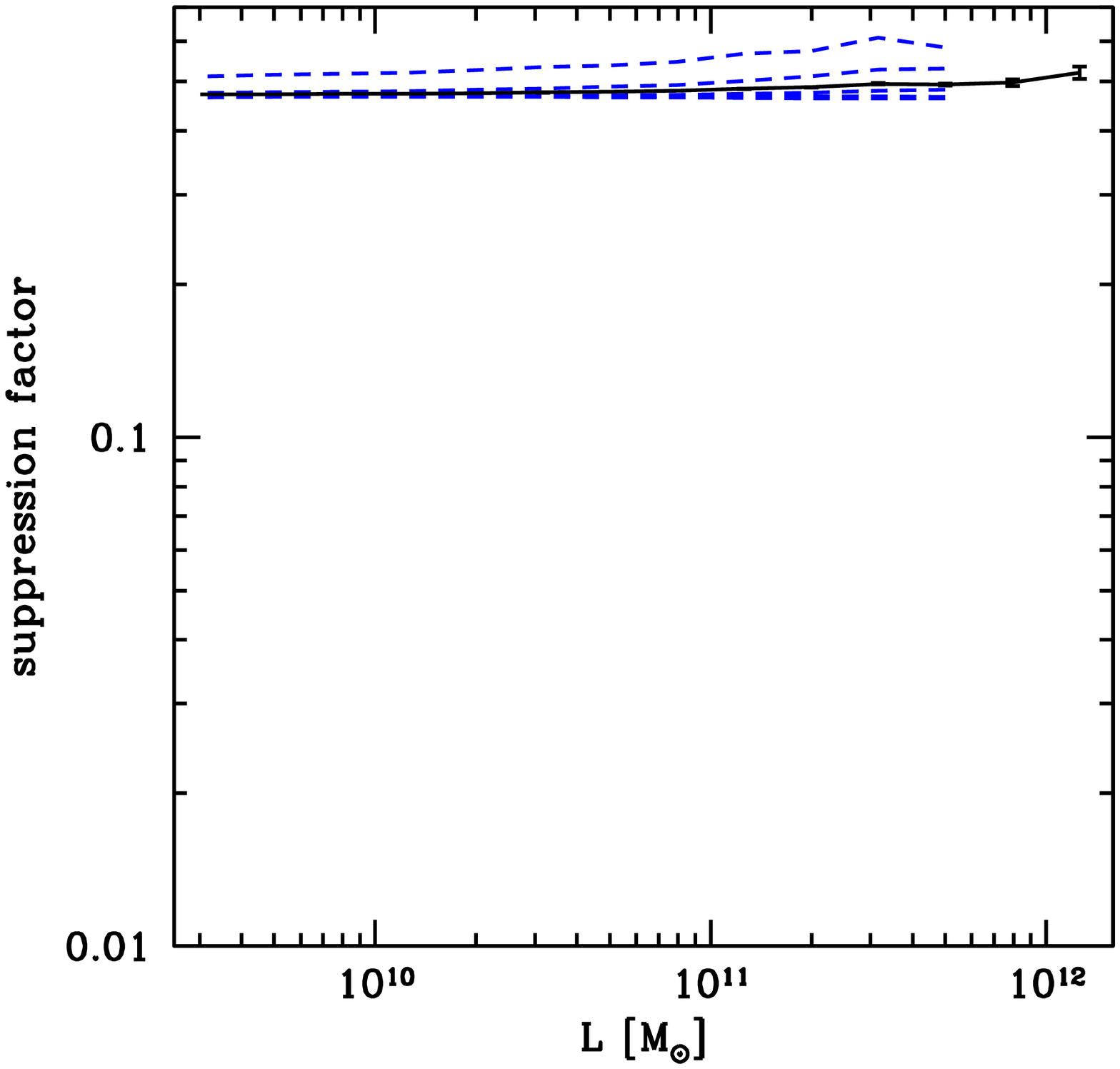}
\caption{Same as Fig.~\ref{sup_dist:fig}, but ignoring any peculiar 
         velocities of the halos and the IGM. 
\label{sup_dist_nopecv:fig}}
\end{figure*}
\begin{figure*}
  \includegraphics[width=2.3in]{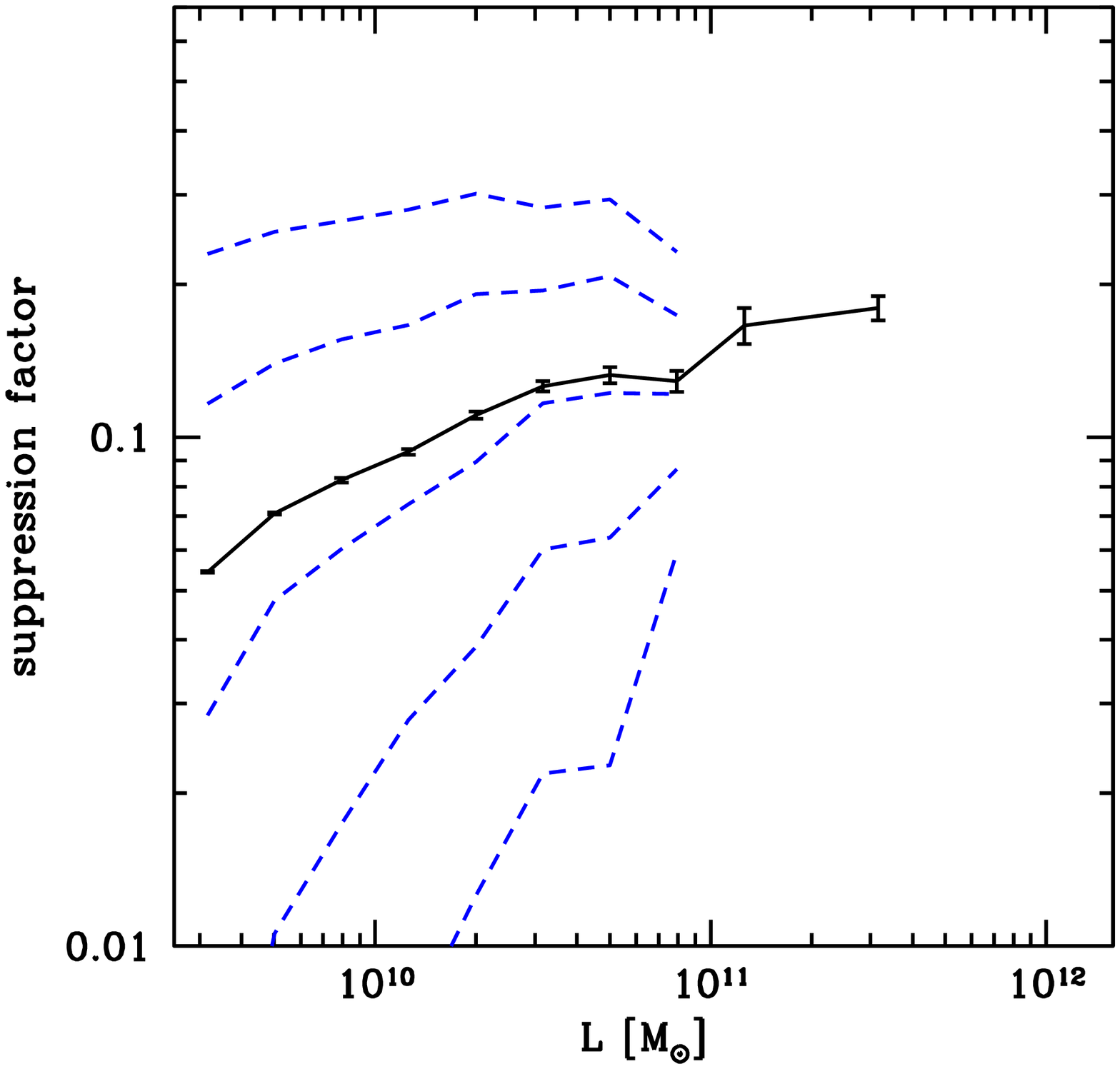}
  \includegraphics[width=2.3in]{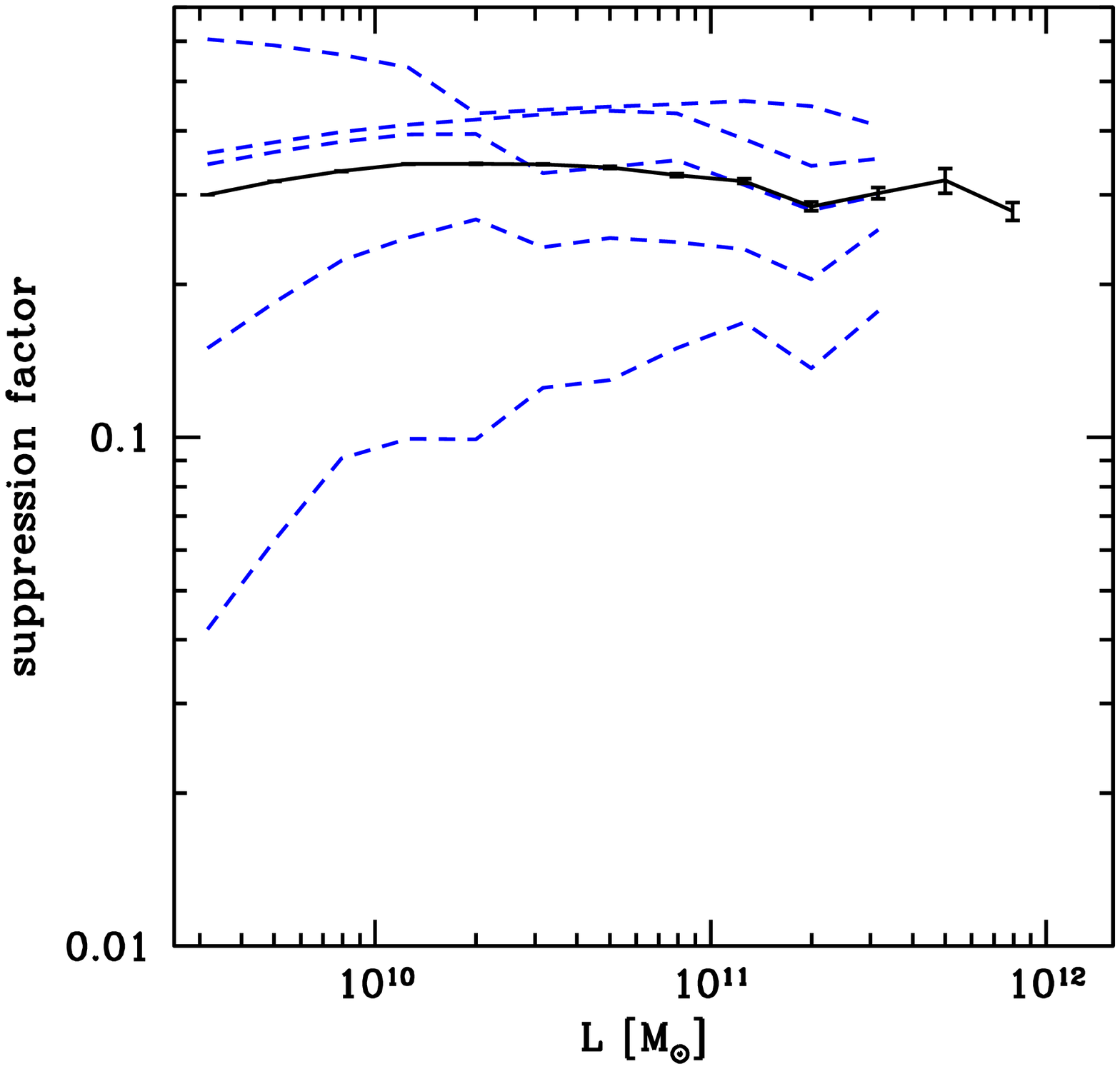}
  \includegraphics[width=2.3in]{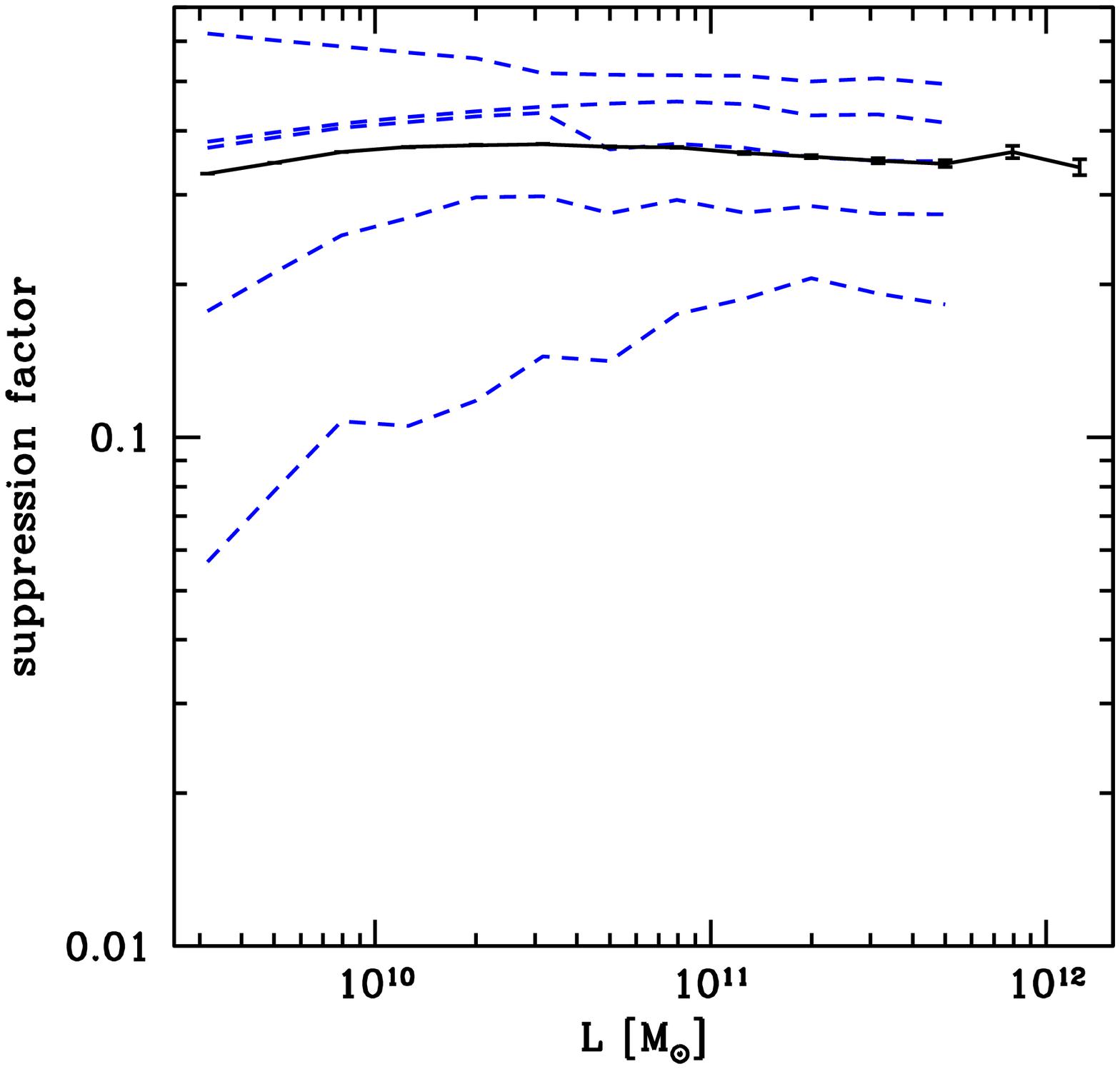}
\caption{Same as Fig.~\ref{sup_dist:fig}, but for a variable Ly-$\alpha$
  emission line width, as discussed in the text.
\label{sup_dist_varwid:fig}}
\end{figure*}

How important is our assumption that all sources have same rms width of 
their intrinsic emission line profile? To check this, we replaced this 
assumption with one where the rms line width varies with the halo mass as
$133\,\rm km/s (M/10^{11}M_{\odot})^{1/3}$ \citep{2007MNRAS.377.1175D}.
 Results, again 
at $z=6$ and in head-to-head comparison with our fiducial case of fixed
line width, are shown in Figure~\ref{lum_funct_varwid:fig}. The faint end
of the luminosity function proves insensitive to the line width, which is 
easy to understand. As we have shown above, on average the IGM completely
absorbs the blue half of the line for the weaker sources and completely
transmits the red half. However, the variable line width has some effect 
on the absorption of bright sources. Their lines become wider under this 
assumption and thus are less affected by the absorption due to the 
infalling gas, resulting in higher transmission by a factor of $\sim2$.

The probability distribution of the transmission fraction per source of a 
given mass/luminosity is shown in Fig.~\ref{sup_dist:fig}. There are 10 
random LOS per source and we required at least 50 LOS in each bin for 
sampling the distribution properly. We also plot the bin-by-bin average 
transmission. For the luminosity bins which do not contain our minimum 
number of LOS we plot only the mean. Several interesting trends emerge. 
The distributions are fairly wide at all times and for all sources, 
reflecting the large variations in opacity from source to source and from
LOS to LOS. The former is due to the different environments sources are 
found in, while the later reflects the anisotropies around each source.
At all redshifts the mean and the median curves are very similar for all
bins, and are essentially identical at the bright end. 

The distribution itself changes its character as the evolution progresses. 
At early times the distribution is much wider for the fainter sources and 
the mean and median are gently rising towards the bright end, reflecting 
the fact that bright sources are found in the middle of larger H~II 
regions, while fainter sources are found in a variety of environments.
Thus, during the early evolution the main factor shaping the distribution
is the the local variation of the neutral fraction around each source.
At late times ($z<7$), however, the situation changes to the opposite, 
with the distribution becoming wider at the bright end and the mean and 
median decreasing there as well. By that time the IGM is already largely 
ionized and the main environmental dependence is due to the anisotropies 
of the density field and, even more so, of the infall around the bright 
peaks, as discussed above. This is clearly demonstrated in 
Fig.~\ref{sup_dist_nopecv:fig}, where we show the distributions as they 
would be if there were no peculiar velocities. At early times the results
are largely unchanged, while at late times the variations between the 
different LOS essentially disappear and all the curves become flat, showing 
that the distribution is shaped mainly by the effects of the peculiar 
velocities.

Finally, in Fig.~\ref{sup_dist_varwid:fig} we show the effect of varying 
intrinsic line width on the distributions. Compared to our fiducial case of
constant line width, at high redshift the distributions are hardly affected,
except for slightly higher absorption of the faintest sources. However, at
later times the varying line width has a more significant effects. At $z=7$
the mean and the median values for the majority of sources decrease from
$\sim40-45\%$ down to $\sim30\%$, but the brightest sources are affected much
less. As a result the curves for the mean and median become largely flat
rather than decreasing towards the bright end. Furthermore, the distribution
becomes wider at the faint end, with many more LOS being absorbed by factor of
10 or more. A similar effect is seen at $z=6$, except in this case the
brightest sources are even less absorbed due to their wider emission lines, in
agreement with what we observed in the luminosity functions.  

\begin{figure}
  \includegraphics[width=3.2in]{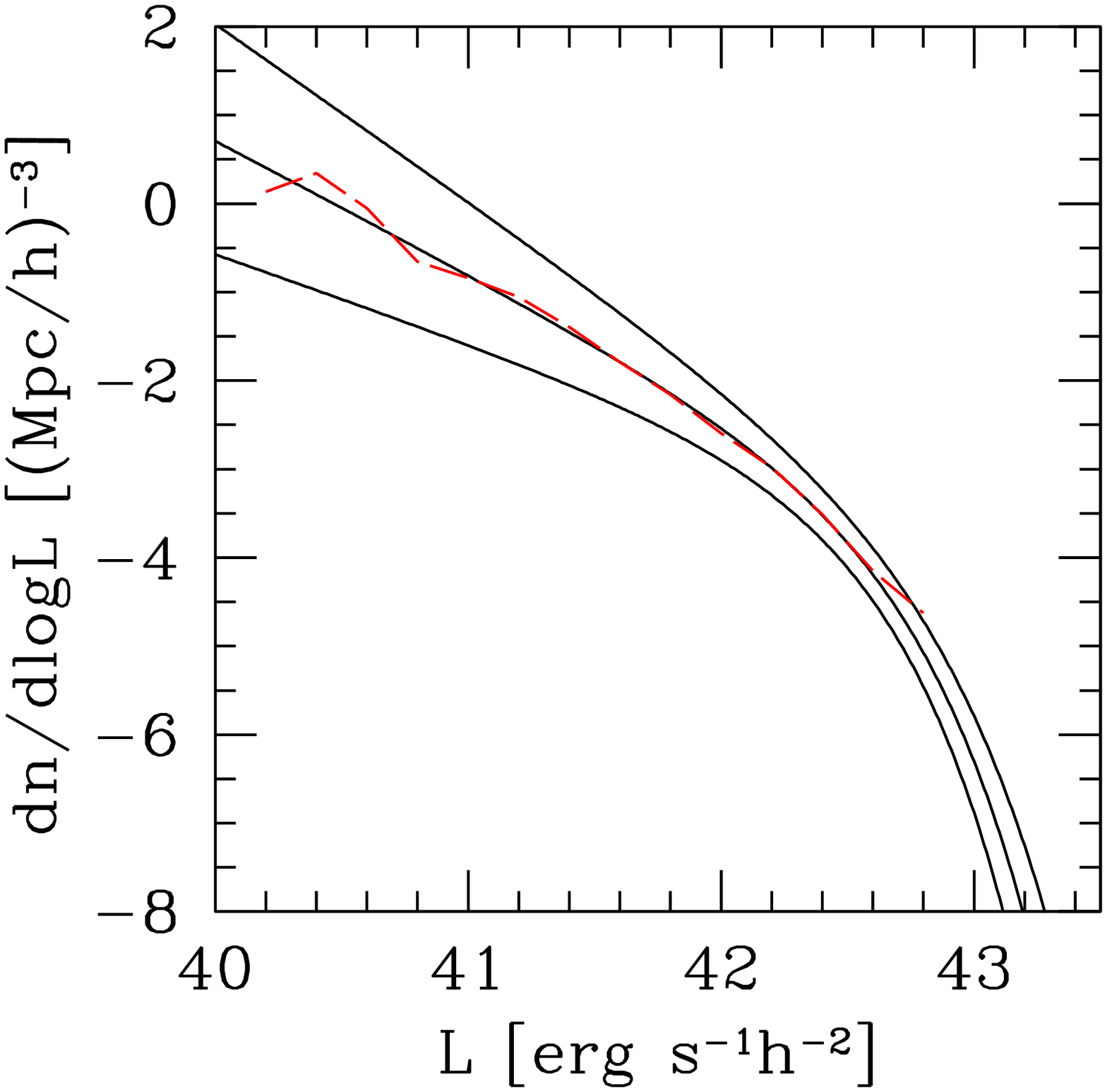}
\caption{Simulated luminosity function at $z=6.6$ (dashed; $x_m=0.99$) vs. 
  best fits of
  \citet{2006ApJ...637..631K} (at $z=6.56$) for (top to bottom) faint-end
  slopes of $\alpha=-2,-1.5,-1)$ (solid). 
\label{lum_funct_vs_obs:fig}}
\end{figure}

Our derived luminosity functions assume a constant mass-to-light ratio and are
in arbitrary units (halo mass $\times$ absorption by the IGM), proportional to
a yet undetermined mass-to-observed light ratio. We can roughly determine the
latter by comparing the number densities of observed and simulated objects. 
\citet{2006ApJ...637..631K} currently provide the best set of data at $z>6$.
They have provided fits to a Schechter function:
\be
\phi(L)dL=\phi_*\left(\frac{L}{L_*}\right)^\alpha
           \exp\left(-\frac{L}{L_*}\right)\frac{dL}{L_*} 
\ee
Since the high-redshift data still has large uncertainties, particularly in
terms of the faint-end slope, the data is fit by assuming
$\alpha=(-2,-1.5,-1)$, with best fit parameters at $z=6.56$ given by
$log(L_*/\rm  h^{-2}_{70} erg/s)=(42.74,42.60,42.48)$  and $log(\phi_*/\rm
Mpc^{-3}h^{2}_{70})=(-3.14,-2.88,-2.74)$, respectively. We plot these fits in
Figure~\ref{lum_funct_vs_obs:fig} against our derived luminosity function. 
The latter was obtained by rescaling our arbitrary luminosity units to
physical ones using a constant ratio, 
\be
L=L(M_\odot)\times10^{30.9}
\label{ml_equ}
\ee 
so as to match it to the observed luminosity function for the same number
densities of objects.  

The fit assuming $\alpha=-1.5$ provides by far the best match to our
luminosity function. The two agree in both amplitude and shape over the whole
available range. The differences at both the luminous and the faint end should
both be expected, the former due to cosmic variance and the latter due to both 
numerical resolution and lack of reliable observational data. Matching the
other two faint end slopes would require us to relax our constant
mass-to-light ratio assumption.

Taking equation~\ref{ml_equ} at face value, we can now make an approximate
correspondence between observed luminosities and masses of the underlying
halos. The sources observed with Subaru at $z=6.56$
\citet{2006ApJ...637..631K} have luminosities $\sim 10^{42}-10^{42.7}\rm
erg\,s^{-1}h^{-2}$. This corresponds to the luminous end of our LF, for
effective masses (halo mass $\times$ absorption by the IGM) of order
$10^{11}M_\odot$ or larger, or relatively rare halos. The observations are not
yet sufficiently sensitive to detect the faint end, which contributes most of
the ionizing emissivity during reionization. Hence, claims that observations
show that there are not enough ionizing photons at $z\sim6$ to reionize the
universe appear premature.

\section{Correlation Functions}

\begin{figure*}
\vspace{-5cm}
  \includegraphics[width=4.2in]{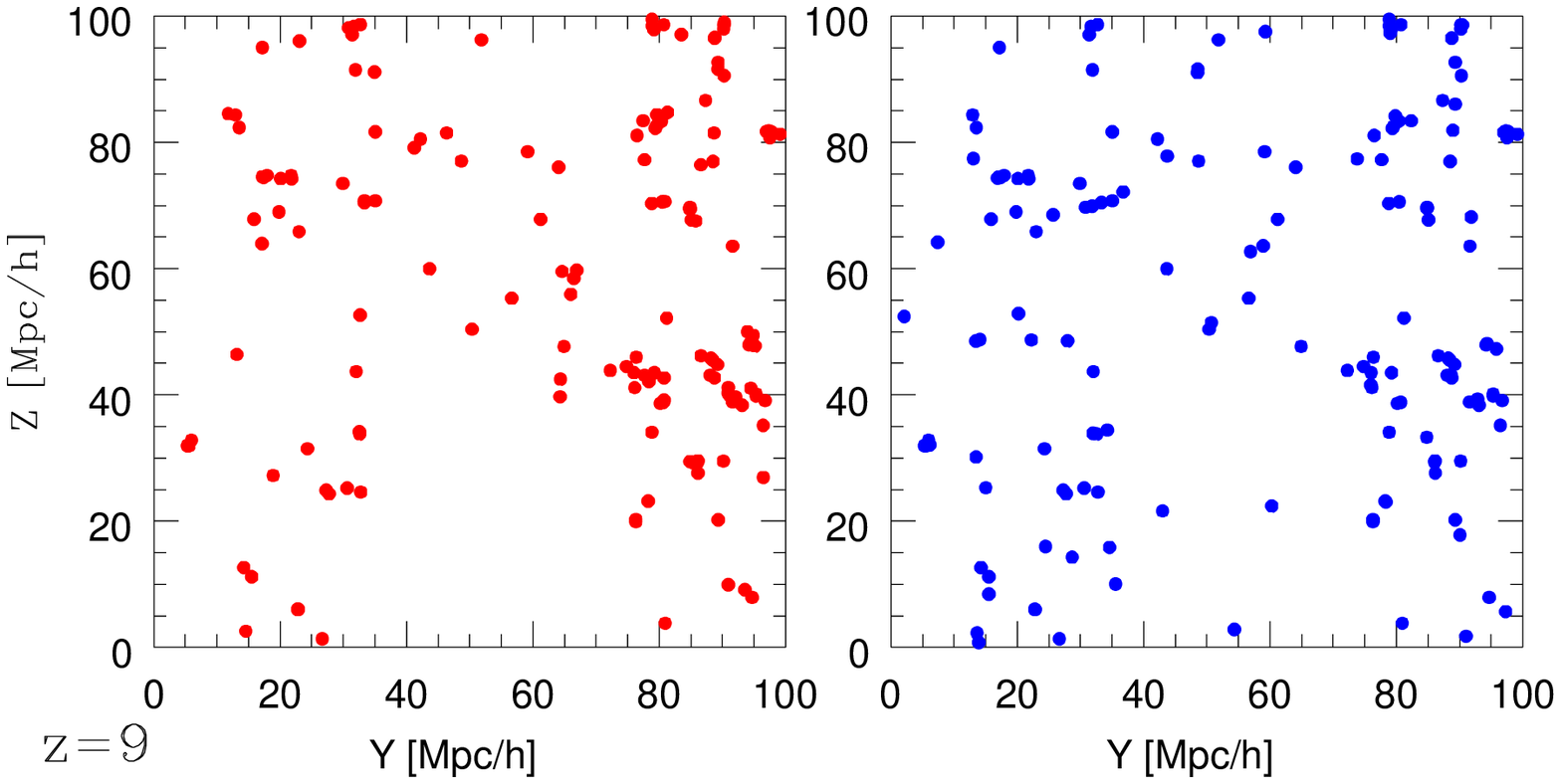}
  \includegraphics[width=2.3in]{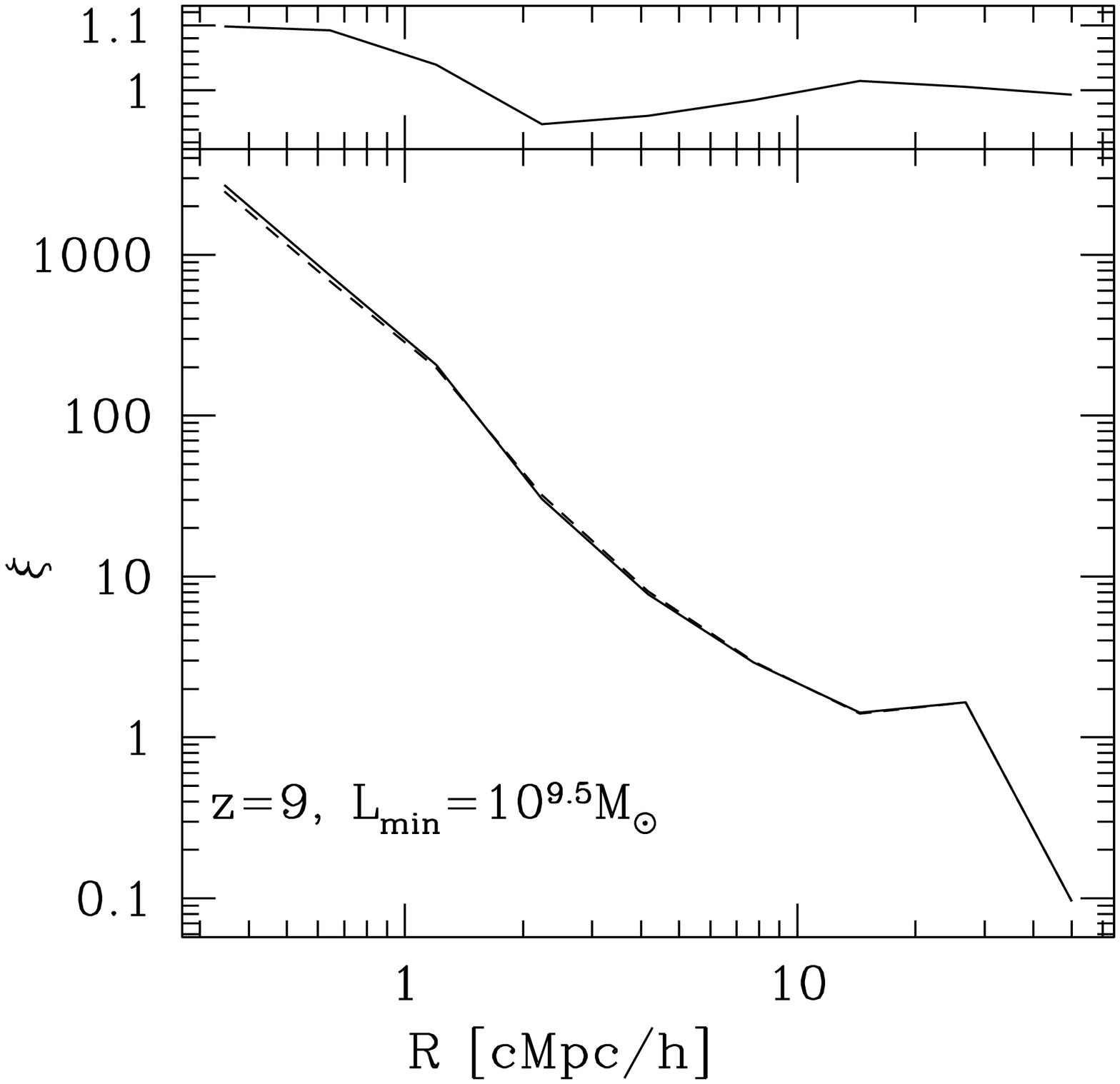}
\caption{Projection of the sources as seen by a mock flux-limited
  survey with $L>10^{9.5}M_\odot$ (198 sources in total) at $z=9$ (left panel; 
  $x_m=0.28$) and sources with the same number density if the IGM 
  absorption were ignored (middle panel) and the 2-point 3D 
  correlation functions (right panels) of the distribution with IGM absorption 
  (solid) and without (dashed) and their ratio (top). 
\label{corr_z9}}
\end{figure*}

\begin{figure*}
\vspace{-5cm}
  \includegraphics[width=4.2in]{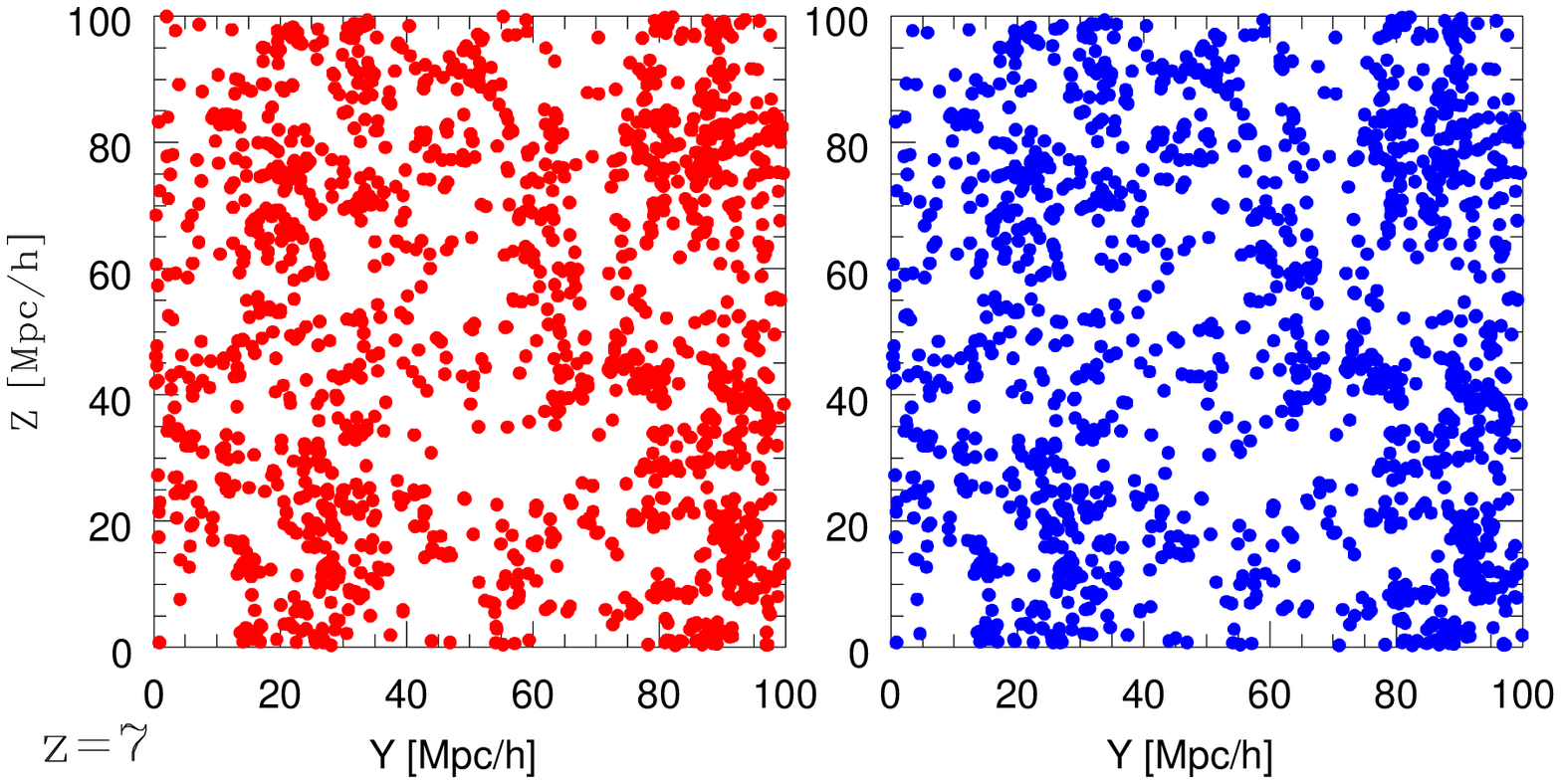}
  \includegraphics[width=2.3in]{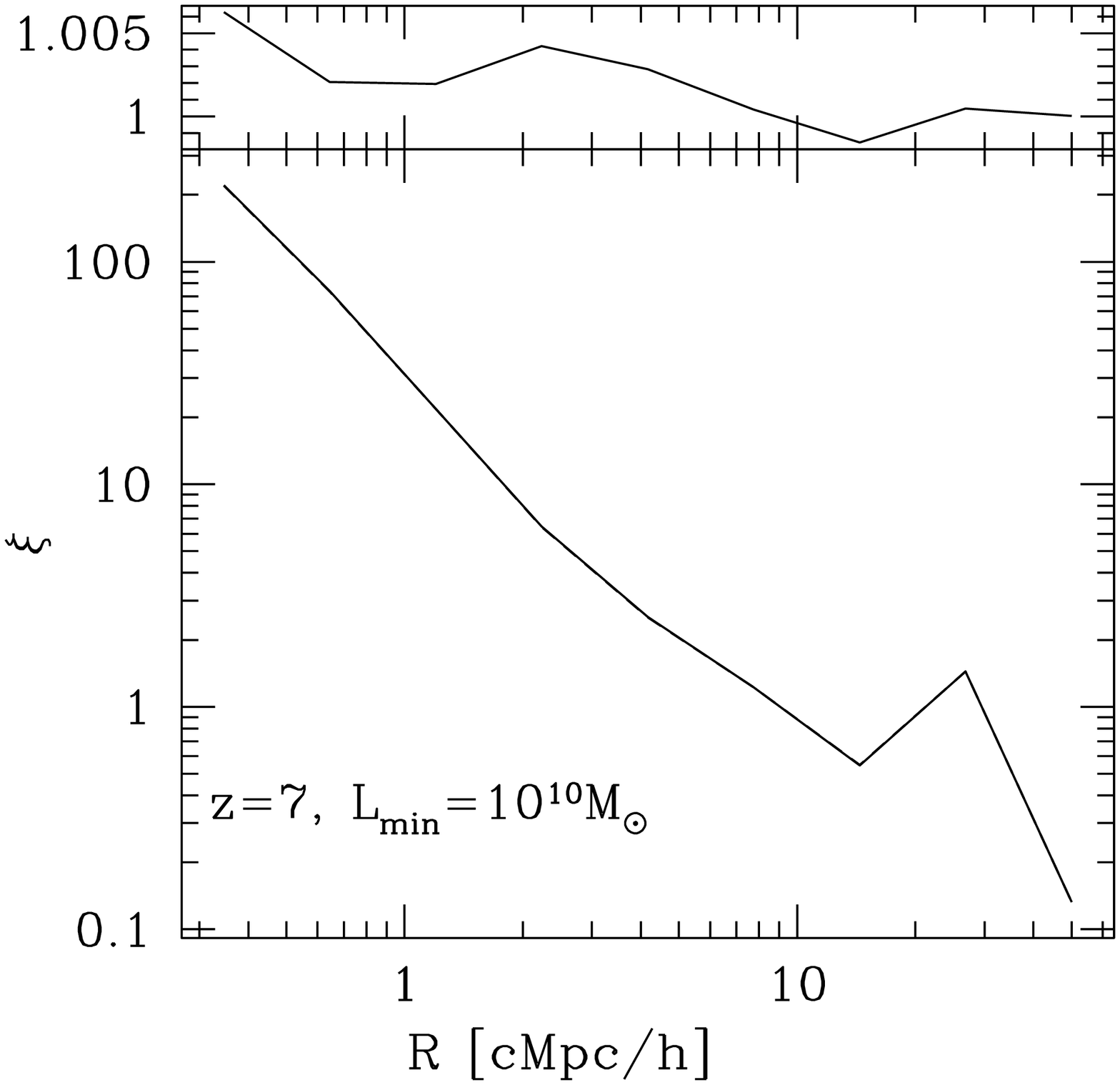}
\caption{Projection of the sources as seen by a mock flux-limited
  survey with $L>10^{10}M_\odot$ (1617 sources in total) at $z=7$ (left panel; 
  $x_m=0.94$) and sources with the same number density if the IGM 
  absorption were ignored (middle panel) and the 2-point 3D 
  correlation functions (right panels) of the distribution with IGM absorption 
  (solid) and without (dashed) and their ratio (top). 
\label{corr_z7}}
\end{figure*}

As we discussed above, the high-redshift haloes are highly clustered in
space, and Ly-$\alpha$ sources should be clustered as well. The latter
clustering has been recently observed at $z\sim5.7$
\citep{2007ApJS..172..523M}. An interesting question to ask is if the
absorption due to the surrounding IGM affects this clustering. If this were
the case, then measuring the correlation function of high-redshift Ly-$\alpha$
sources can give us information about the state of the IGM at that time.

We derive the correlation functions as follows. First we calculate the total
luminosity of each source with and without IGM absorption using the same
method as above, but instead of random LOS directions we only consider
parallel LOS, as would be seen by far away observer. For simplicity we
consider LOS parallel to the axes of our computational box. 
We mock a flux-limited survey by imposing a cutoff on the observed
luminosity. We compare the resulting correlation function to the one obtained
for the same number, but now of the brightest sources based on their intrinsic 
luminosity (i.e. the ones hosted by the most massive halos, thus ignoring IGM 
absorption in this latter case). We calculate the 3D correlation functions 
by direct summation over all pairs of halos as described in
\citet{1991ApJ...366..353M}. The results at redshift $z=9$ (with cutoff
$L_{\rm min}(M)=10^{9.5}M_\odot$) and $z=7$ (with cutoff
$L_{\rm min}(M)=10^{10}M_\odot$) are shown in Figs.~\ref{corr_z9} and
\ref{corr_z7}. In both cases the luminosity cutoffs were chosen so as to
maximize the difference in the correlation functions with and without IGM
absorption, while at the same time allowing for sufficient number of halos
above the cutoff to reduce the noise of the correlation.

In Figure~\ref{corr_z9} we show the projection at $z=9$ of the two source
distributions onto the Y-Z plane (left), and the corresponding 3D correlation
functions and their ratio (right). The projection shows that the two source
populations differ, but cluster in the same spatial regions. This visual
impression is confirmed by the correlation functions. We find that IGM
absorption introduces only small variations in the correlation of sources. The
difference is largest at small scales, for separations below 1 comoving
Mpc. Even there they never exceed 10\%. At intermediate scales, $R\sim2-10$~Mpc
the departures are up to 5\%. There are no appreciable differences at large
scales.  

At redshift $z=7$, close to overlap (Figure~\ref{corr_z7}) there are many 
more sources, even with the low luminosity cutoff raised to $10^{10}M_\odot$. 
The two source distributions remain different, but are still clustered in a 
very similar way, resulting in largely identical correlation functions, 
which never differ by more than 0.7\%.

The small effect of IGM absorption on the correlation function appears
counter-intuitive. The reason for this is that the sources at high redshift,
especially the most massive/luminous ones are strongly clustered and the
ionization field is closely correlated with the galaxy field. The luminous
sources are typically found in the inner parts of the largest ionized bubbles,
where they are typically unaffected by damping from the remaining (few)
neutral patches. This is also supported by the profiles in
Fig.~\ref{meanF_fig} which show the damping wing effect being important over
the same time interval for both luminous and average sources. The damping
affects average sources more strongly, since these are more often found closer
to a neutral patch than the more massive sources. However, this does not
change the correlation function significantly. Basically, the reionization
patchiness has little effect on the source clustering properties. The reason
is that while the IGM absorption diminishes the flux from all sources, the
same source clusters (although not necessarily the same individual sources)
are seen as would be without IGM absorption.  

At late times any clearly defined ionized regions have already disappeared and
the effect of IGM absorption is to replace some sources above the luminosity
cutoff with other ones essentially at random, due to small local variations of
the residual neutral fraction and the gas velocities. Therefore, the
reionization patchiness ultimately has little effect on the correlation
function. That of course does not mean that the source population is not
modified by the IGM - a flux-limited survey will see many fewer sources 
than if IGM absorption were not present, but the clustering properties of
those sources are almost the same as with no absorption for the same number
density of sources.  

Recently, \citet{2007MNRAS.381...75M} claimed that reionization patchiness 
has a significant effect on observed source clustering, in apparent
contradiction to our results 
\citep[see also][]{2006MNRAS.365.1012F,2008MNRAS.386.1990M}. However, they 
compared the clustering properties of Ly-$\alpha$ sources with and without 
(i.e. intrinsic) IGM absorption for {\it a fixed} luminosity cutoff, rather 
than at fixed number density of sources, as we did. When the IGM absorption 
is accounted for, the sources remaining above the imposed flux limit are 
many fewer than in the case without absorption and they are also much 
brighter on average, hosted by more massive halos. This naturally results 
in higher bias of the observed sources compared to all sources with 
intrinsic luminosity above that cutoff. However, it is not necessarily 
related to reionization patchiness. E.g. source bias will increase also 
if the small residual neutral fraction in the ionized IGM increases, 
boosting the IGM opacity and causing dimmer, less clustered sources
to fall below the luminosity cutoff, thus no neutral patches are required 
for this to happen. In fact, our simulation results show almost no neutral 
patches below $z\sim6.6$ (at which time the global mean ionized fraction 
by mass is below 1\%). Therefore, while our conclusions disagree with the 
ones of \citet{2007MNRAS.381...75M}, at least some of the differences 
could be attributed to the different comparisons we make, but some of the 
variations are possibly real, since \citet{2007MNRAS.381...75M} claimed
that the trend is present even for fixed number densities, albeit with 
no quantitative details. We would like to stress, however, that we observe
the same qualitative trend of enhanced clustering of Ly-$\alpha$ sources 
due to patchiness during the early stages of reionization, but the 
quantitative level of the effect is different, being much weaker in our 
case.  

There are a number of possible explanations. In particular, in our 
simulations the pre-absorption clustering between ionized regions and 
sources appears to be more important than in the simulations of 
\citet{2007MNRAS.381...75M}. There are also notable differences between 
our and their modeling of the Ly-$\alpha$ sources and the IGM absorption. 
\citet{2007MNRAS.381...75M} (as many other current studies do) assume 
complete resonance absorption of the blue side of the Ly-$\alpha$ line, 
and only study the suppression due to the IGM damping wing and do not 
include velocity effects in their analysis. They also assume a particular 
duty cycle for their Ly-$\alpha$ emitters (that only 25\% of halos host 
emitters), which we do not do in this work. \citet{2007MNRAS.381...75M} 
also ran simulations with different minimum source halo mass cutoffs, 
either significantly higher ($M_{\rm min}=4\times10^{10}M_\odot$) or 
significantly lower ($M_{\rm min}=10^{8}M_\odot$) than the one we have 
here ($M_{\rm min}=2.2\times10^{9}M_\odot$). It is possible, therefore, 
that our results are more relevant than these previous works, if the 
low-mass sources missing in our current simulation are in fact strongly
suppressed during the late stages of reionization due to Jeans-mass 
filtering. This can only be resolved conclusively by detailed simulations 
which actually follow the complicated radiative feedback effects on 
low-mass halos, which is well beyond the scope of this work. We have run 
several tests in order to try to understand these differences, following 
suggestions by the referee. One possibility we investigated was that 
our H~II regions 
size distribution is more strongly peaked and narrower than the one found 
in the above works, which, if it were the case might have explained some 
of the clustering differences. We found, however, that the H~II region 
size distributions derived from our simulations is in fair agreement with
the one from the \citet{2007MNRAS.381...75M} simulations, and hence this
offers no plausible explanation of the differences. We also compared the
Ly-$\alpha$ damping wing optical depth distributions for sources of 
different mass at a range of reionization stages, as derived by  
\citet{2008MNRAS.386.1990M} (their Figures 2 and 3) and again found no 
significant differences between our results and theirs. We conclude that
more detailed and direct comparisons will be required in order to evaluate 
and understand any differences between our results.  

\section{Summary and Conclusions}

We considered the effects which the reionizing IGM has on the observations of
high-redshift Ly-$\alpha$ sources. For this we utilized detailed structure
formation and radiative transfer simulations, which allowed us to evaluate 
many features which can only be studied by detailed simulations, as well as to
quantify better a number of previously-proposed effects. We followed the full
reionization history self-consistently and accounted for the actual source
distribution, neutral fraction, density and velocity fields.

We find that the density, neutral fraction and velocity fields are all highly
anisotropic, which results in large variations in the IGM transmission and
source visibility among different LOS. The velocity effects, both gas infall
and source peculiar velocity are most important for massive, luminous sources.
The most luminous sources are found in highest peaks of the density field,
which at late times are significantly overdense out to $\sim10$~comoving Mpc
(cMpc) and are surrounded by infall extending to $\sim20$~cMpc. 
The infall of gas blueshifts it in frequency space and results in significant
absorption on the red side of the line center, while the peculiar velocity of
the source itself can either alleviate or exacerbate this effect, depending on
the halo- and infall velocity alignment.

The spherically-averaged local density enhancement and gas infall have been
modelled analytically in approximate ways \citep{2004MNRAS.347...59B}, and
thus can be incorporated in semi-analytical models
\citep[e.g.][]{2007MNRAS.377.1175D}. However, such models are unable to
account for the strong intrinsic anisotropies of the neutral fraction, density
and velocity fields. The analytical and semianalytical models typically assume
spherical symmetry and full ionization inside the H~II regions, both of which
assumptions are quite unrealistic. 

The Ly-$\alpha$ lines we derive are generally asymmetric and vary hugely from
LOS to LOS. The luminous sources form at the highest density peaks and as a
consequence their line centers are always highly-absorbed even though their
proximity regions are very highly ionized, with typical neutral fractions
$x_{\rm HI}\sim10^{-5}-10^{-6}$. The luminous sources also more affected by
infall and exhibit more pronounced proximity region with higher transmission
of the blue wing of the line.  

High-redshift sources are strongly clustered around the high peaks of the
density field. The central source contributes the majority of the ionizing
flux only in its immediate vicinity, within 1-2 comoving Mpc. Beyond 
that distance the
ionizing flux is dominated by the fainter sources clustered around it. This
dominance is particularly strong at late times, when both many more sources
form and the ionized regions become larger, resulting in the fainter sources
contributing up to 2 orders of magnitude more photons than the central source.

Compared to single-source ionized bubbles, the larger H~II regions from
clustered sources diminish the effects from the damping wing of the
line. Nevertheless, these remain significant until fairly late (ionized mass
fraction $x_m=0.3-0.7$, which for the simulation considered here corresponds
to redshifts $z\sim9-8$). Interestingly, the average damping wing effect is
similar for luminous and typical sources, even though naiively one might
expect that damping could be weaker for the former, since they are typically
in the middle of large bubbles, away from the neutral patches, unlike the
fainter sources, which are more evenly distributed.

Both the mean IGM transmission and the typical photoionization rates we find
are high compared to observations at $z\sim6$, indicating that our adopted
source efficiencies are also high. The mean IGM transmissivity
decreases only slowly towards the higher redshifts and the GP transparency
occurs significantly after the actual overlap epoch. For the simulation
considered here overlap (defined as 1\% neutral fraction) occurs at $z=6.6$,
while average neutral fraction of $10^{-4}$ is reached only by $z=6$ and even
then relatively small fraction (few to 10\%) of the flux is transmitted.
By overlap the spectra start showing significant transmission gaps in the mean
IGM (i.e. away from the proximity region of a luminous source). 

We find that for a given number density of sources (e.g. as determined by
observations) the clustering of these sources depends only weakly on the IGM
absorption during reionization. As a consequence, the reionization patchiness
has little effect on the observed Ly-$\alpha$ source clustering, which implies
that source clustering is not a good indicator for reionization patchiness.    

Our derived luminosity function assuming constant mass-to-light ratio provides
an excellent match to the shape of the observed luminosity function at $z=6.6$
with faint-end slope of $\alpha=-1.5$. The resulting mass-to-light ratio 
implies that the majority of sources responsible for reionization are too
faint to be observed by the current surveys.  

\section*{Acknowledgments} 

We thank Hugo Martel for letting us use and modify his correlation function
code and X. Fan for useful discussions. This work was partially supported by
NASA Astrophysical Theory Program grants NAG5-10825 and NNG04G177G, Swiss 
National Science Foundation grant 200021-116696/1, and Swedish Research 
Council grant 60336701.

\bibliography{lum}

\end{document}